# Electron-Electron Interactions, Magnetism, and Superconductivity: Lecture Notes for Introduction to Solid State Physics

Daniel C. Ralph

Physics Department, Cornell University, Ithaca NY
and Kavli Institute at Cornell for Nanoscale Science, Ithaca, NY

These lecture notes were developed for a portion of the Cornell University course Phys 7635, a one-semester graduate-level introduction to solid state physics. Relative to standard textbook treatments, I have attempted to unify and update the discussion of magnetism and superconductivity, and to include connections to recent research. The level is meant to be comparable to that of the textbook by Ashcroft & Mermin.

The notes contain seven sections:
  I. An Attempt to Solve the Interacting-Electron Problem Exactly
  II. The Stoner Instability and the Origin of Ferromagnetism
  III. More About Magnets + Their Applications
  IV. Even More About Magnets + Their Applications
  V. Attractive effective electron-electron interactions and how they can destabilize
      the Fermi liquid state (the Cooper Instability)
  VI. Superconductivity: Basic Properties
  VII. More About Superconductors + Their Applications

A few related homework problems are attached at the end.

The contents of the first section set the stage for the model Hamiltonians to be analyzed in the later sections, but this content is not actually essential to understanding the rest if one is happy just to assume the applicability of the model Hamiltonians. Students interested in getting right to the properties of magnets and/or superconductors are welcome to jump ahead.

Given that these are lecture notes, I have tried to make them self-contained, keeping the number of external references to a minimum.

Previously I have posted the lecture notes from a different portion the same course, presenting an introduction to Berry curvature, semiclassical electron dynamics, and topological materials.[1]

I will be grateful for corrections or suggestions for improvement. Please send them to me at dcr14@cornell.edu.

**Some Preliminaries**:

The lectures in these notes are generally given after a brief, qualitative discussion of Fermi Liquid Theory for metals. I ask students to envision solving for the low-lying energy excitations above the ground state for a many-electron metal in the absence of any electron-electron interactions, and then imagine that the students have the ability to turn a knob and gradually increase the strength of the interactions from zero. What happens?

In general, there are two classes of possibilities. One possible outcome is that not much changes – it is possible that the low-lying excitations might just evolve in some continuous and adiabatic way. Some energy levels might cross or undergo avoided crossings, but still it might be possible to make a one-to-one correspondence between the low-lying excited states of the non-interacting system and the excited states of the interacting one. In this case the interacting system will still behave like a Fermi liquid, and the low-energy excitations can be described in terms of effectively weakly-interacting screened quasiparticles. The properties of the quasiparticles might be renormalized relative to completely non-interacting electrons, possibly by a large amount (e.g., effective mass, compressibility, spin susceptibility, etc. might change), but the overall system will still behave like an ordinary metal with specific heat $\propto T$ at low temperature, Pauli paramagnetism independent of $T$, non-zero resistivity associated with quasiparticle scattering, a quasiparticle-quasiparticle scattering rate $\propto T^2$, etc.

The other possibility is that the low-lying excited states might *not* evolve continuously. As the strength of interactions is tuned, at some particular value of the interaction strength there might be a sudden change in the excitation spectrum, so that there is no way to identify all of the low-lying excited states of the interacting system as part of some smooth evolution from a non-interacting system. There could be a proliferation of energy levels, the appearance of gaps where there were none before, or other weirdness. Typically, this type of sudden change in system properties reflects an instability in which the ground state of the many-electron system changes into some new electronic phase which might have properties very different from a non-interacting Fermi liquid. Understanding these new electronic phases will require theoretical approaches very different from just doing an accounting based on the filling of effectively independent, non-interacting electron states.

These lecture notes explore two common types of instabilities in metals due to electron-electron interactions, the instabilities that lead to the formation of ferromagnetism and superconductivity. I will also discuss some of the useful physical properties associated with these very interesting many-electron quantum states.

## I. An Attempt to Solve the Interacting-Electron Problem Exactly[2]

*Topic: To motivate an approximate model Hamiltonian that we will use later for analyzing magnetic and superconducting instabilities.*

What sort of instabilities can be caused by electron-electron interactions in a metal, and what are the properties of the resulting new electronic ground states of the interacting electrons?

The approach we will attempt in this lecture is to try to solve the interacting electron property exactly by brute-force diagonalization of the Hamiltonian. It will turn out that this is impractical, but it will provide us with insights that will allow further progress. Please think of this process as a "thought calculation" in analogy with similarly-impractical "thought experiments" which can nevertheless provide useful insights.

We will try to solve the interacting-electron problem by just following the usual procedures in quantum mechanics:

1. **Set up a complete set of states to use as a basis.**
2. **Evaluate the matrix elements of the Hamiltonian in this basis.**
3. **Diagonalize the Hamiltonian matrix.**

It will turn out that we will be able to get almost all of the way through this process. With some work, steps 1 and 2 can be completed, and the only difficulty that we will not be able to overcome in the final step 3 is that the matrices we will need to diagonalize will be too large. Nevertheless, we will learn a lot about the consequences of electron interactions. In particular, we will see that the matrix elements of the Hamiltonian can be classified into a hierarchy according to their size – almost all of the matrix elements will be exactly zero, most of the rest will be small, and only a very tiny subset will be bigger than all of the others. Remarkably, in the case of a good metal, only three parameters will be required to specify all of the large matrix elements, in what I will call a "universal Hamiltonian"! This hierarchy will provide a way to make systematic approximations, and to understand the most important consequences of magnetic and superconducting instabilities using a simple model Hamiltonian that takes a universal form.

As a consequence, if we think about the problem in terms of the Landau-Fermi Liquid picture, we will see that for a good metal that remains spatially homogeneous there will be only a relatively small number of ways that electron-electron interactions among free-electron-like states can destabilize the Fermi Liquid ground state and cause phase transitions to very different non-Fermi-liquid states.

### Step 1. A complete set of states to use as a basis -- Slater Determinants and 2nd-Quantized Notation:

Due to electron-electron interactions, the electrons in a metal are not independent, so one cannot consider only simple independent single-electron states. The quantum-mechanical wavefunctions should include a description of all of the electrons in the sample simultaneously, so the wavefunctions must be functions of the coordinates of all of the electrons. Moreover, because electrons are fermions, the many-electron state must be antisymmetric upon exchange of electron coordinates. These two requirements can be satisfied by choosing the many-electron basis states to be Slater determinants composed of non-interacting single-electron basis states. To do so, we first need to define a complete set of single-electron wavefunctions $\{\psi_{k\sigma_k}(r)\}$, where $\sigma_k$ is a label denoting spin up or spin down so that $\psi_{k\sigma_k}(r) = \psi_k(r)|\!\uparrow\rangle$ or $\psi_k(r)|\!\downarrow\rangle$. (Note for experts: this form assumes that spin-orbit interactions are negligible, but the arguments I will give can also be rephrased in ways that allow spin-orbit interactions to be included.) We will imagine making this definition such that our single-electron basis states provide the best possible solution to the one-particle effective Hamiltonian, treating the electronic interactions in effectively a mean field approximation. (For example, using states derived using density functional theory, or solutions of a tight-binding model with parameters chosen to fit ARPES measurements.) For metals, once we begin treating the electronic interactions beyond mean field these basis states will correspond to wavefunctions for weakly-interacting quasiparticles with short-ranged screened interactions. For a good crystal with lattice translation symmetry, natural choices for the single-electron basis states might include either Bloch states with periodic boundary conditions, so that $k$ is a wavevector (and we will use these basis states later when we discuss superconductivity) or perhaps localized Wannier states associated with the single-particle band structure. However, to keep today's discussion as simple as possible, one can just as well imagine that the sample we are dealing with is a finite and disordered piece of metal so that the single-electron basis states will be standing-wave states that are effectively randomly fluctuating as a function of position. If spin-orbit coupling is negligible and there is no applied magnetic field, these standing-wave states can be chosen to be real-valued, so that the effectively-random oscillations are to positive and negative real values (and the oscillations are effectively uncorrelated between different states, except that the states remain mutually orthogonal). In this case, $k$ is just a label for the orbital state. The set of basis states should include all states within a relevant band, not just some subset of filled states, and if interaction strengths are large compared to band gaps one might need to include multiple bands.

After the single-electron basis states are defined, we can define a set of $n$-electron Slater determinant states each of which has the form

$$\Psi_A(r_1, r_2, \ldots, r_n) = \frac{1}{\sqrt{n!}} \begin{vmatrix} \psi_{a\sigma_a}(r_1) & \psi_{a\sigma_a}(r_2) & \ldots & \psi_{a\sigma_a}(r_n) \\ \psi_{b\sigma_b}(r_1) & \psi_{b\sigma_b}(r_2) & \ldots & \psi_{b\sigma_b}(r_n) \\ \ldots & & & \ldots \\ \psi_{g\sigma_g}(r_1) & \psi_{g\sigma_g}(r_2) & \ldots & \psi_{g\sigma_g}(r_n) \end{vmatrix}. \quad (1.1)$$

The $n$ single-electron states $\psi_{a\sigma_a}$, $\psi_{b\sigma_b}$, $\cdots$ $\psi_{g\sigma_g}$, etc., are chosen from among the set of single-electron basis states. The vertical lines on the outside of the matrix in Eq. (1.1) mean that you should take the determinant. The set of all possible $n$-electron Slater determinants formed by combining the single-electron basis states in all possible combinations forms a complete basis set for determining the true $n$-electron wavefunctions -- that is, we will be able to write any $n$-electron wavefunction as a linear superposition of these Slater determinant states.

How many independent Slater determinant basis states will there be? If there are $N$ states in the single-electron basis and $n$ electrons in the system whose interacting wavefunctions you want to specify, then there are $N!/[(N-n)!n!]$ possible ways to choose the combinations of single-particle states out of which to form the Slater determinants. This is a big number! For a problem with 20 single-electron basis states that you might want to fill with 10 electrons, the number of Slater-determinant 10-electron basis states is 184,756. This means that for just these 10 electrons, to solve for the many-body wavefunctions you would need to diagonalize a 184,756 ×184,756 matrix. For 30 single-electron basis states and 15 electrons, the matrix is approximately 160 million × 160 million. For a small piece of metal with volume (1 mm)$^3$, $n \sim 10^{20}$ and assuming $N \sim 2n$ the matrix to diagonalize has size approximately $10^{50,000,000,000,000,000,000}$ × $10^{50,000,000,000,000,000,000}$, which is unimaginably huge (the total number of atoms in the universe is estimated to be less than only about $10^{90}$!). Except for this difficulty, solving the interacting electron problem would be simple.

*A Handy Alternative Notation for Slater Determinant States*:

To within a sign, a Slater determinant state is completely determined just by saying what list of single-electron states goes into the determinant. This suggests a labeling based on raising operators, such that the same Slater determinant state written above might be specified as

$$\Psi_A = c^\dagger_{a\sigma_a} c^\dagger_{b\sigma_b} \cdots c^\dagger_{g\sigma_g} |vacuum\rangle. \tag{1.2}$$

To fix the sign, one normally specifies some conventional order in which the raising operators are applied. In order to function as raising operators, the $c$'s corresponding to the same single-electron state $a$ must satisfy the usual relation

$$c_a c^\dagger_a = 1 - c^\dagger_a c_a. \tag{1.3}$$

(Here, to save writing, I am specifying both the orbital and spin state with the same label.) Furthermore, in order that the many-body wavefunction should change sign upon exchange of electrons, it turns out that the raising operators must anticommute with other raising operators, the lowering operators must anticommute with lowering operators, and a lowering operator corresponding to one state $a$ must anticommute with a raising operator corresponding to any different state $b$.

$$\begin{aligned} c^\dagger_a c^\dagger_b &= -c^\dagger_b c^\dagger_a, \\ c_a c_b &= -c_b c_a, \\ c_a c^\dagger_b &= -c^\dagger_b c_a \text{ if } a \neq b. \end{aligned} \tag{1.4}$$

(For raising and lowering operators corresponding to the same state, Eq. (1.3) applies.) See a nice discussion by Ambegaokar for the details.[3] The reason for writing the Slater determinants in the "2$^{nd}$-quantized" form of Eq. (1.2) is that it is then possible to evaluate the form of the matrix elements between Slater determinant states by simply manipulating the anticommutation rules of the raising and lowering operators, rather than by multiplying the Slater determinants out and solving $n!$ $3n$-dimensional integrals. You will see the virtue of the simpler approach as we go along.

The important thing to keep in mind at this stage is that the many-body basis states of the form written in Eq. (1.2) using the raising operators are the full Slater determinant states. They are not simply a product state of single-electron wavefunctions.

## Step 2. Evaluate the matrix elements of the electron-electron interaction between the Slater determinant basis states

*A Practice Exercise: Matrix elements for a 1-body potential energy*

Ultimately we will care about the matrix elements for the 2-body electron-electron interactions. This will involve summing the interactions between all possible pairs of electrons:

$$\langle \Psi_A | V_2 | \Psi_B \rangle \equiv \langle \Psi_A | \frac{1}{2} \sum_{\substack{i,j=1 \\ i \neq j}}^{n} V_{ee}(r_i - r_j) | \Psi_B \rangle,$$ where the factor of 1/2 is to avoid double-counting.

However, as a warmup exercise, let's think about evaluating the matrix elements of a simpler 1-body interaction (for example, a potential energy term). Since all of the electrons are identical, this must enter in the form $V_1 = \sum_{i=1}^{n} V(r_i)$. The matrix element of this potential between two Slater determinant states $\Psi_A$ and $\Psi_B$ is

$$\langle \Psi_A | V_1 | \Psi_B \rangle = \frac{1}{n!} \sum_{i=1}^{n} \int d^3 r_1 \cdots d^3 r_n \begin{vmatrix} \psi^*_{a\sigma_a}(r_1) & \cdots & \psi^*_{a\sigma_a}(r_n) \\ \cdots & \cdots & \cdots \\ \psi^*_{g\sigma_g}(r_1) & \cdots & \psi^*_{g\sigma_g}(r_n) \end{vmatrix} V(r_i) \begin{vmatrix} \psi_{h\sigma_h}(r_1) & \cdots & \psi_{h\sigma_h}(r_n) \\ \cdots & \cdots & \cdots \\ \psi_{w\sigma_w}(r_1) & \cdots & \psi_{w\sigma_w}(r_n) \end{vmatrix}. \quad (1.5)$$

In writing the integrals over spatial coordinates, I wish to imply that one should also contract over the spin parts of the wavefunctions, too, in computing the matrix element. (Note, if we were to consider expectation values for the sum of the kinetic energies and $V_1$ then $\langle \Psi_A | KE + V_1 | \Psi_A \rangle$ would be just the sum of single-electron energies for the occupied states, but we are not considering the kinetic energies here.)

Let's think about expanding the determinants and doing the integrals over the single-electron wavefunctions. Each of the integrals involves one integration over the variable $r_i$ with an integrand of the form $\psi^\dagger_{a\sigma_a}(r_i) V(r_i) \psi_{q\sigma_q}(r_i)$ and $n-1$ integrations over the other variables for which the integrand is simply a product of wavefunctions, $\psi^\dagger_{b\sigma_b}(r_j) \psi_{k\sigma_k}(r_j)$, with no potential energy term. For the integrals not involving $V(r_i)$, the single-particle wavefunctions within $\Psi_A$ and $\Psi_B$ must all be the same, or else there will be an integration over orthogonal single-particle wavefunctions, which causes the matrix element to be zero. This means that in order for the matrix element to be nonzero, $\Psi_A$ and $\Psi_B$ must be composed of the same set of single particle states, with at most one change, say $\psi_{d\sigma_d} \to \psi_{p\sigma_p}$. This one switch is allowed, because the one integral containing $V(r_i)$ can be non-zero even when the two wavefunctions in the integrand are different. In this case, if we evaluate all of the integrals in Eq. (1.5) except the one over $V(r_i)$, then

$$\langle \Psi_A | V_1 | \Psi_B \rangle = \frac{1}{n} \sum_{i=1}^{n} \int d^3 r_i \psi^\dagger_{p\sigma_p}(r_i) V(r_i) \psi_{d\sigma_d}(r_i), \text{ assuming } d \neq p \quad (1.6)$$

(because all of the other integrals, in (n-1)! combinations, give 1)

$$= \int d^3 r \psi^\dagger_p(r) V(r) \psi_d(r) \langle \sigma_p | \sigma_d \rangle \equiv V_{dp} \delta_{\sigma_d \sigma_p}. \quad (1.7)$$

In this last line I have explicitly written down the contraction over spin that must be included to evaluate the matrix element, in addition to the integral over the spatial part of the wavefunction.

I have also assumed that the interaction $V(r)$ is not spin-dependent, which means that the spin of the state $p$ must be the same as the spin of the state $d$ if the matrix element is to be nonzero. The last equality defines the coefficient $V_{dp}$.

If $\Psi_A$ and $\Psi_B$ contain exactly the same single-particle states, then by the same sort of analysis one can see that for the diagonal matrix elements

$$\langle \Psi_A | V_1 | \Psi_A \rangle = \sum_{\substack{\text{occupied} \\ \text{states } a}} V_{aa}, \tag{1.8}$$

where the sum is over all of the single-electron states that are occupied in $\Psi_A$ and $V_{aa}$ is defined in analogy with the definition in Eq. (1.7).

*Expressing the same answers in 2nd quantized notation*:

Next I wish to show that these results are easier to think about using 2nd quantized notation. The key point is that it is possible to express the 1-body interaction in an operator form

$$V_1 = \sum_{\substack{j,k \\ \sigma_j, \sigma_k}} V_{jk} c^\dagger_{j\sigma_j} c_{k\sigma_k}. \tag{1.9}$$

What I mean by "express the 1-body interaction in an operator form" is that if we use this form for $V_1$ and apply the anticommutation relations for the raising and lowering operators when evaluating the matrix elements between Slater determinant states, we can get exactly the same final answers as we do by expanding out the full Slater determinants. There is also a nice intuitive way to think about this form of the interaction -- it describes scattering events. The operator destroys an electron in state $k$ and creates one in state $j$, or in other words scatters an electron from state $k$ to state $j$ (while summing over all states $j$ and $k$ with a weighting given by the coefficient $V_{jk}$).

Let us verify that this form of the interaction does indeed give the same matrix elements between Slater determinant states as before. In 2nd quantized notation,

$$\langle \Psi_A | V_1 | \Psi_B \rangle = \langle \text{vacuum} | c_{g\sigma_g} \cdots c_{b\sigma_b} c_{a\sigma_a} \left( \sum_{\substack{j,k \\ \sigma_j, \sigma_k}} V_{jk} c^\dagger_{j\sigma_j} c_{k\sigma_k} \right) c^\dagger_{h\sigma_h} \cdots c^\dagger_{w\sigma_w} | \text{vacuum} \rangle \tag{1.10}$$

$$= \sum_{\substack{j,k \\ \sigma_j, \sigma_k}} V_{jk} \langle \text{vacuum} | c_{g\sigma_g} \cdots c_{b\sigma_b} c_{a\sigma_a} c^\dagger_{j\sigma_j} c_{k\sigma_k} c^\dagger_{h\sigma_h} \cdots c^\dagger_{w\sigma_w} | \text{vacuum} \rangle.$$

The matrix elements inside the sum can be nonzero only if each lowering operator is matched with a raising operator, state by state. If there is a lowering operator that is unmatched, we can anticommute it through all the other operators to act on the vacuum state, and this gives zero. Since the operators contributed by the 1-body interaction change at most one state (destroying state $k$ and creating $j$), the single particle levels contributing to the many body states $\Psi_A$ and $\Psi_B$ can differ by at most one state. If $\Psi_A$ contains state $d$ while $\Psi_B$ contains state $p$, but otherwise the two many-body states are the same, then the only way to have a nonzero matrix element is to have $k = p$ and $d = j$ (and $\sigma_k = \sigma_p$, $\sigma_d = \sigma_j$). We can eliminate pairs of operators by anticommuting and using Eq. (1.3) (I will talk more about doing this in class), so that if $\Psi_A$ and $\Psi_B$ are not exactly the same states,

$$\langle \Psi_A | V_1 | \Psi_B \rangle = \sum_{\substack{j,k \\ \sigma_j, \sigma_k}} V_{jk} \langle vacuum | c_{d,\sigma_d} c^\dagger_{j,\sigma_j} c_{k,\sigma_k} c^\dagger_{p,\sigma_p} | vacuum \rangle = \sum_{\substack{j,k \\ \sigma_j, \sigma_k}} V_{jk} \delta_{dj} \delta_{kp} \delta_{\sigma_d \sigma_j} \delta_{\sigma_k \sigma_p} = V_{dp} \delta_{\sigma_d \sigma_p}. \quad (1.11)$$

Likewise, if $\Psi_A$ and $\Psi_B$ contain exactly the same states, we have for the diagonal matrix elements

$$\langle \Psi_A | V_1 | \Psi_A \rangle = \sum_{\substack{j,k \\ \sigma_j, \sigma_k}} V_{jk} \langle vacuum | c_{g,\sigma_g} \cdots c_{a,\sigma_a} c^\dagger_{j,\sigma_j} c_{k,\sigma_k} c^\dagger_{a,\sigma_a} \cdots c^\dagger_{g,\sigma_g} | vacuum \rangle. \quad (1.12)$$

This will be nonzero only if $k$ and $\sigma_k$ match one of the states in $\Psi_A$, because otherwise we could anticommute $c_{k,\sigma_k}$ past all of the raising operators to the right of $c_{k,\sigma_k}$ in Eq. (1.12), until we have the lowering operator acting directly on the vacuum state, giving zero. Also, only when $j = k$ and $\sigma_j = \sigma_k$ will all of the raising and lowering operators in Eq. (1.12) be matched in pairs, so as to allow for a nonzero matrix element. Consequently,

$$\langle \Psi_A | V_1 | \Psi_A \rangle = \sum_{\substack{occupied \\ states\ k}} V_{jk} \delta_{jk} \cdot 1 = \sum_{\substack{occupied \\ states\ k}} V_{kk}. \quad (1.13)$$

Therefore, when using the 2nd-quantized operators, we do indeed get the same matrix elements between Slater determinant states as before. (Compare Equations (1.11) and (1.13) to Equations (1.7) and (1.8).)

*Matrix elements for electron-electron interactions:*

Next let's consider the case that we really care about, that of 2-body electron-electron interactions:

$$V_2 = \frac{1}{2} \sum_{\substack{i,j=1 \\ i \neq j}}^n V_{ee}(r_i - r_j). \quad (1.14)$$

We will think about this interaction being the residual interaction between screened quasiparticles, so that it will be short-ranged and spin-independent.

If we write out the matrix element for this interaction between the two Slater determinant states $\Psi_A$ and $\Psi_B$ in the fully-expanded (clumsy) notation, we have

$$\langle \Psi_A | V_2 | \Psi_B \rangle = \frac{1}{n!} \sum_{\substack{i,j=1 \\ i \neq j}}^n \int d^3 r_1 \cdots d^3 r_n \begin{vmatrix} \psi^*_{a,\sigma_a}(r_1) & \cdots & \psi^*_{a,\sigma_a}(r_n) \\ \cdots & & \cdots \\ \psi^*_{g,\sigma_g}(r_1) & \cdots & \psi^*_{g,\sigma_g}(r_n) \end{vmatrix} V_{ee}(r_i - r_j) \begin{vmatrix} \psi_{h,\sigma_h}(r_1) & \cdots & \psi_{h,\sigma_h}(r_n) \\ \cdots & & \cdots \\ \psi_{w,\sigma_w}(r_1) & \cdots & \psi_{w,\sigma_w}(r_n) \end{vmatrix}. \quad (1.15)$$

Now the integrals will be nonzero only if $\Psi_A$ and $\Psi_B$ are the same except for at most **2** single-particle states different, since the two integrals over $r_i$ and $r_j$ can be nonzero even when evaluated between orthogonal single-particle states.

In 2nd quantized form, this interaction can therefore be written

$$V_2 = \frac{1}{2} \sum_{\substack{j,k,m,n \\ \sigma_j \sigma_k \sigma_m \sigma_n}} V_{jkmn} c^\dagger_{k,\sigma_k} c^\dagger_{j,\sigma_j} c_{m,\sigma_m} c_{n,\sigma_n}. \tag{1.16}$$

The full proof of this identification is given in the write-up by Ambegaokar. The form of the interaction in Eq. (1.16) can be understood intuitively in terms of scattering events in which two electrons initially in states *m* and *n* scatter into states *j* and *k* due to the electron-electron interaction. The coefficient in Eq. (1.16) is defined as

$$V_{jkmn} = \int d^3r_1 d^3r_2 \psi_k^*(r_2)\psi_j^*(r_1) V_{e-e}(r_1 - r_2) \psi_m(r_1)\psi_n(r_2) \langle \sigma_j | \sigma_m \rangle \langle \sigma_k | \sigma_n \rangle. \tag{1.17}$$

In this expression I have written out the spin contractions explicitly, and I have used that the electron-electron interacting is independent of the electron's spin, so that the matrix element is non zero only if both electron states associated with coordinate $r_1$ have the same spin ($\sigma_j = \sigma_m$, let us call this spin *s*) and likewise both electron states associated with coordinate $r_2$ have the same spin ($\sigma_k = \sigma_n$, let us call this spin $\sigma$). The values of *s* and $\sigma$ can be the same or different.

If we include these spin indices explicitly and write out the orbital parts of the single-electron wave functions, we have

$$V_2 = \frac{1}{2} \sum_{\substack{j,k,m,n \\ s,\sigma}} V_{jkmn} c^\dagger_{k,\sigma} c^\dagger_{j,s} c_{m,s} c_{n,\sigma} \tag{1.18}$$

and

$$V_{jkmn} = \int d^3r_1 d^3r_2 \psi_k^*(r_2)\psi_j^*(r_1) V_{e-e}(r_1 - r_2) \psi_m(r_1)\psi_n(r_2). \tag{1.19}$$

From this last expression, we can begin to see why the values of the matrix elements $\langle \Psi_A | V_2 | \Psi_B \rangle$ take on a hierarchy of values. For almost all choices of $\Psi_A$ and $\Psi_B$, the two many-body wavefunctions will differ in terms of their single-particle content by more than two single-electron states, so the 2-body matrix element will be identically zero. Even when $\Psi_A$ and $\Psi_B$ differ in their single-particle content by only two states or less, the integrand in Eq. (1.19) will generally be highly oscillatory, oscillating strongly to both positive and negative values, so when the integrals are performed there will be lots of cancellations and the end result will be small. However, if the states $\psi_j$, $\psi_k$, $\psi_m$, and $\psi_n$ are chosen so that they occur only in pairs of the form $\psi_a^*(r)\psi_a(r) = |\psi_a(r)|^2$, then it is possible to make the overall product in the integrand in Eq. (1.19) assume a definite sign, and for these special cases we will not have the same sorts of cancellations between positive and negative contributions when performing the integrals. In the end, this small subset of matrix elements will have *much* larger values, and therefore they will generally be much more important, than all of the other matrix elements for which the wavefunctions are not matched pairwise.

It is easiest to see how this comes about by assuming the limit of a very short-ranged screened interaction within a metal, so that $V_{e-e}(r_1 - r_2) \approx V_0 \delta(r_1 - r_2)$, where $V_0$ is a numerical prefactor. (This is the first place in our discussion where the form of the interaction potential enters at all.) This delta-function form is not a crazy approximation, since the screening length in a metal is typically $\sim 1/k_F$, the same scale as the electron wavelength, so to a reasonable approximation the range of the interaction is short compared to the length scale over which the wavefunctions vary. For a $\delta$-function interaction, Eq. (1.19) becomes

$$V_{jkmn} = V_0 \int d^3r_1 \psi_k^*(r_1)\psi_j^*(r_1)\psi_m(r_1)\psi_n(r_1). \qquad (1.20)$$

From this we can see clearly that if we choose the electron wavefunctions pairwise, we can be guaranteed to make the integrand positive definite, and get a large matrix element. It turns out that there are 3 different pairwise combinations that accomplish this, each with a different physical interpretation:

(a) We can choose $m = j$, $n = k$, so that the matrix element becomes
$V_{jkjk} = V_0 \int d^3r |\psi_j(r)|^2 |\psi_k(r)|^2 \equiv K_{jk}$. We will call these the **direct**, or **Hartree** terms.

(b) We can choose $n = j$, $m = k$ so that the matrix element is
$V_{jkkj} = V_0 \int d^3r |\psi_j(r)|^2 |\psi_k(r)|^2 = K_{jk}$. (The same value as in (a) for a $\delta$-function interaction but not the same for a general interaction.) We will call these the **exchange**, or **Fock** terms.

(c) If states $j$ and $k$ are chosen in a pair related by time-reversal symmetry and likewise states $m$ and $n$ are a pair related by time-reversal symmetry then
$\psi_k^*(r) = \psi_j(r)$ and $\psi_m(r) = \psi_n^*(r)$, so that $V_{jjmm} = V_0 \int d^3r |\psi_j(r)|^2 |\psi_m(r)|^2 = K_{jm}$.
(Again the same as in (a) and (b) for a $\delta$-function interaction, but not the same in general.) We will call these the **pairing** or **BCS** terms, for reasons to be explained later (BCS for Bardeen-Cooper-Schrieffer). If one chooses to use single-electron basis states that are standing waves, then (when there is no applied magnetic field or spin orbit interactions) the wavefunctions are real, so that the time-reversed pairs consist of spin-up and spin-down electrons with the same orbital wavefunction. In the presence of spin-orbit interactions, the wavefunctions are complex, and it is important to keep track of the complex conjugations, but time-reversal will still be a good symmetry and consequently the arguments that I give will still be valid. When there is an applied magnetic field that breaks time-reversal invariance, the pairing matrix elements can be forced to be very small.

Before moving on to consider the physical consequences of these three classes of large matrix elements, it is important to emphasize that these different classes of matrix elements all come from the same source -- the spin-independent screened electron-electron interaction. Nevertheless, we will see that we get some non-intuitive consequences, for instance we will find that the spin-independent interaction can lead to eigenstates whose energy nevertheless depends on their total spin. This comes about as a consequence of taking matrix elements of the same interaction between different Slater determinant states.

Let us consider in more detail these three different classes of large matrix elements, for the general case of a short-ranged screened Coulomb interaction within a metal (but not necessarily a $\delta$-function interaction):

### *Direct, or Hartree Terms*:

With the choice $m = j$, $n = k$, Eqs. (1.18) and (1.19) become

$$V_{direct} = \frac{1}{2} \sum_{\substack{j,k \\ s,\sigma}} V_{jkjk} c^{\dagger}_{k,\sigma} c^{\dagger}_{j,s} c_{j,s} c_{k,\sigma} \tag{1.21}$$

$$\begin{aligned} V_{jkjk} &= \int d^3r_1 d^3r_2 \psi^*_k(r_2) \psi^*_j(r_1) V_{e-e}(r_1 - r_2) \psi_j(r_1) \psi_k(r_2) \\ &= \int d^3r_1 d^3r_2 |\psi_j(r_1)|^2 V_{e-e}(r_1 - r_2) |\psi_k(r_2)|^2. \end{aligned} \tag{1.22}$$

This is the matrix element one might write down simply by considering a Coulomb interaction between the local charge densities of two interacting electrons.

Anticommuting the operators in Eq. (1.21) twice, it can be rewritten as

$$V_{direct} = \frac{1}{2} \sum_{\substack{j,k \\ s,\sigma}} V_{jkjk} c^{\dagger}_{k,\sigma} c_{k,\sigma} c^{\dagger}_{j,s} c_{j,s} = \frac{1}{2} \sum_{\substack{j,k \\ s,\sigma}} V_{jkjk} n_{k,\sigma} n_{j,s}, \tag{1.23}$$

where the $n$'s are number operators. Since $V_{direct}$ can be written entirely in terms of number operators, it contributes only to diagonal matrix elements of the form $\langle \Psi_A | V_{direct} | \Psi_A \rangle$, not to any off-diagonal matrix elements between different Slater determinant states.

If these Hartree terms are strong enough, they can cause a breakdown in Fermi Liquid theory in the form of localization and the formation of a Mott insulator.

### *Exchange, or Fock Terms*:

With the choice $n = j$, $m = k$, Eqs. (1.18) and (1.19) become

$$V_{exchange} = \frac{1}{2} \sum_{\substack{j,k \\ s,\sigma}} V_{jkkj} c^{\dagger}_{k,\sigma} c^{\dagger}_{j,s} c_{k,s} c_{j,\sigma} . \tag{1.24}$$

$$\begin{aligned} V_{jkkj} &= \int d^3r_1 d^3r_2 \psi^*_k(r_2) \psi^*_j(r_1) V_{e-e}(r_1 - r_2) \psi_k(r_1) \psi_j(r_2) \\ &= \int d^3r_1 d^3r_2 \psi^*_j(r_1) \psi_k(r_1) V_{e-e}(r_1 - r_2) \psi^*_k(r_2) \psi_j(r_2) \end{aligned} \tag{1.25}$$

When $V_{e-e}$ is not exactly a $\delta$-function interaction this can no longer be written as simply an interaction between local charge densities, as in the Hartree form. The coordinates of two of the wavefunctions have been exchanged. Nevertheless, this still has a large value when evaluating the matrix elements of the Coulomb interaction between Slater determinant states.

With one anticommutation of operators in Eq. (1.24),

$$V_{exchange} = -\frac{1}{2} \sum_{\substack{j,k \\ s,\sigma}} V_{jkkj} c^{\dagger}_{k,\sigma} c_{k,s} c^{\dagger}_{j,s} c_{j,\sigma} . \tag{1.26}$$

The exchange terms therefore contribute to both diagonal matrix elements in the Hamiltonian when $s = \sigma$ (because in this case the combinations of raising and lowering operators reduce to simple number operators), and off-diagonal matrix elements when $s \neq \sigma$.

The neatest thing about the exchange matrix elements is that they lead to spin-dependent energies, even though the underlying Coulomb interaction is spin-independent. To see how this happens, consider the diagonal matrix element

$$\langle \Psi_A | V_{exchange} | \Psi_A \rangle = -\frac{1}{2}\sum_{\substack{j,k\\s}} \langle \Psi_A | V_{jkkj} c^\dagger_{k,s} c_{k,s} c^\dagger_{j,s} c_{j,s} | \Psi_A \rangle = -\frac{1}{2}\sum_{\substack{j,k\\s}} V_{jkkj} \langle \Psi_A | n_{k,s} n_{j,s} | \Psi_A \rangle. \quad (1.27)$$

For repulsive short-ranged electron-electron interactions in a metallic sample, $V_{jkkj} > 0$. If $\Psi_A$ contains an up electron in the orbital state $j$ and an up electron in state $k$, but no down electrons in either of these states, (or down electrons but no up electrons) this configuration will lower the overall energy because $\langle \Psi_A | n_{k,\uparrow} n_{j,\uparrow} | \Psi_A \rangle + \langle \Psi_A | n_{k,\downarrow} n_{j,\downarrow} | \Psi_A \rangle = 1 + 0 = 1$ in Eq. (1.27). However, if instead $\Psi_A$ contains an up electron in state $j$ and a down electron in state $k$ (instead of the up electron in state $k$), the contribution of these states to the diagonal matrix element will then be zero because $\langle \Psi_A | n_{k,\uparrow} n_{j,\uparrow} | \Psi_A \rangle + \langle \Psi_A | n_{k,\downarrow} n_{j,\downarrow} | \Psi_A \rangle = 0 + 0$, and there will be no energy benefit from the exchange term. Therefore the exchange matrix elements for a repulsive electron interaction in a metal will favor states in which electrons in different orbital states have their spins aligned. Again, this result is in spite of the fact that the exchange interaction is the result of the spin-independent Coulomb interaction. Later (in the next lecture), I will work through in detail the example of the two-electron problem, and will show that the spin-dependent energies result from the fact that if you populate two different orbital states each with one electron, the average distance between the electrons will depend on their relative spin. These orbital correlations can therefore lead to different energies because of the Coulomb interaction.

If the exchange matrix elements are sufficiently large, they can lead to an instability of the Fermi Liquid state in which the spin orientations of a majority of the electrons become aligned. The state of matter produced by this phase transition is ferromagnetism. (This is in contrast to the Fermi Liquid state, which has equal occupations for spin up and spin down electrons.)

*An Aside: On Combining the Direct and Exchange Terms*:

If we assume that our single-electron basis states are defined for zero spin-orbit coupling, our problem is by assumption symmetric under rotation in spin space. In this case, it can be convenient to express the Direct and Exchange matrix elements in a form that makes clear this symmetry. This can be accomplished by writing them in terms of the operators for the total charge in orbital state $j$ and the total spin in state $j$:

$$n_j = c^\dagger_{j\uparrow} c_{j\uparrow} + c^\dagger_{j\downarrow} c_{j\downarrow}$$
$$\vec{S}_j = \frac{1}{2}\sum_{s\sigma} c^\dagger_{js} \vec{\sigma}_{s\sigma} c_{j\sigma}, \quad (1.28)$$

where $\vec{\sigma}_{s\sigma}$ denotes the Pauli matrices. After a considerable amount of algebra (discussed briefly in the Kurland et al. paper[2]), for the case of a $\delta$-function interaction it is possible to rewrite

$$V_{direct} + V_{exchange} = \sum_{k,j}\left[\frac{1}{2}K_{jk}n_j n_k - 2K_{jk}\vec{S}_j\cdot\vec{S}_k\right]. \tag{1.29}$$

This is transparently invariant with respect to rotations (as it should be under our assumptions), and it also clearly shows how exchange interactions favor states in which spins in different orbital levels are aligned when $K_{jk} > 0$.

### *Pairing, or BCS Terms*:

Finally we come to the third choice of states that gives large matrix elements, the choice in which states *j* and *k* are related by time-reversal symmetry and states *m* and *n* are related by time reversal symmetry. If we choose our single-electron states to be real-valued standing waves (in the absence of spin-orbit interactions or an applied magnetic field), the pairs of states related by time-reversal symmetry have the same orbital state, say $\psi_j(r)$. For this choice, Eqs. (1.18) and (1.19) become

$$V_{pairing} = \frac{1}{2}\sum_{\substack{j,m \\ s,\sigma}} V_{jjmm} c^\dagger_{j,\sigma} c^\dagger_{j,s} c_{m,s} c_{m,\sigma}. \tag{1.30}$$

$$V_{jjmm} = \int d^3r_1 d^3r_2 \psi_j(r_2)\psi_j(r_1)V_{e-e}(r_1-r_2)\psi_m(r_1)\psi_m(r_2). \tag{1.31}$$

These pairing matrix elements contribute explicitly off-diagonal matrix elements in the Hamiltonian matrix within the basis of Slater determinant states -- the terms in Eq. (1.30) cannot be expressed in terms of simple number operators. (Note that we have chosen to use disordered standing wave states here purely for the mathematical convenience of being able to work with real-valued wavefunctions; see the Appendix to Section V for the analogous analysis using more general Bloch states. The main loss of using disordered standing-wave basis states is that this can obscure effects of angle-dependent electron-electron scattering that can exist within some crystals, and can be important in generating certain kinds of unconventional superconductivity.)

If we expand Eq. (1.30),

$$V_{pairing} = \frac{1}{2}\sum_{j,m} V_{jjmm}\left(c^\dagger_{j\uparrow}c^\dagger_{j\uparrow}c_{m\uparrow}c_{m\uparrow} + c^\dagger_{j\downarrow}c^\dagger_{j\downarrow}c_{m\downarrow}c_{m\downarrow} + c^\dagger_{j\uparrow}c^\dagger_{j\downarrow}c_{m\downarrow}c_{m\uparrow} + c^\dagger_{j\downarrow}c^\dagger_{j\uparrow}c_{m\uparrow}c_{m\downarrow}\right). \tag{1.32}$$

The first two terms inside the parentheses will give zero, since double occupancy of Fermion states is not allowed (or by the operator algebra, $c^\dagger_{j\uparrow}c^\dagger_{j\uparrow} = -c^\dagger_{j\uparrow}c^\dagger_{j\uparrow}$ by anticommutation, so the combination can only give zero). The third and 4th terms in Eq. (1.32) are the same, since with two anticommutations $c^\dagger_{j\downarrow}c^\dagger_{j\uparrow}c_{m\uparrow}c_{m\downarrow} = c^\dagger_{j\uparrow}c^\dagger_{j\downarrow}c_{m\downarrow}c_{m\uparrow}$. Therefore we can cancel the factor of 1/2 in front of Eq. (1.32) and write the form of the pairing interaction simply as

$$V_{pairing} = \sum_{j,m} V_{jjmm} c^\dagger_{j\uparrow}c^\dagger_{j\downarrow}c_{m\downarrow}c_{m\uparrow}. \tag{1.33}$$

When written in this way, it is clear why this is called the "pairing" contribution. This contribution takes two electrons that are originally in a pair of time-reversed states ($m\uparrow,m\downarrow$) and scatters them into a different pair of time-reversed states ($j\uparrow,j\downarrow$). If the effective electron-electron interactions are attractive ($V_{jjmm} < 0$), this interaction can lead to an instability in the Fermi Liquid ground state beyond which the ground-state many-body wavefunction will consist of a linear superposition of Slater determinants made up of these pair-correlated states. This pairing, as we will see, will turn out to be the microscopic explanation for superconductivity. If the effective electron-electron interactions are repulsive, the pairing matrix elements do not lead to any instability in the Fermi Liquid state.

**Evaluating the Coefficients $K_{jk}$:**

For the limit in which the screened Coulomb interaction in a metal can be taken to be a $\delta$-function, we have seen that the coefficients which describe the strength of the direct, exchange, and pairing matrix elements are all exactly the same:

$$K_{jk} = V_0 \int d^3r |\psi_j(r)|^2 |\psi_k(r)|^2. \tag{1.34}$$

Let's think about this quantity in a bit more detail. If the sample size is much greater than the Fermi wavelength of the electrons, $\psi_j(r)$ and $\psi_k(r)$ will undergo many independent (uncorrelated) oscillations within the sample volume, and therefore to a good approximation

$$K_{jk} = V_0 \int d^3r \langle |\psi_j(r)|^2 \rangle \langle |\psi_k(r)|^2 \rangle \tag{1.35}$$

where the brackets $\langle \ \rangle$ denote independent averages of the two oscillating functions. If the wavefunctions are normalized, these two averages are each equal to just 1/(sample volume), so that

$$K_{jk} = V_0 \int d^3r \left(\frac{1}{volume}\right)^2 = \frac{V_0}{volume} \tag{1.36}$$

The fact that these coefficients are independent of $j$ and $k$ represents a huge simplification, in that to a very good approximation we can say that the matrix elements of the short-ranged screened electron-electron interaction should be the same regardless of which single-electron states go into the Slater determinants. Let us call this one coefficient $K$. In our simple theory of a delta-function-like interaction, we have seen that the same value of $K$ contributes to all of the direct, exchange, and pairing terms, so that we can write the large matrix elements of the electron-electron interaction in a "universal" form:

$$V_{direct} + V_{exchange} + V_{pairing} = \sum_{k,j}\left[\frac{1}{2}Kn_j n_k - 2K\vec{S}_j \cdot \vec{S}_k\right] + \sum_{j,m} K c^\dagger_{j\uparrow} c^\dagger_{j\downarrow} c_{m\downarrow} c_{m\uparrow},$$

$$= \frac{1}{2}K\hat{n}^2 - 2K\hat{\vec{S}}^2 + K\sum_{j,m} c^\dagger_{j\uparrow} c^\dagger_{j\downarrow} c_{m\downarrow} c_{m\uparrow} \tag{1.37}$$

with only the one adjustable parameter $K$, that will depend on the choice of material. Here $\hat{n}$ is the operator for the total number of electrons and $\hat{\vec{S}}$ is the operator for their total spin.

In a slightly more sophisticated theory, one can think about performing a renormalization-group-type analysis so that the effective Hamiltonian is meant to describe just the states close to the Fermi energy. This involves integrating out the high-energy states away from the Fermi energy (their contribution is to screen the bare Coulomb interaction and make it short-ranged), and in the process it turns out that the strengths of the direct, exchange, and pairing matrix elements can renormalize to different values. However, even in these more sophisticated theories, as long as the states near the Fermi energy are dominated by a single band and the delta-function interaction is a reasonable approximation (due to screening of the Coulomb interaction), then the matrix elements still don't depend on the choice of the single-electron states. This means that under these conditions we can still use a model Hamiltonian for the electron-electron interactions with only 3 parameters characterizing the strength of all of the important matrix elements (I will call them $E_C$, $J$, and $\lambda_{BCS}$). Later when we wish to build simple models for magnetism and superconductivity, we will incorporate this expression for the interaction matrix elements into a sort of approximate "Universal Hamiltonian" which includes a sum over single-electron energies as well as the electron-electron interaction contributions:

$$H = H_{non-interacting} + V_{e-e} = \sum_{i,s} \epsilon^0_{i,s} c^\dagger_{i,s} c_{i,s} + E_C \hat{n}^2 - J\hat{\vec{S}}^2 - \lambda_{BCS} \sum_{j \neq m} c^\dagger_{j\uparrow} c^\dagger_{j\downarrow} c_{m\downarrow} c_{m\uparrow}. \quad (1.38)$$

Here $V_{e-e} = V_{direct} + V_{exchange} + V_{pairing}$ and I have excluded the $j = m$ terms from the pairing sum because they can be incorporated into the single-electron energy.

For repulsive interactions in good metals, $J$ is greater than 0 (favoring high-spin, or ferromagnetic states) and generally $\lambda_{BCS}$ renormalizes to small values so that the pairing term does not need to be considered when analyzing ferromagnetic states. If there is an effectively attractive electron-electron interaction (as can arise on account of phonon-mediated interactions, for instance, as we will discuss in a later lecture), then $J$ is less than 0 and $\lambda_{BCS}$ is greater than 0, given the conventions with which minus signs are introduced into Eq. (1.38). That is the regime of superconducting instabilities. Since the $J$ and $\lambda_{BCS}$ terms give rise to different types of instabilities, it is common that model Hamiltonians might include just one or the other, even though both originate from the same spin-independent electron-electron interaction.

For a more realistic form of the electron-electron interaction (beyond the δ-function approximation), in a single-band model of a disordered quantum dot there can be small fluctuations between the matrix elements for different single-electron states, but typically for good metal samples these fluctuations are smaller than 1/100 of the average values and they can almost always be ignored. For metals with 2 or more different electron bands near the Fermi energy, the model can be generalized, but in this case more than 3 parameters are needed to characterize the different possible matrix elements.

*What this model Hamiltonian does not include*: I also want to make clear that there are additional modes of instability that can be generated by electron-electron interactions, but which are not captured by this treatment. In the above, we have focused on the assumption of effectively randomly-oscillating single-electron basis states from within a single electronic band. This obscures any angle-dependent factors in the electron-electron scattering that might exist in well-ordered crystalline materials, and as a consequence the only instabilities that are apparent in this model are independent of angle relative to the crystal axes. In a well-ordered pure crystal, it would be more appropriate to use as our complete set of single-electron basis states the Bloch states that respect the crystal symmetries of the material. In that case, angle-dependent factors in the matrix elements for the electron interactions can generate instabilities that are not spherically symmetric. Girvin and Yang's textbook discuss these briefly in section 15.13.2. For non-superconducting materials, the results are known as a type of Pomeranchuk instability, giving rise to a nematic electronic state. For superconducting materials, the results can be anisotropic *s*-wave or non-*s*-wave superconductors. For magnetic systems, various forms of antiferromagnetism can exist in addition to the ferromagnetism we will consider. In what follows I will also generally assume that I start with single-electron basis states that come from topologically-trivial electron bands. Therefore this treatment will not be directly relevant to Chern insulators (also known as quantized anomalous Hall states) or to fractional quantized Hall states that can result from electron interactions within topological bands.

Even more types of electronic-state instabilities can occur when electrons interact with other low-energy excitations, for example phonons (leading to charge density waves), magnons (leading to spin density waves), or electrons in other bands (e.g., instabilities associated with heavy fermions). These states are not present in our model because we are considering the properties of interacting electrons from single bands in isolation (except for simple screening).

**Step 3. Diagonalizing the Hamiltonian**

We have shown that when we evaluate the matrix elements of the electron-electron interaction between Slater determinant states, only a few of the matrix elements will be non-negligible. For randomly-oscillating single-electron wavefunctions in a disordered good metal coming from a single band, we can express these largest matrix elements in terms of an approximate Universal Hamiltonian, with remarkably few adjustable parameters (only 3). All that is left for us to do then is to write down the full Hamiltonian matrix, (including also the sum of the single-electron energies associated with each Slater determinant basis state), and then diagonalize the matrix to determine our answer. In general, the true energy eigenstates will be linear superpositions of different Slater determinant states (rather than single Slater determinant states), because the Hamiltonian matrix does contain off-diagonal elements from the exchange and pairing contributions. The only difficulty in completing the diagonalization is the exponentially-large size of the matrix required even when considering just tens of electrons.

**What have we learned?:**
- The direct, exchange, and pairing matrix elements all result from the same spin-independent 2-body Coulomb interaction, evaluated between Slater determinant states.
- Spin-independent interactions produce spin-dependent many-body energies because of spatial correlations within the Slater determinant states. We will see next time that the root cause of this is the Pauli principle – electrons with the same spin can't get arbitrarily close to each other in space, so high-spin states exhibit less Coulomb energy cost.
- There are not arbitrarily many ways for electron-electron interactions to produce phase transitions that destroy the Fermi Liquid state of metallic systems. Instead, the instabilities that lead to phase transitions can be categorized by which type of matrix elements leads to the instability: ferromagnetism can result from the exchange terms, superconductivity from the pairing terms, and localization can result primarily from the direct terms (exchange does play a role here, too). [In a more detailed analysis that takes into account angle-dependent electron-electron scattering, non-isotropic electronic instabilities might also appear due to these types of interactions, but because of the simple assumptions of our model, only isotropic instabilities will be apparent.]

- Superconductivity is produced by pairing matrix elements that are explicitly off-diagonal in the basis of Slater determinant states. Therefore, a description of the superconducting state will require that we consider linear superpositions of different Slater determinant states – there is no way for a single Slater determinant state to include the types of pair correlations that are important for superconductivity.

[As a final side note, I will make a few comments about what one can do theoretically given that exact diagonalization of the many-electron Hamiltonian is not possible. This is an area of active research that I introduce in another lecture not included in these notes. The most popular workhorse technique is known as Density Functional Theory (DFT), which aims to calculate ground-state electronic properties by focusing on the total electron density function $n(\vec{r})$, which is a function of only 3 spatial variables and is therefore much simpler than the total many-electron wavefunction (although by the Hohenberg-Kohn Theorem there is a one-to-one mapping between $n(\vec{r})$ and the many-body ground-state wavefunction so that all of the information present in one is in principle also contained in the other). DFT has various levels of sophistication depending on the choice of functional used to approximate the exchange-correlation part of the electron-electron interactions, but it generally gives a reasonable account of ground state electronic properties, including quantities like strengths of bonding, crystal structures, phonon dispersion curves, and magnetic states. However, it is not designed to calculate excited electronic states and is not capable of doing this well. More sophisticated approaches to approximate interacting-electron wavefunctions include "Monte Carlo" techniques, the "Density Matrix Renormalization Group", and others, and these remain somewhat limited in the number of atoms or electrons they can consider at once (although still much better than our brute-force diagonalization "thought calculation"). A paper by the Simons Collaboration on the Many-Electron Problem recently published a comparison of different forefront many-electron calculational techniques in K. T. Williams et al., *Phys. Rev. X* **10**, 011041 (2020).]

## II. The Stoner Instability and the Origin of Ferromagnetism

*Topics: 1. A more intuitive understanding about why spin-independent electron-electron interactions can yield spin-dependent energies.*

*2. Analysis of the simplest toy model for a ferromagnetic instability: the Stoner Model.*

Ferromagnetism is the first example of a non-Fermi liquid state that we will explore – that is, we will analyze metals with a non-zero equilibrium magnetic moment. This is in contrast with Fermi-liquid metals which have zero magnetic moment in equilibrium because they have equal numbers of spin-up and spin-down electron states occupied. Fe, Co, Ni, and Gd are the elements which are magnetic metals at room temperature. For technological applications, magnetic alloys can often have superior properties. There also exist a great many insulating magnetic compounds. I will say a little about them in the next lecture.

Ferromagnets have been known since ancient times, but the microscopic explanation was elusive until well into the 1900's. There is quite a bit of interesting history. In the early 1900's one hypothesis, promoted by Einstein among others, was that ferromagnetism might be due to circulating electron currents within the material, in other words orbital magnetism. This led to a prediction that magnetism should be closely related to angular momentum. If one considers a circular electron orbit of radius r, the magnetic moment is $M =$ (current)(area enclosed) $= (ev/2\pi r)(\pi r^2) = evr/2$ and the associated orbital angular momentum is $L = rmv$, so one expects a simple ratio $M/L = e/(2m)$ (here $e$ is the magnitude of the electron charge, $m$ is the electron mass, and $v$ is a velocity). This speculation was tested by the Einstein-de Haas experiment (1915) in which an iron sample was suspended from a fiber, the magnetization was periodically reversed by using an external magnetic field, and the resulting rotation of the magnet was measured. The results came out as Einstein and de Haas expected. In fact, they reported an experimental value of $M/L$ very close to $e/(2m)$ within 10% uncertainty. It turns out that this value was wrong – off by a factor of 2, and the result serves as a useful cautionary tale about confirmation bias – that in experimental physics expecting a given result can make it more likely that you will seem to find that result, even if it is wrong. The correct result published a few months later by Samuel Jackson Barnett is that magnetism and angular momentum are indeed closely connected within iron, but to good accuracy with $M/L = e/m$. The factor of 2 difference arises because the magnetization in iron (and the other room-temperature magnets) is due primarily to the intrinsic spin of electrons, rather than orbital motion, and for the electron spin the ratio of magnetic moment to angular momentum is approximately a factor of 2 greater than for orbital motion – this is the famous electron "g-factor".

Before we return to a discussion of the exchange interaction, it is useful to briefly consider one bit of physics that you might think should be important to the formation of magnetism, but actually is not– the magnetic dipole interaction. Every electron acts like a tiny dipole magnet with a magnetization equal to the Bohr magneton, and the magnetic field produced by one electron will interact with other electrons nearby. Let's estimate the scale of this dipole interaction energy for two electron spins separated by roughly $d = 0.3$ nm, or an atomic spacing:

$$\Delta E \approx \frac{\mu_0 (\mu_B)^2}{2\pi d^3} = \frac{(4\pi \times 10^{-7} \text{ N/A}^2)(9.27 \times 10^{-24} \text{ J/T})^2}{2\pi (3 \times 10^{-10} \text{ m})^3} \approx 6 \times 10^{-25} \text{ J} = (0.046 \text{ K}) k_B \quad (2.1)$$

(here $\mu_B$ is the Bohr magneton and $k_B$ is the Boltzmann constant). The critical temperature of iron is above 1000 K. From this comparison it is clear that the dipole interaction is far too weak (by at least a factor of $10^4$) to keep the spins in iron aligned against thermal fluctuations up to 1000 K. The exchange interactions that actually stabilize ferromagnets are much, much stronger than dipole fields on atomic scales – remarkably so. Nevertheless, dipole fields will re-enter our discussion at a later stage. Dipole fields are long-ranged (falling off as a power law $\propto 1/d^3$) compared to exchange interactions (which fall off exponentially on atomic scales) so when integrating over longer length scales the two effects can come into competition. This will become important later when we consider the formation and spatial distributions of magnetic domains within a ferromagnetic sample.

Let us therefore refocus on exchange interactions, and consider in greater detail why it is that spin-independent electron-electron interactions can give spin-dependent exchange energies. It is easiest to analyze the simplest possible case: 2 electrons, with one in an orbital state $u_a(r)$ (capable of occupying either spin up or spin down) and the other electron is a different orbital state $u_b(r)$ (also capable of occupying either spin up or spin down). What are the two-electron energy eigenstates and eigenvalues in this situation? The problem has rotational symmetry is spin space, so we can know immediately that the energy eigenstates can be chosen to also be eigenstates of the total spin $S$ and also the z-component $S_z$. For two electrons, that completely determines the form of the four energy eigenstates. I'll write things out in both 2nd-quantized notation and (equivalently) also expanding out the full Slater determinants:

<u>$S = 1$, $S_z = 1$</u>:
$$c_{a\uparrow}^\dagger c_{b\uparrow}^\dagger |\text{vacuum}> = \frac{1}{\sqrt{2}} [u_a(r_1) u_b(r_2) - u_b(r_1) u_a(r_2)] \uparrow_1 \uparrow_2$$

<u>$S = 1$, $S_z = -1$</u>:
$$c_{a\downarrow}^\dagger c_{b\downarrow}^\dagger |\text{vacuum}> = \frac{1}{\sqrt{2}} [u_a(r_1) u_b(r_2) - u_b(r_1) u_a(r_2)] \downarrow_1 \downarrow_2$$

<u>$S = 1$, $S_z = 0$</u>:
$$\frac{1}{\sqrt{2}} (c_{a\uparrow}^\dagger c_{b\downarrow}^\dagger + c_{a\downarrow}^\dagger c_{b\uparrow}^\dagger) |\text{vacuum}>$$
$$= \frac{1}{\sqrt{2}} \Big[ \frac{1}{\sqrt{2}} (u_a(r_1) \uparrow_1 u_b(r_2) \downarrow_2 - u_b(r_1) \downarrow_1 u_a(r_2) \uparrow_2)$$
$$+ \frac{1}{\sqrt{2}} (u_a(r_1) \downarrow_1 u_b(r_2) \uparrow_2 - u_b(r_1) \uparrow_1 u_a(r_2) \downarrow_2) \Big]$$
$$= \frac{1}{\sqrt{2}} [u_a(r_1) u_b(r_2) - u_b(r_1) u_a(r_2)] \frac{1}{\sqrt{2}} [\uparrow_1 \downarrow_2 + \downarrow_1 \uparrow_2]$$

<u>$S = 0$, $S_z = 0$</u>:
$$\frac{1}{\sqrt{2}} (c_{a\uparrow}^\dagger c_{b\downarrow}^\dagger - c_{a\downarrow}^\dagger c_{b\uparrow}^\dagger) |\text{vacuum}> = \frac{1}{\sqrt{2}} [u_a(r_1) u_b(r_2) + u_b(r_1) u_a(r_2)] \frac{1}{\sqrt{2}} [\uparrow_1 \downarrow_2 - \downarrow_1 \uparrow_2].$$

The spin-triplet $S = 1$ states have a symmetric spin state with an antisymmetric orbital state (I'll call this orbital state $\Psi_1(r_1, r_2)$), while the spin-singlet $S = 0$ has an antisymmetric spin state with a symmetric orbital state ($\Psi_0(r_1, r_2)$).

Now let's evaluate the eigenvalue of the electron-electron interaction energy in these states. For any of the triplet states this energy is

$$\langle \Psi_1 | V_{e-e}(r_1 - r_2) | \Psi_1 \rangle$$
$$= \int d^3r_1 d^3r_2 \frac{1}{\sqrt{2}}[u_a^*(r_1)u_b^*(r_2) - u_b^*(r_1)u_a^*(r_2)]V_{e-e}(r_1 - r_2)\frac{1}{\sqrt{2}}[u_a(r_1)u_b(r_2) - u_b(r_1)u_a(r_2)].$$
(2.2)

If we multiply this out, combine the two diagonal terms together, and also combine the two cross terms together, we get for the $S = 1$ states

$$\langle \Psi_1 | V_{e-e}(r_1 - r_2) | \Psi_1 \rangle$$
$$= \int d^3r_1 d^3r_2 \, |u_a(r_1)|^2 |u_b(r_2)|^2 V_{e-e}(r_1 - r_2) - \int d^3r_1 d^3r_2 \, u_a^*(r_1)u_b(r_1)u_b^*(r_2)u_a(r_2)V_{e-e}(r_1 - r_2).$$
(2.3)

Going through the same process for the $S = 0$ state gives

$$\langle \Psi_0 | V_{e-e}(r_1 - r_2) | \Psi_0 \rangle$$
$$= \int d^3r_1 d^3r_2 \, |u_a(r_1)|^2 |u_b(r_2)|^2 V_{e-e}(r_1 - r_2) + \int d^3r_1 d^3r_2 \, u_a^*(r_1)u_b(r_1)u_b^*(r_2)u_a(r_2)V_{e-e}(r_1 - r_2)$$
(2.4)

(the same except for a plus sign in front of the cross-terms integral). If the cross-terms integral is not equal to zero, the two energies are not the same, even though we have explicitly assumed that the electron-electron interaction is spin-independent. In a metal, for an interaction that is repulsive ($V_{e-e} > 0$) and short ranged due to screening, we can approximate $V_{e-e}(r_1 - r_2) = V_0 \delta(r_1 - r_2)$ and see that the cross-term integrand is positive-definite. This means that the spin-triplet state has a lower energy than the spin singlet state, and so in a metal repulsive electron-electron interactions will favor parallel alignment of electron spins.

The reason for this difference in energy is that the requirement of a totally-antisymmetric many-electron wavefunction leads to spin-dependent differences in spatial correlations between electrons. For the $S = 1$ states, the orbital part of the wavefunction is $\Psi_1(r_1, r_2) = \frac{1}{\sqrt{2}}[u_a(r_1)u_b(r_2) - u_b(r_1)u_a(r_2)]$, so the wavefunction goes to zero whenever $r_1$ approaches $r_2$. In contrast, for the $S = 0$ state we have $\Psi_0(r_1, r_2) = \frac{1}{\sqrt{2}}[u_a(r_1)u_b(r_2) + u_b(r_1)u_a(r_2)]$, which approaches the value $\sqrt{2} \, u_a(r_2)u_b(r_2)$ as $r_1 \to r_2$, which can be non-zero. As a consequence, the electrons in the $S = 1$ states are on average farther apart than in the $S = 0$ state, and so they have a lower energy because they experience weaker electron-electron repulsion on average. This same physics is generalizable to states with more than two electrons, as the Pauli exclusion principle forbids having a non-zero probability for finding two electrons with the same spin at the same location. These effects also play a central role in atomic physics as the basis for Hund's rules.

**The Stoner Model**

With this understanding of exchange interactions, we are now ready to consider the simplest model of metallic ferromagnetism, the somewhat unfortunately-named Stoner Model (after Edmund Clifton Stoner, 1899-1968). The basic idea of the model is that exchange interactions can induce a process of positive feedback, in that if some electrons align their spins this can make it more favorable for other electrons to align their spins, too. (Think peer pressure.) For strong enough exchange interactions, you can get a runaway process, and you can end up with an electronic ground state in which macroscopic fraction of electrons have their spins aligned even in the absence of any applied magnetic field. The key ingredient of the analysis will be a competition between single-electron level spacings and the exchange energy.

Our starting point is a Hamiltonian that includes a sum over the energies of a non-interacting set of single-electron basis states plus our "Universal Hamiltonian" to account for the electron-electron interactions in a metal. We'll assume that the electron number stays fixed so that the overall charging energy $E_C \hat{n}^2$ is a constant that can be ignored, and we will also ignore the BCS pairing term assuming that it renormalizes to a small value for repulsive electron-electron interactions. We'll allow for an applied magnetic field in the $\hat{z}$ direction, which interacts with the electron spins via a Zeeman interaction. The full Hamiltonian of the model is therefore

$$H = \sum_\alpha \epsilon_\alpha^0 (\hat{n}_{\alpha\uparrow} + \hat{n}_{\alpha\downarrow}) - J\left(\hat{\vec{S}}\right)^2 - g\mu_B S_z B. \tag{2.5}$$

Here $\epsilon_\alpha^0$ are the single-electron energies, $\hat{n}_{\alpha\uparrow}$ and $\hat{n}_{\alpha\downarrow}$ are number operators accounting for occupation of the spin-up and spin-down states for the orbital state $\alpha$, and $g \approx 2$ is the electronic g factor. For the simplest possible argument that gets at the relevant physics, we'll consider the case of a finite quantum dot with equally-spaced single-electron energy levels with level spacing $\delta$. The question at hand is: What value of the total spin $S$ gives the lowest overall energy?

A single Slater determinant can be an eigenstate of this Hamiltonian for a state with maximal $S_z = S$, which are also the states favored by an applied magnetic field. So we'll consider the energies of this type of single Slater determinant and won't worry about the off-diagonal matrix elements coupling different Slater determinants. We can then solve for the lowest-energy state corresponding to any given value of total $S$ using simple cartoons:

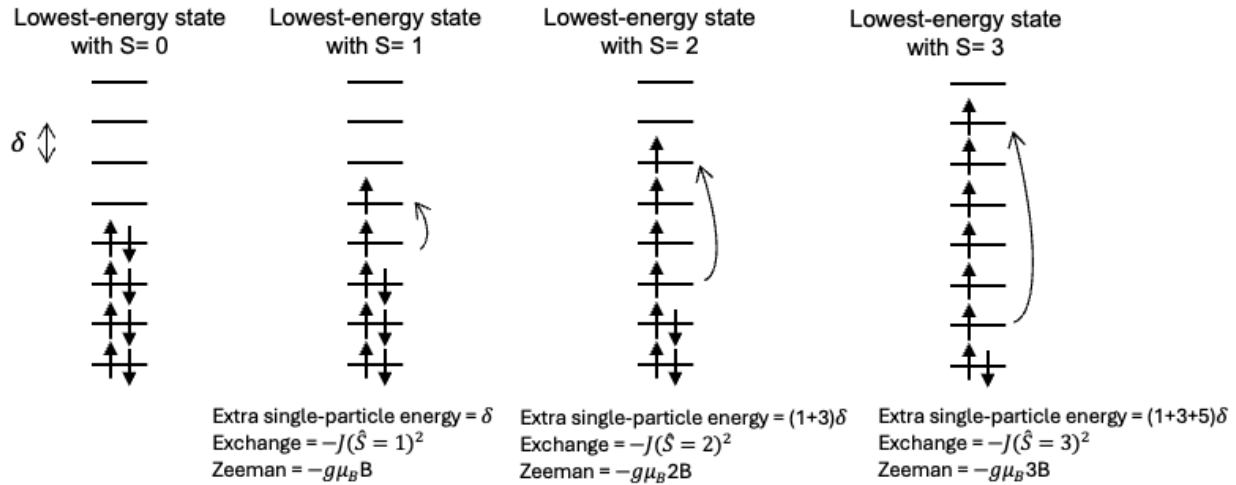

Relative to the energy of the $S = 0$ state, the ground-state energy as a function of $S$ therefore follows a simple pattern:

$$\begin{aligned} E(S) &= \delta S^2 - J\left(\hat{S}\right)^2 - g\mu_B SB \\ &= \delta S^2 - JS(S+1) - g\mu_B SB. \end{aligned} \tag{2.6}$$

If we only care about macroscopic samples, so that either $S = 0$ or $S$ is much bigger than one, we can take a macroscopic limit $S(S + 1) \rightarrow S^2$, and we have

$$E(S) = \delta S^2 - JS^2 - g\mu_B SB. \tag{2.7}$$

We can then easily minimize $E(S)$ for a given value of $B$:

$$\text{set } \frac{dE(S)}{dS} = 0 \rightarrow 2S(\delta - J) - g\mu_B B = 0 \rightarrow S = \frac{g\mu_B B}{2(\delta - J)} = \frac{\mu_B B}{\delta - J} \tag{2.8}$$

since $g = 2$ for electrons.

If we consider the non-interacting case $J = 0$, the value $S = \mu_B B/\delta$ corresponds to the standard result for Pauli paramagnetism of a non-interacting electron gas. (Check: this is often written as $dM/dB = \mu_B^2 g(\varepsilon_F)$ with $M = 2\mu_B S/(volume)$ and $g(\varepsilon_F)(volume) = 2/\delta$.)

For weak interactions, $0 < J < \delta$, the lowest-energy spin state is $S = 0$ when $B = 0$, but the susceptibility $dS/dB = \mu_B/(\delta - J)$ is enhanced relative to $J = 0$. This reflects the positive feedback – the aligning some spins helps additional spins to align, too. As $J \rightarrow \delta$, the spin susceptibility diverges, which is an indication that the Fermi liquid state becomes unstable and the true ground state is some non-Fermi-liquid state.

For stronger interactions with $J > \delta$, the spin of the ground state is not zero anymore even when $B = 0$. With $B = 0$, the formula for $E(S)$ in the macroscopic limit is simply $E(S) = S^2(\delta - J)$, so in our toy model of equally-spaced levels the energy is minimized with S takes its maximum possible value – all of the spins will align to give the maximum possible spin state!

All this is connected in a direct way to our previous consideration of what happens to low-energy excitations above the ground state as a function of tuning the interaction strength. In this toy model, the electronic system remains a Fermi liquid for interaction strengths in the range $0 \leq J < \delta$. In this range, the low-energy excitations involve changes in the occupation of the low-energy independent electron states. For non-zero $J$, the value of the excitation energies will be different than for $J = 0$ because of the exchange contributions, but one could still make a one-to-one correspondence between the low-lying excitations for zero and non-zero $J$, with a continuous evolution between them. This changes when $J$ increases through $\delta$, because at that point there will be a discontinuous change to a ground state with non-zero total spin. The low-lying excitations of this magnetic state will be very different than for a Fermi liquid, in part because they include a new type of collective excitation, known as spin waves or magnons, that is not present at all in Fermi liquids.

One additional lesson of our toy model comes from the result that the condition for the formation of the magnetic state is that $J \geq \delta$. The level spacing $\delta$ will be inversely related to the electronic bandwidth, because narrow bands will have lots of closely-space energy levels and therefore a small value of $\delta$. Narrow-band materials like magic-angle twisted bilayer graphene are of significant interest because the small value of $\delta$ affects the competition between single-electron excitation energies and electron-electron interactions. In narrow bands, changes in single-electron energies will be relatively unimportant relative to electron-electron interactions so that it is easier to generate non-Fermi-liquid states – magnetism in the case we have just considered but also other types of exotic electronic states, too.

I'll wind up this section with a few tangential but hopefully-interesting side notes:

*Non-Extrinsic Magnetism*: In our discussion of our toy Stoner Model, I made the approximation $S(S+1) \approx S^2$, as I had in mind only the case of a macroscopic sample where observable values of $S$ are on the order of Avogadro's number. However, this assumption is not necessary, and one can just as well solve for the lowest energy spin states corresponding to our full expression (here with no applied magnetic field):

$$E(S) = \delta S^2 - JS(S+1) \tag{2.9}$$

It turns out that as a function of increasing $J$ this model has a cascade of instabilities in which the spin changes by one unit at a time, rather than just a single instability:

$$\begin{aligned} &E(1) < E(0) \text{ when } J > \delta/2 \quad \text{(rather than } \delta\text{)} \\ &E(2) < E(1) \text{ when } J > 3\delta/4 \\ &E(3) < E(2) \text{ when } J > 5\delta/6 \\ &\ldots. \\ &S \to \infty \text{ when } J > \delta. \end{aligned} \tag{2.10}$$

This means that in the range $\delta/2 < J < \delta$ within the toy model it is possible to have a ground state with a small value of the spin that is non-extrinsic, in that the total spin doesn't scale with the volume of the sample. (By our definitions, $J$ and $\delta$ both scale the same way with sample volume, $\propto 1/(volume)$, so a macroscopic sample with $\delta/2 < J < \delta$ might have a total spin with just $S = 1, 2,$ or $3$.) For a macroscopic sample, this type of extremely weak magnetism is unmeasurable, but it can be detected in nanoscale quantum dots where a spin 1 state gives a clearly different electronic spectrum than spin 0. Furthermore, in real quantum-dot samples, the level spacing won't be uniform. If a few energy levels near the Fermi energy happen to be more closely spaced than on average, this will make this type of weak non-extrinsic magnetism more likely. The upshot is that nanoscale samples made of non-magnetic materials can sometimes still be weakly magnetic if $J$ is non-zero but still $< \delta$, and that this result can vary from sample to sample depending on the particular pattern of level spacing near the Fermi energy.

*Orbital Magnetism*: In my discussion of the Einstein-de Haas experiment, I mentioned that the correct value of the ratio of magnetic moment to angular momentum, $M/L = e/m$ (rather than $e/2m$), indicates that the origin of magnetism in metals like iron is primarily due to ordering of electron spins rather than electron orbital motion. However, recent experiments on moiré lattices made from van der Waals materials have also now found evidence of magnetized states primarily due to orbital moments, rather than spin. This evidence is the existence of anomalous Hall effects in the absence of applied magnetic field in materials where orbital magnetic moments are associated with states located near particular valleys in the electron bandstructure. This is very much an active research field. For those interested, some of the initial experiments include A. L. Sharpe et al., *Science* **365**, 605 (2019); M. Serlin et al., *Science* **367**, 900 (2020); and G. Chen et al., *Nature* **579**, 56 (2020).

*The Stoner Model For a Continuum Band Picture*: I like solving the Stoner Model assuming discrete energy levels as we have above because it is very simple and captures the essential physics. However, most textbooks do the calculation assuming a continuous density of states. There is a subtlety in doing so that many textbooks get incorrect. For those who might be interested, I have included a correct calculation in an appendix to this section for the Stoner Model in a continuum band picture.

**What have we learned?:**

- Spin-independent electron-electron interactions produce spin-dependent energies because the spatial correlations between electrons depend on their relative spin. $\Psi(r_1, r_2) \to 0$ as $r_1 \to r_2$ for electrons with the same spin, but not for electrons with opposite spin.

- For repulsive electron-electron interactions in a metal, there can be an instability to ferromagnetism (in which a macroscopic fraction of the electron spins are aligned) when the exchange interaction strength ($J$) exceeds the single-electron level spacing ($\delta$).

## Appendix to Section II. Stoner Instability for a Continuum Band Picture

Assume a single free-electron band with an exchange interaction $-J\langle S^2 \rangle \approx -JS_z^2$ and with a magnetic field applied to give a Zeeman interaction, $-2\mu_B B S_z$. (I use that the electron g-factor is approximately 2.) What will be the induced magnetization?

That is, take the Hamiltonian to be

$$H = \sum_\alpha \varepsilon_\alpha^0 (n_{\alpha\uparrow} + n_{\alpha\downarrow}) - JS_z^2 - g\mu_B B S_z. \qquad (2.11)$$

Summing over occupied states, we have $S_z = \frac{1}{2}\sum_\alpha (n_{\alpha\uparrow} - n_{\alpha\downarrow}) = \frac{1}{2}(N_\uparrow - N_\downarrow)$, where $N_\uparrow$ and $N_\downarrow$ are the total numbers of electrons with up and down spins. If we then use each of the expressions for $S_z$ once to write

$$S_z^2 = \frac{1}{4}\sum_\alpha (n_{\alpha\uparrow} - n_{\alpha\downarrow})(N_\uparrow - N_\downarrow) \qquad (2.12)$$

then the total energy of the interacting electrons can be re-written in a form that can be understood as a sum over single-electron states whose energies are shifted by the exchange and Zeeman interactions:

$$\begin{aligned}
E &= \sum_\alpha \left\{ n_{\alpha\uparrow}\left(\varepsilon_\alpha^0 - \frac{J}{4}(N_\uparrow - N_\downarrow) - \mu_B B\right) + n_{\alpha\downarrow}\left(\varepsilon_\alpha^0 + \frac{J}{4}(N_\uparrow - N_\downarrow) + \mu_B B\right) \right\} \\
&= \sum_\alpha \left\{ n_{\alpha\uparrow}\left(\varepsilon_\alpha^0 - \frac{J}{2}S - \mu_B B\right) + n_{\alpha\downarrow}\left(\varepsilon_\alpha^0 + \frac{J}{2}S + \mu_B B\right) \right\}
\end{aligned} \qquad (2.13)$$

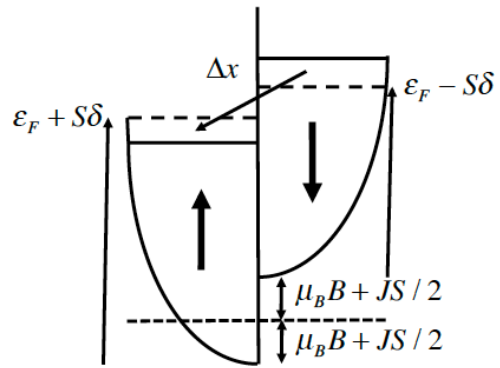

Now, assume that we start with the electrons in their $S = 0$ ground state in the absence of any applied magnetic field, then keep the electrons frozen in these states when the field is applied to shift by the Zeeman energy, and finally allow the highest energy spin-down electrons within an energy interval $\Delta x = S/[g(\varepsilon_F) Vol/2] = S\delta$ to flip their spins (giving a total spin equal to S) to try to lower the total energy of the system. (Here $\delta$ is the level spacing.)

I wish to estimate what is the energy of the system as a function of the total spin S that is produced, and then minimize this energy $E(S,B)$ with respect to S to determine what is the value of total spin that minimizes the total energy for the given value of applied magnetic field.

Taking the difference $E(S,B)-E(0,0)$, and calculating the contributions from spin-up and spin-down electrons separately, I have

$$[E(S,B) - E(0,0)]/Vol = \int_0^{\varepsilon_F + S\delta} \frac{g(\varepsilon)}{2}\left(\varepsilon - \frac{J}{2}S - \mu_B B\right)d\varepsilon - \int_0^{\varepsilon_F} \frac{g(\varepsilon)}{2}\varepsilon\,d\varepsilon$$
$$+ \int_0^{\varepsilon_F - S\delta} \frac{g(\varepsilon)}{2}\left(\varepsilon + \frac{J}{2}S + \mu_B B\right)d\varepsilon - \int_0^{\varepsilon_F} \frac{g(\varepsilon)}{2}\varepsilon\,d\varepsilon \tag{2.14}$$

(The dummy variable of integration $\varepsilon$ here is not the full energy – I use it just to integrate over all states from the bottom of the band to the highest occupied levels.)

If I add $\int_0^{\varepsilon_F} \frac{g(\varepsilon)}{2}\left(\frac{J}{2}S + \mu_B B\right)d\varepsilon$ to the top line and subtract the same quantity from the bottom line, I get

$$\frac{E(S,B) - E(0,0)}{Vol} = \int_0^{\varepsilon_F + S\delta} \frac{g(\varepsilon)}{2}\left(\varepsilon - \frac{J}{2}S - \mu_B B\right)d\varepsilon - \int_0^{\varepsilon_F} \frac{g(\varepsilon)}{2}\varepsilon\,d\varepsilon + \int_0^{\varepsilon_F} \frac{g(\varepsilon)}{2}\left(\frac{J}{2}S + \mu_B B\right)d\varepsilon$$
$$+ \int_0^{\varepsilon_F - S\delta} \frac{g(\varepsilon)}{2}\left(\varepsilon + \frac{J}{2}S + \mu_B B\right)d\varepsilon - \int_0^{\varepsilon_F} \frac{g(\varepsilon)}{2}\varepsilon\,d\varepsilon - \int_0^{\varepsilon_F} \frac{g(\varepsilon)}{2}\left(\frac{J}{2}S + \mu_B B\right)d\varepsilon \tag{2.15}$$

$$\frac{E(S,B) - E(0,0)}{Vol} = \int_0^{\varepsilon_F + S\delta} \frac{g(\varepsilon)}{2}\left(\varepsilon - \frac{J}{2}S - \mu_B B\right)d\varepsilon - \int_0^{\varepsilon_F} \frac{g(\varepsilon)}{2}\left(\varepsilon - \frac{J}{2}S - \mu_B B\right)d\varepsilon$$
$$+ \int_0^{\varepsilon_F - S\delta} \frac{g(\varepsilon)}{2}\left(\varepsilon + \frac{J}{2}S + \mu_B B\right)d\varepsilon - \int_0^{\varepsilon_F} \frac{g(\varepsilon)}{2}\left(\varepsilon + \frac{J}{2}S + \mu_B B\right)d\varepsilon \tag{2.16}$$

I can then combine the two integrals on each line to give

$$\frac{E(S,B) - E(0,0)}{Vol} = \int_{\varepsilon_F}^{\varepsilon_F + S\delta} \frac{g(\varepsilon)}{2}\left(\varepsilon - \frac{J}{2}S - \mu_B B\right)d\varepsilon - \int_{\varepsilon_F - S\delta}^{\varepsilon_F} \frac{g(\varepsilon)}{2}\left(\varepsilon + \frac{J}{2}S + \mu_B B\right)d\varepsilon. \tag{2.17}$$

If I make the assumption that $S\delta$ is small compared to the energy scale on which the density of state evolves (here $S$ is the total spin and $\delta$ is the level spacing, as in class), I can approximate $g(\varepsilon)Vol/2 \approx g(\varepsilon_F)Vol/2 = 1/\delta$ and take this factor outside of the integrals. Then it is trivial to solve the remaining integrals;

$$E(S,B) - E(0,0) = \left(\frac{1}{\delta}\right)\left\{\frac{1}{2}\left[(\varepsilon_F + S\delta)^2 - (\varepsilon_F)^2\right] - \left(\frac{J}{2}S + \mu_B B\right)S\delta\right.$$
$$\left. - \frac{1}{2}\left[(\varepsilon_F)^2 - (\varepsilon_F - S\delta)^2\right] - \left(\frac{J}{2}S + \mu_B B\right)S\delta\right\}$$
$$= \left(\frac{1}{\delta}\right)\left\{S^2\delta^2 - (JS + 2\mu_B B)S\delta\right\} \tag{2.18}$$
$$= S^2(\delta - J) - 2\mu_B B S.$$

This is the same expression we got considering a discrete set of energy levels, rather than a continuum of states. Minimizing $E(S,B)$ with respect to $S$ gives that the spin that gives the minimum energy is $S = \dfrac{\mu_B B}{\delta - J}$. For small values of $J$ ($J < \delta$), the effect of interactions is just to increase the ground state spin slightly above the non-interacting value $S = \mu_B B / \delta$. When $J$ approaches $\delta$, the ground state spin diverges, which is a sign of a proximity to an instability for forming the ferromagnetic state. For $J > \delta$, the assumptions of our calculation break down, because $S\delta$ will no longer be small, but the crux of the physics is that the electron interactions cause a spin-polarized ground state to be stable even in the absence of an applied magnetic field.

[If instead of doing the calculation correctly, you assumed incorrectly (as is done in some textbooks) that the spin-up shifted states and spin-down shifted states should be occupied up to the same value of some common effective single-electron Fermi energy, you would get the wrong answer by a factor of two, $S = \mu_B B / [\delta - (J/2)]$. The cause of the error is that it is incorrect to assume that there is a single overall Fermi energy which describes the filling of the effective single-electron energy levels. Instead, the correct condition of equilibrium is that the total electronic energy (including all interactions) should be minimized. Considering only the energies of states near the Fermi level neglects that shifting the numbers of spin-up versus spin-down states affects the energies of all the states, even those well below the Fermi level, because of the exchange interaction.]

## III. More About Magnets + Their Applications

*Topics: 1. The ways in which real metallic magnets differ from the simple Stoner Model, and the physics behind giant magnetoresistance in metallic magnets.*

*2. A mean-field model of insulating magnets and an analysis of the temperature dependence of their magnetic susceptibility well above the critical temperature.*

*3. Factors governing the formation and configurations of magnetic domains.*

The Stoner Model provides useful insights into the origin of ferromagnetism, but real magnetic metals differ in several ways from the toy model we have considered so far. None of the actual room-temperature elemental magnets (Fe, Ni, Co, Gd) are single-band metals – they all have occupation of both narrow-band ("*d,f*-like") states and broad-band ("*s,p*-like") states. The electrons in these bands are partially spin-polarized, rather than the 100% polarization that resulted in the Stoner Model with a uniform level spacing. The real magnets all also possess substantial spin-orbit coupling which in our presentation of the Stoner Model we have ignored.

To develop a better intuitive understanding of the properties of the metallic magnets, we can begin by thinking separately about the properties of the *d,f*-like and *s,p*-like bands. These bands will couple and be mixed to varying degrees in different parts of the Brillouin Zone, but near the Fermi level it is often a good approximation to think of them being mostly decoupled, giving separate properties. The *d,f*-like bands will be formed from atomic orbitals that are more localized near the atomic core (think tight-binding model with a small coupling between neighboring atoms). This results in bands that are narrow in energy, with a heavy effective mass, small group velocities, and large exchange parameters. Because these bands are narrow, they can have a large density of states near the Fermi level, but nevertheless their effective masses are sufficiently heavy that they often contribute little to current flow. The *s,p*-like bands will act more like nearly-free electrons, with wide bands and large contributions to current flow. They typically have smaller exchange parameters than the *d,f*-like bands, however, so they are only slightly polarized in equilibrium. A cartoon version of the spin-split densities of states versus energy for a *d,f*-like band and an *s,p*-like band (drawn separately) is shown in Fig. III.1.

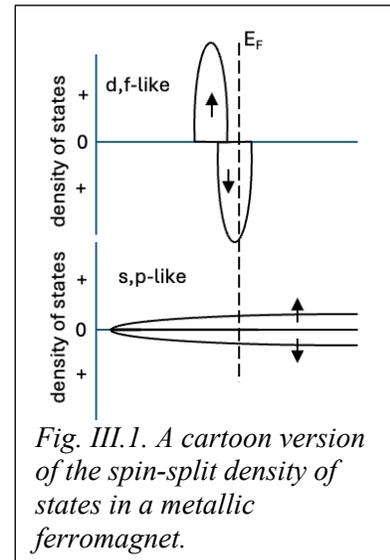

*Fig. III.1. A cartoon version of the spin-split density of states in a metallic ferromagnet.*

More realistic calculations of the band structure for Fe, Co, and Ni are shown in Fig. III.2, taken from ref. 4. These graphs plot the density of states versus energy separately for the two spin species. The majority-spin density of states is shown in the top half of each figure for the different metals – these are the bands that are shifted down in energy by the exchange interaction so that the states are occupied by more electrons compared to the minority-spin states. For all 3 materials one sees a contribution to the density of states with high peaks within about 5 eV from the Fermi level due to multiple *d*-like bands, together with a much smaller background from *s,p*-like states extending over almost the full 15 eV energy range shown in the panels. The difference between the majority and minority bands corresponds reasonably well to a rigid shift in energy, which is what one would expect for our "universal" Hamiltonian with an exchange parameter $J$ that does not vary from state to state within a band. However, unlike our toy model the real band structures do not have a completely empty minority bands. One aspect of the band structures that might be non-intuitive is that for Co and Ni the minority electron bands have a larger density of states at the Fermi level, despite the fact that they are minority bands. For Fe it is the reverse.

**Giant MagnetoResistance (GMR) using metallic magnets**

Magnetoresistance refers to a change in the resistance of a device as a function of the strength and/or direction of an applied magnetic field. Since magnetic orientation can depend on the applied magnetic field, magnetoresistance can also be used to describe a change in resistance of a magnetic device as a function of magnetization orientation.

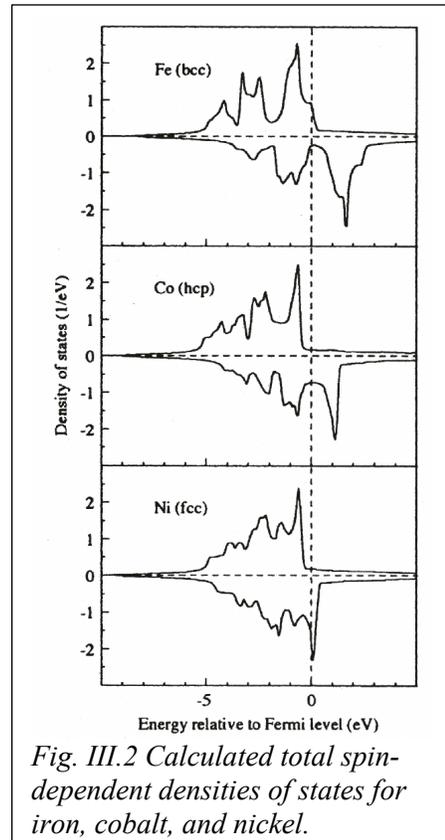

Fig. III.2 Calculated total spin-dependent densities of states for iron, cobalt, and nickel.

The magnetoresistance within a single magnetic material or alloy is generally pretty weak – the resistance will change if the magnetization is rotated between parallel to the current flow versus perpendicular, but typically by only 1% or less of the overall resistance (this effect is known as anisotropic magnetoresistance and it arises from spin-orbit coupling). However if two magnetic layers are present in the same device, there can be much larger changes in resistance as a function of the relative orientations of the magnetizations in different layers. This can be understood in terms of the scattering properties of electrons in the $s,p$-like bands.

Because $s,p$-like bands have much lower effective masses and higher group velocities relative to $d$-like bands, the $s,p$-like bands will carry most of the current in response to an applied electric field and will therefore dominate the resistance properties. A key concept is that the majority and minority-spin $s,p$-like states can have different scattering rates and mean free paths. This comes about because most of the electron scattering processes in a metallic magnet do not alter the spin state. $s,p$-like electrons can scatter into empty $d$-like states with the same spin (but only with the same spin) if such states are available near the Fermi energy. Since the density of states near the Fermi energy can be very different for majority and minority $d$-like states, this means that the mean free path of the $s,p$-like states can also be very different for majority and minority electrons.

Consider the case of Ni or Co for which the minority $d$-like states have a higher density of states near the Fermi level than the majority $d$-like states, and for simplicity of notation let me call majority-spin electrons "spin-↑" and minority-spin electrons as "spin-↓". Spin-↓ $s,p$-like states will scatter more strongly into the available high density of $d$-like spin-↓ states, and will therefore have a shorter mean-free path. This means that a thin film of a magnetic material like

Ni or Co can act like a partial filter for spins. Spin-↓ electrons incident on a thin film of Ni or Co will scatter more strongly within the magnet (and can also scatter more strongly at the interfaces of the film) compared to spin-↑ electrons, with the result that fewer of the spin-↓ electrons will transmit through the film. The end result is that if a spin-unpolarized current is incident onto a thin Ni or Co film from a non-magnetic metal like Cu and passes through into another non-magnetic layer on the other side, the emerging current can be on the order of 30-40%% spin polarized (Fig. III.3).

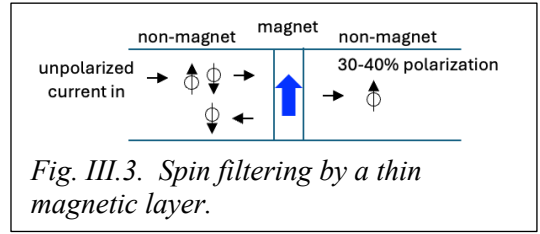

Fig. III.3. *Spin filtering by a thin magnetic layer.*

This spin filtering has particularly dramatic consequences when two magnetic filters are placed in series, for example in a Cu/Co/Cu/Co/Cu 5-layer device (Fig. III.4). If the magnetizations of the two layers are parallel, the spin-↑ electrons can be transmitted with high probability through both layers while spin-↓ electrons will scatter strongly in both layers. Overall, the spin-↑ electrons will "short-out" the device giving a low-resistance state. If, on the other hand, the magnetizations in the two layers are antiparallel, the spin-↑ electrons will scatter strongly in one layer and the spin-↓ electrons will scatter strongly in the other layer. This gives a resistance much higher than for the parallel orientation, some 10's of % change compared to the overall device resistance.

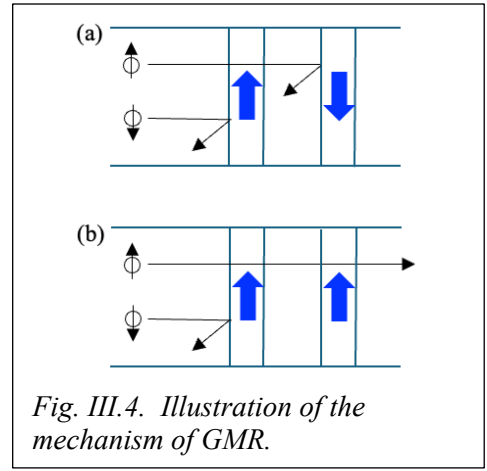

Fig. III.4. *Illustration of the mechanism of GMR.*

This is known as "giant magnetoresistance" (or GMR) because it is much bigger than the magnetoresistance of a single uniform sample of magnetic material. Often, magnetoresistance in magnetic devices is modeled by treating the spin-↑ and spin-↓ electrons as parallel magnetic channels. Within this model if $r_S$ is the small resistance corresponding to the resistance of one layer for a favorable spin state parallel to the magnetization and $R_B$ is the big resistance of one layer for an unfavorable antiparallel spin orientation, the overall resistance of two parallel magnetic layers $R_{\uparrow\uparrow}$ corresponds to $1/R_{\uparrow\uparrow} = 1/(2r_S) + 1/(2R_B)$ and the overall resistance of two antiparallel magnetic layers $R_{\uparrow\downarrow}$ corresponds to $1/R_{\uparrow\downarrow} = 2/(r_S + R_B)$. You can verify that $R_{\uparrow\uparrow} < R_{\uparrow\downarrow}$ when $R_B \neq r_S$.

There is another piece of interesting physics involving closely-spaced magnetic layers separated by a metallic spacer. Because of spin-dependent scattering of electrons at the magnetic interfaces, the spacer layer will contain Friedel oscillations in which the local spin density as well as the charge density undergoes position-dependent oscillations. Each magnetic layer will interact with the spin density near its interface via an exchange interaction. The total energy of both layers interacting with the Friedel oscillations will differ for parallel versus antiparallel magnetic orientations, which means that the

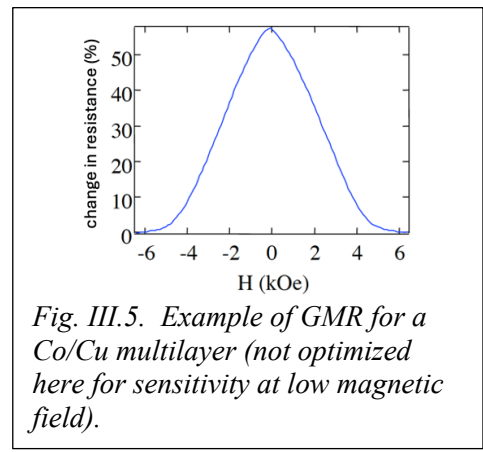

Fig. III.5. *Example of GMR for a Co/Cu multilayer (not optimized here for sensitivity at low magnetic field).*

Friedel oscillations induce an effective interaction between the magnetic layers known as the RKKY interaction (for Ruderman–Kittel–Kasuya–Yosida). The sign of the interaction can be tuned via the thickness of the spacer layer. If the spacer layer is chosen to give a weak antiferromagnetic interaction, the result gives the ingredients to make a very sensitive and small magnetic field detector. At zero applied magnetic field, the antiparallel magnetizations will give a high resistance state, but a small magnetic field can drive them toward a parallel state, lowering the resistance by 10's of % (Fig. III.5). The physics works on very small device scales, down to nanometers, allowing for high spatial resolution and high-sensitivity detection of magnetic fields. These effects were discovered in the late 1980s (for which Albert Fert and Peter Grünberg received the 2007 Nobel Prize) and commercialized quickly for use as the magnetic field sensors in disk drives. I'll have a bit more to say about how magnetic disk drives work next time.

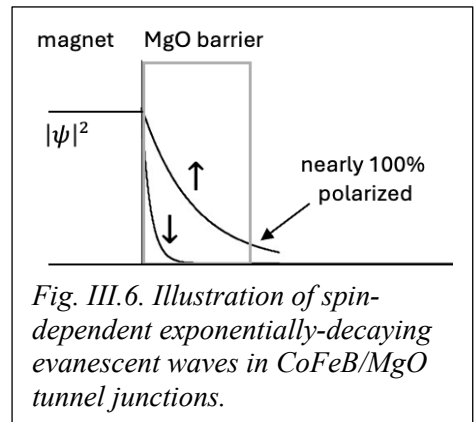

*Fig. III.6. Illustration of spin-dependent exponentially-decaying evanescent waves in CoFeB/MgO tunnel junctions.*

It is possible to get even larger magnetoresistance changes in multilayer magnetic devices by using tunnel barriers rather than metallic metal spacers. An important example is an MgO tunnel barrier sandwiched between CoFeB magnetic alloy electrodes. Because of the symmetries of the CoFeB and MgO band structures, the majority and minority bands in the magnetic electrodes couple to different types of decaying evanescent states within the MgO that have different decay lengths (see Fig. III.6). If the MgO thickness is larger than the shorter of the two decay lengths, you can have a situation in which the tunneling current is almost 100% spin polarized. This leads to large resistance changes between parallel and antiparallel orientations – at least a factor of two change is common in commercial devices. Such tunnel junctions are used primarily in magnetic memory technologies, with the low-resistance parallel configuration and the high-resistance antiparallel configuration serving as the two states for storing a bit of information. The virtues of magnetic memories is that they are naturally non-volatile (not requiring any power to retain information) and they have very high endurance compared to competing types of non-volatile memory because changing the memory state involves only reversing a magnetization without any rearrangements of atoms or any applications of high voltage that can lead to wear-out processes.

**Insulating magnets**

Insulating magnets are far more common that metallic magnets, and they also have a wide variety of important technological applications, including for use in transformers, magnetic recording, electrical generation, and motors. For describing the physics of insulating magnets, the choice of single-particle basis states that we have used so far for metals (delocalized particle-in-a-disordered-box states with no particle symmetry) is not the most illuminating. Instead, it is more convenient to choose basis states localized near particular atoms – e.g., either atomic states or Wannier orbitals. One can work through a very similar argument to the one we did before to analyze which electron-electron interaction matrix elements are large, and therefore most important. By the same logic as before, the dominant contributions to exchange are

$$V_{\text{exchange}} = -\frac{1}{2} \sum_{i,j,s,\sigma} K_{ij} c_{is}^\dagger c_{i\sigma}^\dagger c_{j\sigma} c_{js}$$
$$= -\sum_{i,j} J_{ij} \hat{\vec{S}}_i \cdot \hat{\vec{S}}_j + \text{spin-independent terms.} \quad (3.1)$$

The terms in the sum which give large contributions are those from atoms in close proximity (nearest neighbors, next-nearest neighbors, etc.). This form of a contribution is known as the Heisenberg Hamiltonian. Ashcroft and Mermin have a nice discussion.

However, compared to our previous "universal Hamiltonian" for a disordered metal with delocalized states, the Heisenberg Hamiltonian in insulating magnets has two important differences. The exchange parameter $J_{ij}$ is not the same for all pairs of basis states $i$ and $j$ – it is largest for near-neighbor atoms. Different contributions can also be of either sign, with positive $J_{ij}$ favoring parallel spins and negative $J_{ij}$ favoring antiparallel spins. This can lead either ferromagnetic or antiferromagnetic ordering, or to more complicated types of ordering when not all exchange coefficients within a material have the same sign or the interactions are frustrated, with no spin configuration that can optimize all of the interactions simultaneously. The various possibilities for ferromagnetic or antiferromagnetic effective exchange interactions between two nearby atoms are summarized by the "Goodenough-Kanamori rules", which depend on whether the atomic orbitals involved are filled, half-filled or empty, and on the relative bonding angles when the virtual processes which generate the interaction involve a bridging atom. In the homework problem "The Hubbard Model and the Mott Transition" you will work through a case which gives an antiferromagnetic effective exchange interaction between two atoms.

**Temperature dependence of magnets (including Curie Law)**

Magnets have lower entropy than a corresponding magnetically-disordered state (this is true for both metallic and insulating magnets), so at temperatures higher than some critical magnetic temperature (known as the Curie temperature for ferromagnets, or the Néel temperature for antiferromagnets) there is a magnet→nonmagnet transition (i.e, at high temperature the free energy $E - TS$ becomes lower for the nonmagnetic state which has the higher entropy $S$).

Let's analyze the temperature dependence of the magnetic susceptibility $d\langle M_z \rangle / dB$ where $\langle M_z \rangle$ is the thermal average magnetization of one atom in a magnetic crystal (the component in the direction of an applied magnetic field $B$). It is easiest to do this in a Heisenberg Hamiltonian picture. First, for an isolated atom of total angular momentum $S$ (including any non-zero orbital angular momentum) this is a straightforward stat-mech calculation:

$$\langle M_z \rangle = \langle S_z \rangle g \mu_B = \frac{\sum_{S_z=-S}^{S} g \mu_B S_z e^{g\mu_B S_z B / k_B T}}{\sum_{S_z=-S}^{S} e^{g\mu_B S_z B / k_B T}} = k_B T \frac{d}{dB} \ln\left(\sum_{S_z=-S}^{S} e^{g\mu_B S_z B / k_B T}\right). \quad (3.2)$$

The remaining sum is a simple finite geometric sum, giving

$$\sum_{S_z=-S}^{S} e^{g\mu_B S_z B / k_B T} = \frac{e^{g\mu_B B \left(S+\frac{1}{2}\right)/k_B T} - e^{-g\mu_B B \left(S+\frac{1}{2}\right)/k_B T}}{e^{g\mu_B B / 2 k_B T} - e^{-g\mu_B B / 2 k_B T}}. \quad (3.3)$$

The end result for the thermal-average magnetization is therefore

$$\langle M_z \rangle = g\mu_B \left[ \left(S + \tfrac{1}{2}\right) \coth\left(\frac{g\mu_B B \left(S+\tfrac{1}{2}\right)}{k_B T}\right) - \tfrac{1}{2} \coth\left(\frac{g\mu_B B}{2k_B T}\right) \right]. \tag{3.4}$$

The full temperature dependence for the examples of $S=1/2$ and $S=5/2$ are graphed here

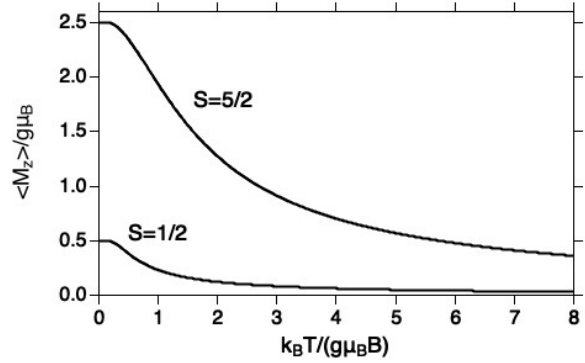

We'll be mostly concerned mostly with the high-temperature behavior. Using a Taylor expansion for $g\mu_B S_z B / k_B T \ll 1$,

$$\langle M_z \rangle \approx \frac{(g\mu_B)^2 S(S+1) B}{3 k_B T}. \tag{3.5}$$

Therefore the susceptibility is

$$\frac{d\langle M_z \rangle}{dB} \approx \frac{(g\mu_B)^2 S(S+1)}{3 k_B T} \propto \frac{1}{T}. \tag{3.6}$$

This is known as the Curie Law, and I'll define $\chi_{\text{bare}} = (g\mu_B)^2 S(S+1)/3 k_B T$.

Now let's consider an atom within a crystal containing exchange interactions, so the moments on neighboring atoms produce an effective magnetic field acting on the atom we are analyzing

$$B_{\text{eff}} = \lambda \bar{M}_z \tag{3.7}$$

with $\bar{M}_z$ equal to the average moment of the near neighbors. Adding this effective field to our previous analysis, we have

$$\langle M_z \rangle = \chi_{\text{bare}}(B + \lambda \bar{M}_z). \tag{3.8}$$

This equation contains two different averages: $\langle M_z \rangle$ is a thermal average at one site and $\bar{M}_z$ is an average value for a specific configuration of the neighbors. For qualitative conclusions, it is natural to try making a "mean field" approximation that $\langle M_z \rangle = \bar{M}_z$. This turns out to be a surprisingly bad approximation for quantitative purposes, but it is OK for the qualitative arguments that I will make. Within the mean field approximation, then

$$\langle M_z \rangle = \chi_{\text{bare}}(B + \lambda \langle M_z \rangle) \text{ or } \langle M_z \rangle = \frac{\chi_{\text{bare}}}{1 - \chi_{\text{bare}} \lambda} B. \tag{3.9}$$

For weak ferromagnetic interactions ($\lambda > 0$), the average magnetization is therefore enhanced compared to the non-interacting value, and as the interaction parameter $\lambda$ is increased one can expect an instability to ferromagnetic ordering when the susceptibility diverges, at $\chi_{\text{bare}} \lambda = 1$.

In the high-temperature limit we have $\chi_{\text{bare}} = Q/T$ with $Q$ a constant, so in this regime

$$\langle M_z \rangle = \frac{Q}{T-Q\lambda} B, \tag{3.10}$$

and the magnetic susceptibility is proportional to $1/(T - T_0)$, where $T_0$ is a constant, rather than just $1/T$. This dependence can be used to detect a tendency toward ferromagnetism or antiferromagnetism even at temperatures well above the transition temperature. If one does a fit to high temperature magnetic susceptibility data and finds a nonzero value of $T_0 > 0$, this indicates the material has a tendency toward ferromagnetism and you can expect a Curie temperature of order $T_c \sim T_0$. If one finds instead from a fit that $T_0 < 0$, this indicates a tendency toward antiferromagnetism, again with a Néel temperature of order $T_N \sim |T_0|$. These are only order-of-magnitude estimates for the critical temperatures, however, both because the high-temperature limit generally is not accurate near the ordering temperature and because the mean-field approximation is not exact.

**Magnetic Domains**

In our discussion of magnets so far, we've assumed that the magnetic sample is uniformly polarized – that the spins point in the same direction everywhere as a function of position. This is accurate for sufficiently small ferromagnetic samples, but for samples larger than 100 nm to 10 μm (depending on the material) the ground-state configuration will generally be spatially non-uniform at zero applied magnetic field. This brings us to the topic of magnetic domains and other magnetic textures.

The direction in which spins will point within a magnet is determined by several competing forces. Importantly, these forces have differing characteristic length scales. Situations in which there are competing forces having differing length scales are common in condensed matter physics, and the result very generally is pattern formation. What are these competing forces within magnets?:

1. Exchange interaction. We have already seen that this favors spins aligned. For non-relativistic exchange, this interaction is rotationally symmetric so it does not favor any one particular orientation. In both metals and insulators it is short-ranged, acting only over atomic length scales. Any local misalignment of the magnetic moment costs exchange energy, leading to an energy increase usually modeled as $\Delta E_{\text{exchange}} \propto \int d^3r \left[ (\nabla M_x)^2 + (\nabla M_y)^2 + (\nabla M_z)^2 \right]$.

2. Magneto-crystalline anisotropy. This type of anisotropy results from spin-orbit interactions that break rotational symmetry in spin space. It is also short ranged. Magneto-crystalline anisotropy can favor alignment of the magnetization along particular crystal axes. An example is Co with a hexagonal crystal structure which has an easy axis in the $\pm \hat{c}$ direction, perpendicular to the hexagonal plane. Anisotropy can also favor alignment perpendicular or parallel to interfaces. This is very important for thin-film devices, where interfacial anisotropy can strongly favor out-of-plane magnetic configurations and thereby allow tiny magnetic memory devices (down to 10 nm or so) to be stable against thermal fluctuations.

3. Magnetic dipole interactions. We've seen that for nearest-neighbor interactions the dipole energy is very weak, ~$10^{-4}$ eV compared to 0.1-1 eV for exchange. However, dipole interactions are long-ranged. If you integrate over a big enough sample, the total dipole energy is no longer negligible, and it can come into competition with exchange and magneto-crystalline anisotropy forces. Dipole interactions can favor spatially-nonuniform configurations at the expense of exchange and/or anisotropy energies and are the main driving force in the stabilization of domains.

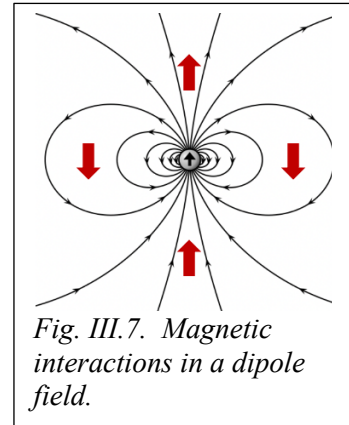

Fig. III.7. Magnetic interactions in a dipole field.

Dipole fields are somewhat complicated, favoring aligned spins in a head-to-tail arrangement but anti-aligned spins when side to side (Fig. III.7). Nevertheless, the relative value of magnetic-dipole energy for different magnetization configurations can be understood in a simple intuitive picture using the concept of magnetic poles.[5] The dipole energy can be written in one form as a double integral over the spins in a sample generating the magnetic field and the spins which feel that field as a Zeeman interaction:

$$E_{\text{dipole}} = \int d^3r \, d^3r' \left[ -\vec{M}(\vec{r}) \cdot \vec{B}(\vec{M}(\vec{r}')) \right], \quad (3.11)$$

where $\vec{B}(\vec{M}(\vec{r}'))$ is the dipole magnetic field generated from the magnetization at position $\vec{r}'$. After integration by parts twice and a bit of vector calculus, this can also be re-written in an alternative, fully equivalent form:

$$E_{\text{dipole}} = \frac{\mu_0}{8\pi} \int d^3r \, d^3r' \left[ \nabla \cdot \vec{M}(\vec{r}) \right] \frac{1}{|\vec{r}-\vec{r}'|} \left[ \nabla' \cdot \vec{M}(\vec{r}') \right]. \quad (3.12)$$

This can be interpreted as analogous to an electrostatic Coulomb energy. The divergences of the magnetization act like effective charge densities (or effective magnetic pole densities) that interact with each other in analogy to a 1/(distance) Coulomb potential. We can use this second expression to visualize the relative energy of different magnetic configurations.

Let's take as a building block a cube of magnetic material small enough that the magnetization is to a good approximation uniform within the cube (Fig III.8(a)). The only divergences of the magnetization are therefore at the cube's surfaces where the magnetization "starts up" (positive magnetic pole density) or "ends" (negative magnetic pole density). Now let's imagine putting these cubes together in different configurations. Suppose we have a long, thin, needle-like sample. If the magnetization is perpendicular to the long axis of the needle (Fig. III.8(b)), this corresponds to forcing our cubes to assemble with effective charges of the same sign on different cubes close together. This means that this type of assembly must overcome strong repulsive forces, or that this orientation of the magnetization has a relatively high energy. On the other hand, if we assemble the cubes with the magnetization pointing along the long direction of the needle (Fig. III.8(c)), positive effective charge density from one cube is immediately adjacent to negative charge density from the neighboring cube. This corresponds to an attractive force between cubes, meaning a relatively low energy. The magnetization stays uniform in both configurations, so there is no difference in exchange energy. The overall lesson here is that dipole interactions can produce a shape-dependent anisotropy. For a needle-shaped sample, a magnetization aligned the long direction of the needle gives the lowest dipole energy and hence the lowest overall magnetic energy.

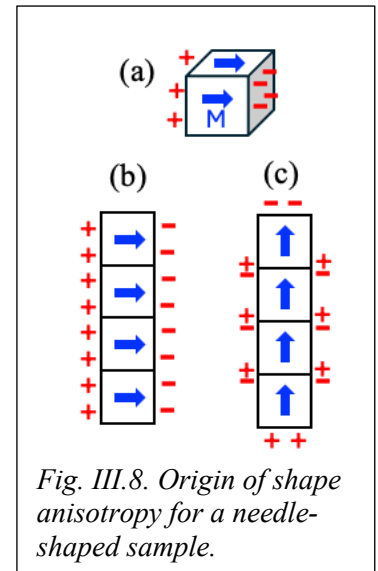

Fig. III.8. Origin of shape anisotropy for a needle-shaped sample.

Analogously, for magnetic thin films dipole interactions produce a shape anisotropy that favors an in-plane magnetization. This can put the shape anisotropy in competition with interface anisotropy. For very thin films of well-chosen heterostructures, the interface anisotropy can overwhelm the dipole energy and produce perpendicular magnetization, but the typical situation is that thin films have in-plane magnetic anisotropy because of the shape anisotropy. (Both in-plane and perpendicular films are useful for different applications.)

4. Dzyalonshinskii-Moriya Interactions (DMI). DMI can be another source of spatially-nonuniform magnetic configurations, in addition to dipole interactions. It corresponds to a local energy density of the form $E_{\text{DMI}} = -\vec{D} \cdot (\vec{M}_1 \times \vec{M}_2)$ where $\vec{D}$ is a material-dependent vector and $\vec{M}_1$ and $\vec{M}_2$ correspond to adjacent spins. It therefore acts as an effective exchange interaction that favors a 90° orientation of neighboring magnetization elements (rather than a parallel alignment), and is short-ranged. DMI can be present only in samples having broken inversion symmetry (which selects a specific direction for $\vec{D}$) and strong spin-orbit coupling. Two examples where DMI can be important are low-symmetry crystals or interfaces between magnets and heavy metals. In both cases, strong DMI can result in spiral magnetic structures or topological vortex-like textures known as skyrmions.

Let's now consider various possible magnetic configurations for a circular thin-film sample. (We'll ignore DMI here.)

A spatially uniform magnetization (Fig III.9(a)) will have the minimum possible values of exchange and magneto-crystalline anisotropy a relatively high dipole energy.

If you form two magnetic domains (Fig. III.9(b)), one in each semi-circle, this will reduce the dipole energy compared to a uniform configuration, but at the cost of higher exchange and anisotropy energies within the domain wall, where the magnetization must gradually rotate between the up and down domains and possibly away from a favored crystal axis. (The width of a domain wall is determined by a competition of exchange and anisotropy energies – see one of the problem sets.)

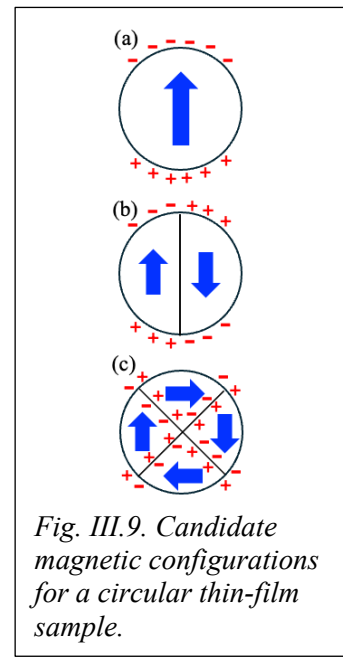

Fig. III.9. Candidate magnetic configurations for a circular thin-film sample.

With four magnetic domains (Fig. III.9(c)), we can minimize the dipole energy because there are almost no uncompensated magnetic poles, but at the cost of even larger exchange and anisotropy energies. This is called a closure domain.

Which state wins? This depends in part on sample size. For sample diameters less than about 100 nm - 1 μm, exchange usually wins and samples are approximately single-domain. For sample sizes greater than 1-10 μm, you almost always have domains in the lowest-energy state at zero applied magnetic field. Complicating matters even further, though, magnetic samples are often not in the lowest-possible energy state because domain walls can become pinned at defects or other non-uniformities in a material so that they do not always immediately move to establish the lowest-possible energy configuration. This means that the domain configuration of samples is usually history dependent and hysteretic. We'll return to this topic next time when we discuss magnetic dynamics.

Sophisticated software now exists to calculate both the static magnetic configuration of samples and magnetic dynamics. This type of numerical modeling is known as "micromagnetics".

**Observing magnetic domains**

There are lots of nice experimental techniques for characterizing magnetic domains, with different sensitivities, spatial resolutions, and ease of use:

- Scanning magnetic force microscopy. Need to be careful that the magnetic field from the tip does not alter the sample being measured.
- Optical techniques – Kerr and Faraday rotation. Usually the spatial resolution is diffraction limited, but good sensitivity and can be ultra-fast to measure magnetic dynamics.
- GMR sensors. The workhorse for sensing magnetic fields within magnetic disk drives.
- Scanning SQUIDs or Hall bars. Can be more sensitive than GMR sensors, but not scalable to give the few-nanometer resolution that can be achieved with GMR sensors.
- Scanning NV center microscopy. Using one or many NV-centers within diamond as quantum sensors can provides both high sensitivity and excellent spatial resolution, but the set-ups are involved and not fast.
- Deflection of electron beams in scanning transmission electron microscopy. Enabled for high performance recently by high-dynamic-range pixel-array detectors developed by David Muller and Sol Gruner at Cornell.
- Scanning electron microscopy with polarization analysis of the secondary electrons that are emitted.
- X-ray microscopy with magnetic dichroism. Neat but requires beamtime.
- Magneto-thermal microscopy. Developed by Greg Fuchs at Cornell.

The Wikipedia page for "Magnetic domain" has nice images of domains obtained by some of these techniques.

**What have we learned?:**

- Different scattering properties for majority and minority electrons within metallic ferromagnets allow these ferromagnets to act like spin-dependent filters. This allows for large magnetoresistance in magnet/spacer/magnet multilayers. Even larger magnetoresistance is possible due to spin-dependent tunneling in magnetic tunnel junctions.

- Measurements of the temperature dependence of magnetic susceptibility even well above any magnetic transition temperature can indicate whether a material will have a transition to a low-temperature ferromagnetic or antiferromagnetic state.

- Competition between exchange interactions, magneto-crystalline anisotropy, dipole interactions, and DMI within ferromagnets can lead to complicated pattern formation in the form of magnetic domains.

# IV. Even More About Magnets + Their Applications

*Topics: 1. Domain wall motion in response to an applied magnetic field, and how this leads to magnetic hysteresis.*

*2. How the dynamics of single-domain magnetic nanoparticles differ from the dynamics of larger magnetic samples, and how this is useful for making hard magnets.*

*3. A brief introduction to spin waves.*

*4. A tour of recent research related to magnetism – the development of improved permanent magnets, magnetic disk drives, and magnetic random access memory.*

## Domain wall motion and hysteresis curves

We have discussed that magnetic samples larger than about 1 μm in size will generally contain domains, while much smaller samples can be single-domain. Next I want to talk about what happens in both regimes when a changing magnetic field is applied.

In both cases, if you apply an external quasi-static magnetic field $B$ to a magnet, the magnetization $M$ is subject to a Zeeman interaction

$$E_{\text{Zeeman}} = -\vec{M} \cdot \vec{B} \tag{4.1}$$

That favors alignment of the magnetization with the magnetic field.

For samples containing domain walls, in response to an applied magnetic field the energy can be lowered in a simple way by shifting domain walls. The idea is that from a microscopic point of view, a domain wall consists of a gradual rotation of the magnetization as a function of position (Fig. IV.1), because an abrupt discontinuity in the magnetization would require a large cost in exchange energy. When a magnetic field is applied, it can be easy for each spin to rotate slightly in the direction of the field. This will have the overall effect of moving the midpoint of the domain wall (Fig. IV.2). In general even small changes in magnetic field will drive changes in domain-wall positions, but the process is not entirely reversible – if the applied magnetic field is returned to its original value a domain wall might not return to its original position. This non-reversibility occurs for two reasons: (1) Domain walls can become "pinned" at sites of defects or material inhomogeneities, and (2) Sometimes domains might be entirely eliminated (or created) by an applied field, and upon removing the field the total number of domains might be different from the original number. There are energy barriers to nucleating domains that make the process of domain creation and destruction not completely reversible. The end result is that magnetization versus magnetic-field curves can be history dependent, or in other words they can exhibit "hysteresis".

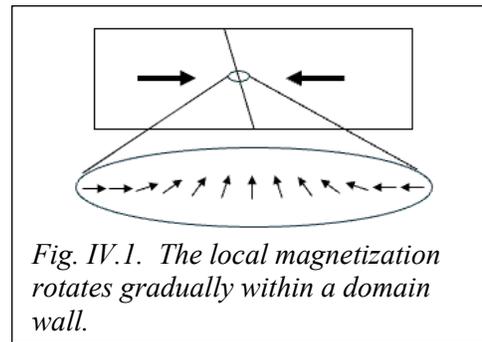

*Fig. IV.1. The local magnetization rotates gradually within a domain wall.*

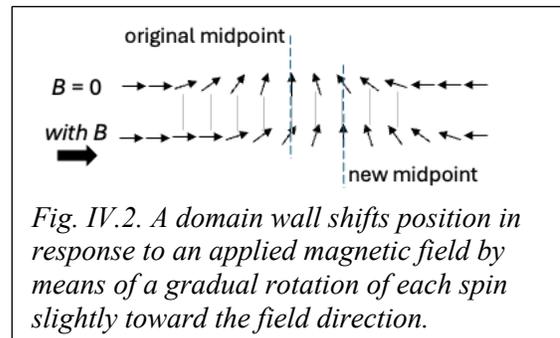

*Fig. IV.2. A domain wall shifts position in response to an applied magnetic field by means of a gradual rotation of each spin slightly toward the field direction.*

A typical hysteresis curve for a sample larger than the micron scale is shown in Fig. IV.3. Let's start at a large positive value of magnetic field, large enough that the Zeeman interaction forces a fully-saturated value of the magnetization ($M_s$). Then imagine sweeping the magnetic field down to decreasing values. Magnetic domains will generally start to form before the applied magnetic field reaches zero, so that the magnetization will begin to decrease from the fully-saturated value. However, typically the average magnetization will not decrease all the way to zero at zero magnetic field, even though this might be the lowest-energy configuration. The magnetization value at $B=0$ is called the remanent magnetization. The magnitude of the reversed magnetic field required to drive the average magnetization to zero is called the coercive field. Further sweeping of the magnetic field to negative values eventually drives the magnet to full saturation in the negative direction. Reversing the direction of the magnetic field sweep will then sweep out the rest of the hysteresis curve.

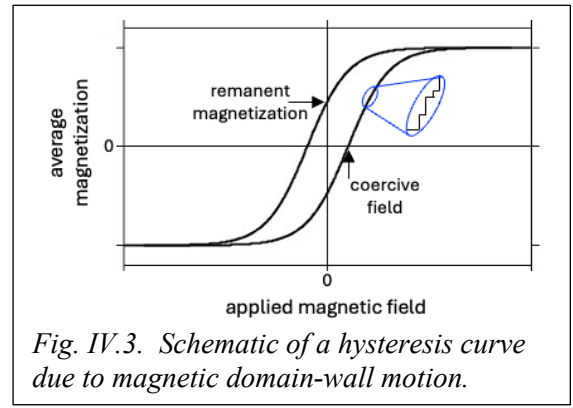

*Fig. IV.3. Schematic of a hysteresis curve due to magnetic domain-wall motion.*

If one looks closely at the magnetization as it undergoes changes, it will not evolve in s completely continuous manner. Instead, it changes in a series of small steps. This is due to domain walls being momentarily stuck at pinning sites and then releasing to be captured by new pinning sites. The result is known as Barkhausen noise. The process is analogous to avalanches, and can exhibit interesting statistics with respect to the size distribution of the magnetization steps.

Different magnetic materials are described as "soft" or "hard" depending on the amount of hysteresis in the hysteresis loop. Soft magnets have little hysteresis, in that their domain-wall pinning is sufficiently weak that the average magnetization decreases almost all the way to zero at zero applied magnetic field, and the area inside the hysteresis loop is small. The area inside the hysteresis loop is identically equal to the heat absorbed by the magnet (per unit volume) during one full cycle around the loop. Soft magnets are therefore useful in electrical transformers for which heat dissipated in the magnet detracts from the energy efficiency.

Hard magnets have lots of hysteresis. An ideal hard magnet would retain the full value of its saturated magnetization at zero applied magnetic field and would also have a large magnitude for its coercive field. A useful figure of merit is therefore (remnant magnetization)×(coercive field), which has units of energy per unit volume. Hard magnets are required for permanent magnets (of the types used, e.g., in motors and electrical generators) and for magnetic information storage.

If you want a hard magnet, it is useful to make it from an assembly of single-domain nanoparticles (i.e., diameters below about 100 nm) because this allows the elimination of domain walls that can allow the magnetization to relax toward zero at low values of applied magnetic field via domain-wall motion. Magnetic reversal in single-domain nanoparticles must happen by a different mechanism than domain-wall motion, and typically requires much larger coercive fields.

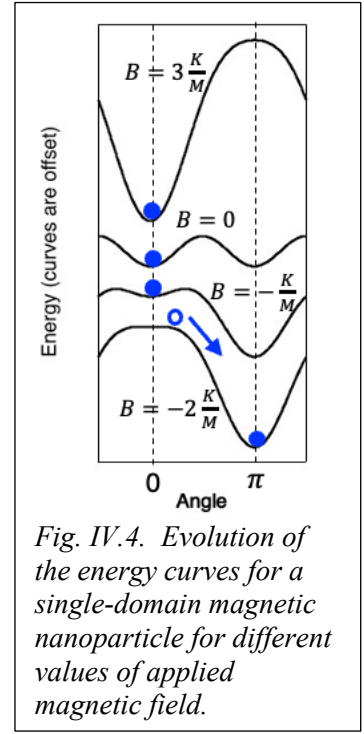

Fig. IV.4. Evolution of the energy curves for a single-domain magnetic nanoparticle for different values of applied magnetic field.

Suppose you have a single-domain magnet, with (for simplicity) an easy anisotropy axis in the $\hat{z}$ direction and with the magnetic field also swept along the same axis. How will the ground-state orientation of the magnetization ($\theta$) vary as the magnetic field is swept? We can analyze the angle-dependent magnetic energy as the sum of a Zeeman term ($-\vec{M}\cdot\vec{B} = -MB\cos\theta$) and a uniaxial anisotropy term ($-K\cos^2\theta$). Suppose that we start with the magnetization pointed in the $+\hat{z}$ direction (i.e., $\theta=0$) with a large positive magnetic field and sweep the magnetic field toward negative values. The evolution of the energy curves are shown in Fig. IV.4. With a large positive magnetic field of $B = 3K/M$, the Zeeman energy dominates and the $\theta=0$ angle is the only stable equilibrium point. If the applied magnetic field is reduced to zero, the $\theta=0$ and $\theta=\pi$ angles become degenerate (as required by time-reversal symmetry) but there is an energy barrier in between them so that the magnet remains at its original $\theta=0$ angle. A large negative magnetic field of $B = -K/M$ can be applied and still the magnet won't reorient, because the $\theta=0$ angle remains metastable even though it is not the global energy minimum. Only when the $\theta=0$ angle becomes fully unstable (for very large fields more negative than $B = -2K/M$ in our model) or when the magnet can escape the metastable energy well by thermal activation or tunneling will the magnet reorient and reach the lower-energy $\theta=\pi$ state. The hysteresis curve in this example will be fully rectangular (Fig. IV.5).

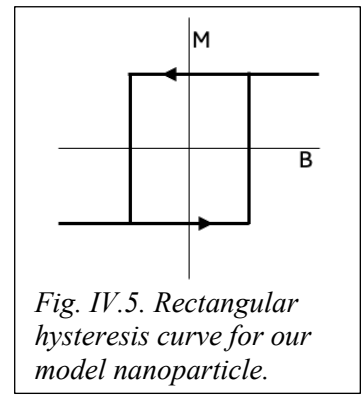

Fig. IV.5. Rectangular hysteresis curve for our model nanoparticle.

**Spin Waves/Magnons**

So far I've just talked about the ground states of magnets and the dynamics in response to quasi-static magnetic fields. What about excitations?

Metallic magnets still have electron-hole-type excitations similar to Fermi liquids (with the slight complication in magnets that you need to account for changes in exchange energy to be precise about their energies). In addition, both metallic and insulating magnets have low-energy excitations corresponding to spatial deviations from uniform magnetization. Classically you can think of these as waves of precessing magnetization (spin waves). Quantum mechanically these are bosonic states with angular momentum $-\hbar$ compared to the uniform magnetic state (magnons).

Spin waves/magnons can transmit angular momentum with no flow of charge. They therefore represent a type of spin-charge separation not present in Fermi liquids. In a crystal,

Bloch's theorem says that the energy eigenstates of spin waves/magnons can be written in a plane-wave form, with dispersion curves dependent on crystal momentum $\vec{k}$. I'll leave the solution of a simple model to the problems. In general, for long wavelengths (small $k$) the dispersion curves go like $\omega(k) \propto k^2$ for ferromagnets and $\propto |k|$ for antiferromagnets. The $k=0$ mode for a ferromagnet corresponds to spatially-uniform magnetization precession, and can be readily excited by a resonant GHz-frequency applied magnetic field. This type of measurement is known as ferromagnetic resonance (FMR), and is a very useful technique for characterizing the magnetic anisotropy and energy-loss processes. The frequencies of shorter-wavelength modes scale into the THz regime. One difference between magnon dispersion curves and phonon dispersion curves is that magnon dispersions are more highly nonlinear. Therefore multi-magnon scattering processes are more important for understanding magnon dynamics compared to the analogous processes for phonons.

There is considerable research activity at present which aims to utilize magnons for low-loss information and energy transfer in advanced microelectronics – a field known as "magnonics". The motivation is that for a given frequency magnons have wavelengths that are orders of magnitude smaller than electromagnetic waves used in microwave or photonic interconnects, allowing for better scaling. Magnons also possess potentially useful non-reciprocity (one-way travel), and their strong nonlinearities could allow many of the tricks used in nonlinear optics to be adapted for magnon manipulation (e.g., 3- and 4-wave mixing, amplification, frequency conversion, etc.). For practical technologies, further progress is needed in developing efficient approaches for exciting and detecting magnons, and also for manipulating them while in transit to make useful switches, modulators, and amplifiers.

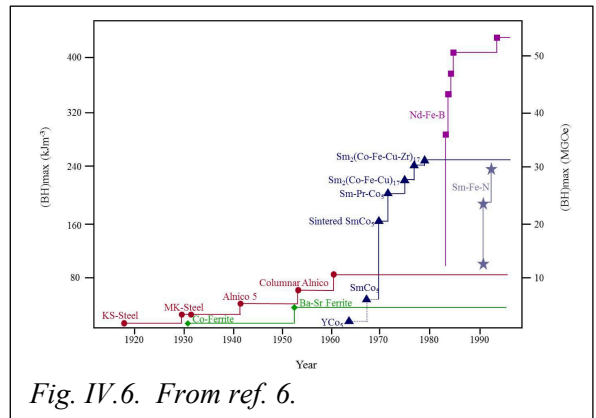

*Fig. IV.6. From ref. 6.*

**Some vignettes about magnetic technologies:**

*Development of improved hard ferromagnets.* Hard ferromagnets are essential components of electric motors, actuators, and generators, with sales of the magnets worth more than $20 billion/year. As noted above, the quality factor used for characterizing hard magnets is their "energy content" equal to the remanent magnetization times the coercive field. Higher energy content allows a similar strength of magnet to be made in a smaller volume with less weight, and is particularly important in automotive markets. A chart of progress in different types of hard magnets during the 20th century is shown in Fig. IV.6. The best-performing class of hard ferromagnets today remains the "rare-earth" NdFeB family developed in the 1980's. Figure IV.7 shows pictorially how these provide higher performance compared to naturally-occurring lodestone or older ferrite magnets.

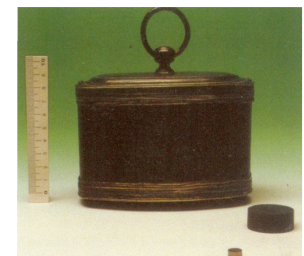

*Fig. IV.7. A lodestone magnet from the 1750's and ferrite and NdFeB magnets used in modern appliances. Each store the same magnetic energy. From ref. 6.*

The key discoveries which enabled NdFeB magnets were made independently by researchers at General Motors and Sumimoto Special Metals, and they only learned of each others work when both teams happened to make presentations at the same magnetism conference in 1983. Optimization of the materials properties required careful control over the crystalline phase (to obtain strong magneto-crystalline anisotropy) and also, critically, the grain size of the material with some magnetic decoupling between grains. If the grain size was too large, on the order of 100 nm or above, domain wall motion resulted in soft magnetic properties. If too small, thermal fluctuations could begin to degrade the magnetic stability. (When magnetic nanoparticles are too small to be stable against thermal fluctuations, the result is called superparamagnetism.) Grain sizes in the correct regime could be controlled by a careful process of sintering (developed by Sumimoto) or by melt spinning in which a molten starting liquid was rapidly cooled against a spinning metal wheel (developed by General Motors). Figure IV.8 shows how dramatic a difference the grain size can make for the magnetic properties of these hard magnets.

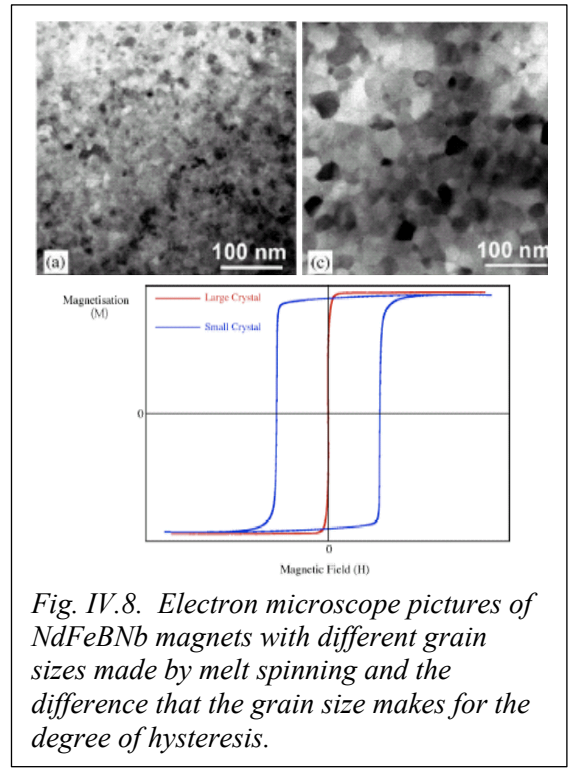

*Fig. IV.8. Electron microscope pictures of NdFeBNb magnets with different grain sizes made by melt spinning and the difference that the grain size makes for the degree of hysteresis.*

*__Magnetic disk drives__*. Magnetic disk drives currently provide the lowest cost per bit for storing large amounts of information in a rapidly-accessible format. For personal computers they have recently been largely replaced by solid state drives which utilize silicon-based flash memory, because flash memory can provide higher speed, better ruggedness, and lower power consumption. However, disk drives remain several times cheaper per bit to produce, and more than 100 million new units continue to be sold per year.

The idea of a magnetic disk drive is that information is stored in the form of magnetic domains in a thin film of material on a spinning platter, and an actuator arm that can swing across the platter is used to write and read information. This form of information storage was first introduced by IBM in the 1950's. Since that time, a series of innovations have improved the storage media, the writing process, and the reading process, so that the storage density of disk drives has improved exponentially with an average rate even faster than Moore's law for electronics. Figure IV.9 shows the evolution of the storage density over time. Overall, the storage density of disk drives has improved by a factor of 100 million since their invention, within the same basic design structure. Other than perhaps the transistor, I don't know of any other human technology with a similar factor of improvement.

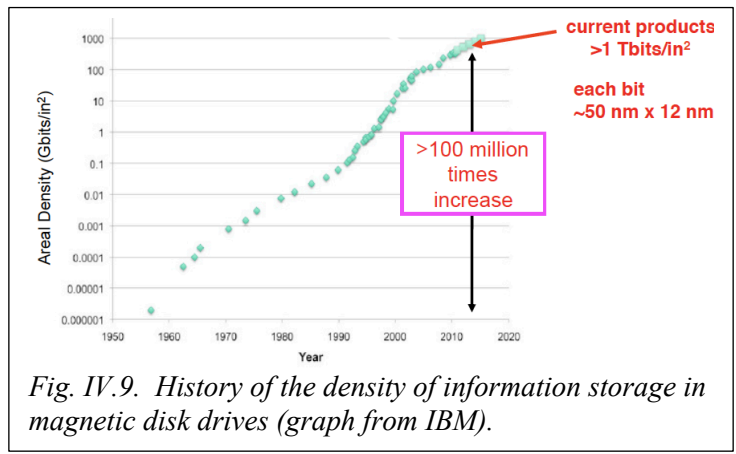

*Fig. IV.9. History of the density of information storage in magnetic disk drives (graph from IBM).*

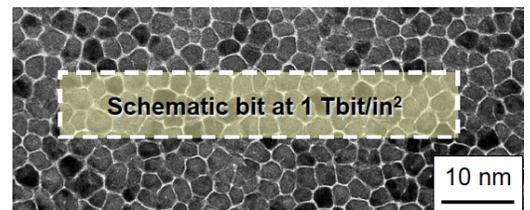

The magnetic thin film which stores information in a disk drive is typically a cobalt-based alloy engineered to have domains on the order of 5 nm in order to behave as good hard magnets (i.e., no domain walls) for data retention (Fig. IV.10). In modern drives, these films have perpendicular magnetic anisotropy. The insides of a disk drive are shown on various length scales in Fig. IV.11. Writing is achieved using an extremely small electromagnet to apply a local magnetic field to the platter. A thin-film wire coil wraps through a U-shaped soft magnetic core and produces a strong Oersted field due to the large magnetic permeability of the soft magnet. The magnetic field lines are guided to the platter by the pole pieces of this iron core. One pole piece is made narrower than the other, producing a focusing of magnetic field lines to a length scale comparable to 10 nm, enabling writing of very small bits. Reading of the magnetic fields emanating from the bits is performed by a nanoscale giant magnetoresistance (GMR) or tunneling magnetoresistance (TMR) device. This is typically engineered with magnetic shields to focus the field lines toward the sensor and a small bias magnet to orient the magnetic layers within the sensor at approximately a 90° angle. This enables the largest resistance change for small deflections of the magnetic layers.

*Fig. IV.10. Electron microscope image of nanoscale grains in the magnetic film for a disk drive.*

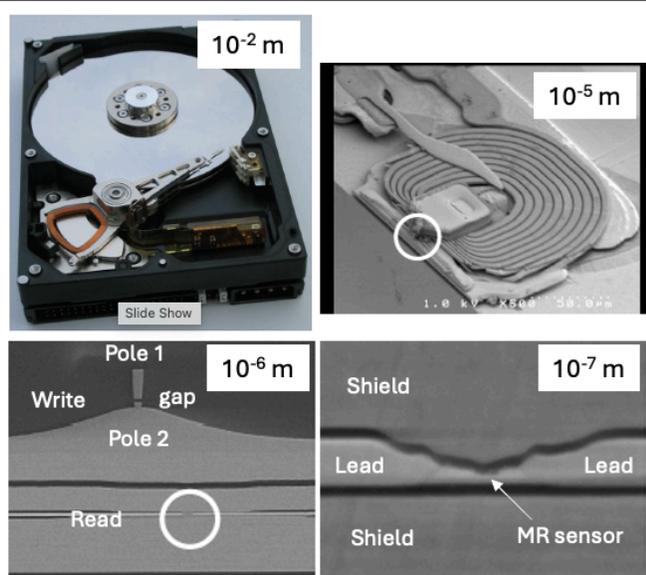

*Fig. IV.11. Pictures of the insides of a magnetic disk drive. Credit: Tuyet Nguyen and Monica Vargas*

Magnetic disk drives have reached the stage where they are not far from the physical limits for their storage density. To maintain stability against thermal fluctuations for smaller bits requires increased magnetic anisotropy, but that then makes the bits harder to write using an applied magnetic field. Recent innovations have incorporated local heating or local microwave fields to assist the applied magnetic field in the writing process.

*Magnetic random access memory (MRAM)*.  Magnetic memory technologies naturally provide the benefits of nonvolatility (information is retained with no power applied) and high endurance. The high endurance results from the fact that in magnetic devices information is written simply by reorienting a magnetization with no movement of atoms required. Other competing types of nonvolatile memories struggle to achieve high endurance because they involve either atomic motion that eventually leads to wear-out (e.g., ferroelectric or phase-change memories) or they require large electric fields that lead to material degradation. Silicon-based flash memories within current solid-state drives, for example, generally wear out after only about 1 million write cycles. This is OK for solid-state drives in which information is written infrequently and generally stored for a long time, but much higher endurance ($\sim 10^{15}$) is desired for "embedded" applications in which nonvolatile memory is tightly integrated with computer processors and written every few nanoseconds.

In magnetic random access memory (MRAM), information is stored using magnet tunnel junctions, each of which can be stable in two different configurations, a low-resistance state in which the electrode magnetizations are parallel versus a high-resistance state in which they are antiparallel. Reading of these bits is performed just by detecting the tunnel-junction resistance, which can vary by about a factor of 2 or 3 between the two configurations. Writing the memories (i.e., switching between the parallel and antiparallel states) is the trickier aspect of getting MRAM to work. You might imagine that this could be achieved using the Oersted magnetic field from current flowing in wires near the sample. However, if you work through the numbers it turns out that it requires a destructive level of current density in a nanoscale wire to create enough magnetic field to switch a nearby thermally-stable nanoscale magnetic device. Instead, to reorient magnets on this scale it is much more efficient to use exchange interactions rather than magnetic fields. The way this is done in the present generation of technology is to use a phenomenon known as spin transfer torque (Fig. IV.12). Consider what happens when a spin-polarized electrical current is incident onto a magnetic layer with a spin direction in the current canted relative to the layer's magnetization. Within the frame of the magnet's magnetization, you can think of the canted spins as a superposition of spin-up and spin-down states. The magnet will act like a filter for spins in the same way we analyzed when we discussed GMR – on average spin-up electrons will be preferentially transmitted and spin-down electrons reflected. After the process of filtering is complete, the incident component of spin that was perpendicular to the magnetization does not come out, which means that as part of the filtering process the magnetic layer must necessarily absorb this transverse angular momentum. (Microscopically, the incoming spin precesses around the magnetization of the magnet, with electrons following different paths precessing by different phases; upon exiting the magnet the electrons are generally fully phase-randomized, so that the average transverse spin exiting the magnet is zero.) By conservation of angular momentum, the magnetic layer therefore experiences a torque which tends to tilt the direction of the magnetization toward the spin direction of the incident current.

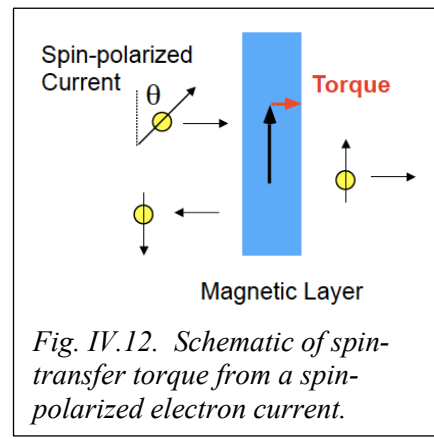

*Fig. IV.12. Schematic of spin-transfer torque from a spin-polarized electron current.*

The simplest form of spin-transfer-torque magnetic random access memory (STT-MRAM) consists of a magnetic tunnel junction with two magnetic layers, one with the magnetization held fixed (generally using interactions with an antiferromagnetic layer) and serving as a spin filter to make spin polarized current and the other free layer with magnetization free to be switched between orientations parallel or antiparallel to the fixed layer via the spin-transfer torque. Switching from antiparallel to parallel orientations is achieved by passing a relatively large flow of electrons from the fixed layer to the free layer, while switching from parallel to antiparallel is achieved by passing a large flow in the opposite direction to reverse the sign of the spin-transfer torque. Reading is achieved by passing a smaller current through the same two electrodes. Because the write and read currents follow the same path, one of the challenges in engineering STT-MRAM to avoid "read-disturbs" in which the read current accidentally produces a switching event. Careful optimization of materials and fabrication processes have allowed this problem to be controlled, and STT-MRAM has begun widespread commercialization for applications in which high-endurance nonvolatile memory is desired. In particular, it has supplanted flash memory at the 28 nm node and below.

Considerable ongoing research is aimed to further improve magnetic memories. STT-MRAM is about 1000 times more efficient than using Oersted magnetic fields for switching, but nevertheless STT-MRAM still requires substantial writing currents – many tens of microamps per bit. The transistors needed to source such currents are much larger than the minimum-area transistors used in highly-scaled CMOS circuits, and each memory bit requires at least one transistor for control, so that in the end it is the transistors which limit the density of STT-MRAM rather than the magnetic components themselves. The basic problem is that spin-transfer torque from a spin-polarized current is subject to a fundamental quantum limit on efficiency – no more than about $\hbar/2$ angular momentum transferred per electron passing through the tunnel junction. Research going on now is investigating alternative physical mechanisms that might allow this limit to be avoided. One possibility is to generate spin currents using spin-orbit interactions in a way that allows each electron passing through the device to transfer angular momentum to a magnetic layer several times, rather than just once, to make spin-orbit torque MRAM, or SOT-MRAM (Fig. IV.13). Other research aims to achieve efficient magnetic switching by applying electric fields with minimal current flow, by electric-field modulation of magnetic anisotropy, exchange interactions, multiferroic materials (i.e., both ferroelectric and magnetic), or strain. Overall, magnetic memories are in a very active competition with several other potential emerging technologies for making useful nonvolatile memories, and it is not clear which alternatives will eventually emerge to be the most successful.

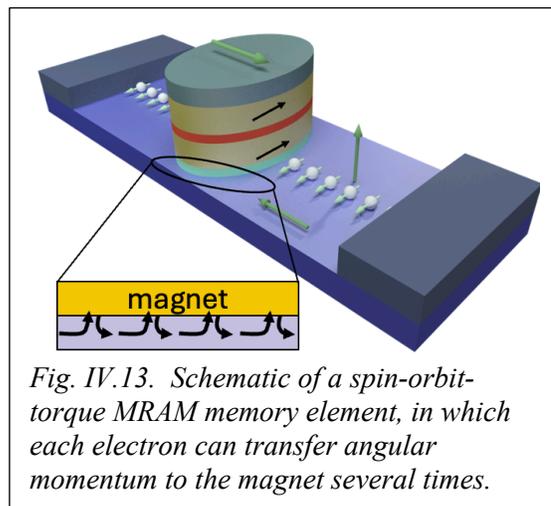

*Fig. IV.13. Schematic of a spin-orbit-torque MRAM memory element, in which each electron can transfer angular momentum to the magnet several times.*

**What have we learned?:**

- Low barriers to domain-wall motion allow for soft ferromagnets with little hysteresis, useful for minimizing energy loss in transformers.
- To make hard magnets with a large amount of hysteresis, a good strategy is to fabricate nanoparticles from a material with strong magnetocrystalline anisotropy, with the nanoparticle size small enough that the particles will not host domain walls that allow the magnetization to be reversed at low coercive fields. Variations of such nanoparticle-based materials are useful for permanent magnets in motors, generators, and actuators, and for information-storage media in disk drives.

- Magnets host low-energy spin-wave excitations (also known as magnons) which can transmit spin angular momentum with no accompanying charge transfer.

- The magnetization direction within nanoscale magnets can be manipulated efficiently via exchange interactions using spin-transfer torque from spin currents.

# V. Attractive effective electron-electron interactions and how they can destabilize the Fermi liquid state (the Cooper Instability)

*Topics: 1. How overscreening by phonons can produce an effective electron-electron interaction that is attractive.*

*2. How an attractive electron-electron interaction produces an instability of the Fermi liquid state in which electrons become correlated within "Cooper-pair" wavefunctions.*

This lecture begins our discussion of superconductivity. There are many excellent references on this subject. A few that I have found particularly useful are:
  A. C. Rose-Innes and E. H. Rhoderick, *Introduction to Superconductivity*
  D. L. Goodstein, *States of Matter*
  M. Tinkham, *Introduction to Superconductivity*

First, let's recall the results for our model "universal Hamiltonian" for a disordered piece of metal with no special symmetries (so that the single-particle basis states are essentially randomly oscillating functions)

$$H = H_{non-interacting} + V_{e-e} = \sum_{i,s} \epsilon_{i,s}^0 c_{i,s}^\dagger c_{i,s} + E_C \hat{n}^2 - J\hat{\vec{S}}^2 - \lambda_{BCS} \sum_{j \neq m} c_{j\uparrow}^\dagger c_{j\downarrow}^\dagger c_{m\downarrow} c_{m\uparrow} \quad (5.1)$$

In the previous lectures, we've seen that for a metal with repulsive electron-electron interactions $J > 0$, while $\lambda_{BCS} < 0$ and is typically small. In this case when a positive $J$ is greater than the single-particle level spacing, one can have an instability of the Fermi liquid state in which the resulting new ground state is ferromagnetic with a large value of the spin $S$. If, on the other hand, one has an effectively attractive electron-electron, the Fermi liquid is subject to a different kind of instability with a qualitatively different ground state. For attractive interactions, one has $J < 0$ and $\lambda_{BCS} > 0$. In this case, the exchange term favors a state with zero spin while the pairing matrix elements (the last sum in the equation above) can produce a non-Fermi-liquid state formed from a superposition of Slater determinant states in which each pair of single-electron states $\{m \uparrow, m \downarrow\}$ is always occupied or unoccupied together.

It is possible to carry through a calculation of superconducting states using the same set of disordered single-electron basis states that we have focused on up to now. (Reference 7 contains a review of this approach.) However, it is more typical to start the analysis of superconductivity using Bloch wavefunctions as the single-electron basis states labeled by a wavevector $\vec{k}$ and spin index, and starting now we will switch to that basis. The analysis of which electron-electron matrix elements are large in this basis proceeds very similarly to what we did before for disordered electron-in-a-box basis states, with the result that the matrix elements most important for determining an electronic ground state at rest can be modeled as:

$$H = \sum_{k,s} \epsilon_{k,s}^0 c_{k,s}^\dagger c_{k,s} + E_C \hat{n}^2 - J\hat{\vec{S}}^2 - \lambda_{BCS} \sum_{k \neq j} c_{j\uparrow}^\dagger c_{-j\downarrow}^\dagger c_{-k\downarrow} c_{k\uparrow} \quad (5.2)$$

(For those wanting to see the details, and some additional matrix elements than can contribute to states with flowing supercurrents or spin-triplet states, see the appendix to this lecture.) The pair scattering in the Bloch-state basis takes two electrons in a Kramers doublet of states that are related by time reversal symmetry, $\{\vec{k} \uparrow, -\vec{k} \downarrow\}$ and scatters them to a different Kramers doublet.

The notation we are using assumes implicitly that spin-orbit scattering is weak so that the eigenstates can be written as purely spin-↑ and spin-↓, but our arguments will be more general and will apply even with strong spin-orbit coupling as long as one defines basis states which are paired into Kramers doublets and therefore related by time reversal symmetry.

## **How is it possible for electron-electron interactions to be attractive?**

Before we get to the nature of the instability in the Fermi liquid state for attractive interactions, I want to address the obvious question of how on earth an electron-electron interaction can be attractive. The interaction between two electrons in vacuum is always repulsive, as both are negatively charged. However, electrons in a solid are not in a vacuum. They exist in a background with a soup of lots of other low-energy excitations, with phonons, other charges, possibly spin excitations, etc. The screening produced by these other excitations can generate an excitation-mediated effective electron-electron interaction that is attractive.

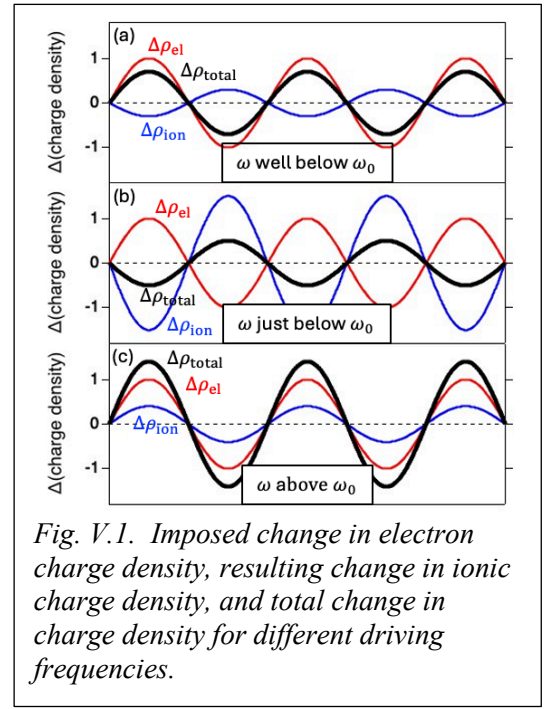

*Fig. V.1. Imposed change in electron charge density, resulting change in ionic charge density, and total change in charge density for different driving frequencies.*

To gain some intuition into one way this can come about, consider a toy model for phonon-mediated screening – imagine that the negatively-charged mobile conducting electrons interact with a background of positively charged ions that act as Einstein oscillators around their equilibrium positions, all with the same frequency $\omega_0$. Suppose you impose a small variation of conduction electron charge density $\Delta\rho_{el}$ relative to the equilibrium value that oscillates in both space and possibly time, with a long wavelength $\lambda$ and frequency $\omega$ (Fig. V.1). For a DC conduction electron modulation ($\omega = 0$), the ions will shift their position as best they can to cancel the charges due to the electrons, producing an oscillation in the ionic charge density $\Delta\rho_{ion}$ relative to the ionic equilibrium value that is opposite in sign to the imposed variation $\Delta\rho_{el}$. However, unless the springs are very weak they will not be able to screen the electronic charges completely (Fig. V.1(a)). With $|\Delta\rho_{ion}| < |\Delta\rho_{el}|$ the total resulting change in local charge density $\Delta\rho_{total} = \Delta\rho_{el} + \Delta\rho_{ion}$ will still have the same sign as the imposed electron density and a test electronic charge will still be driven away from regions of large imposed electron density. This means that the effective electron-electron interaction will still be repulsive, and screening by the ions will simply reduce its strength to give $V_{eff}(\omega = 0) = V_{bare}/\epsilon(\omega = 0)$ where $\epsilon(\omega = 0)$ is the DC dielectric constant from the ions.

Now let's imagine that we force the imposed electron charge density variation to oscillate as a function of time and tune the oscillation frequency through the resonance frequency of the Einstein oscillators. For the Einstein oscillators, their response is just that of a simple harmonic oscillator driven near resonance, so that (assuming negligible damping) the amplitude of the oscillations takes the textbook resonance form

$$A(\omega) = -\frac{F/m}{\omega^2 - \omega_0^2} \qquad (5.3)$$

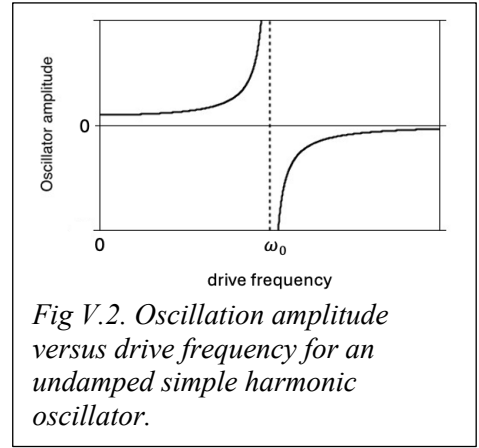

Fig V.2. Oscillation amplitude versus drive frequency for an undamped simple harmonic oscillator.

where $F/m$ is the amplitude of the driving force divided by the mass (see Fig. V.2). For $\omega$ less than the resonance frequency of the oscillator, $\omega_0$, the ionic charge density oscillates in phase with the driving force, meaning that the ionic charge density modulation $\Delta\rho_{ion}$ has the opposite sign from $\Delta\rho_{el}$, just as we saw for the DC response. However, as $\omega$ approaches $\omega_0$ from below, the ionic oscillation grows and eventually diverges. Just below the resonance frequency, it is possible that $|\Delta\rho_{ion}|$ can become greater than the imposed charge density $|\Delta\rho_{el}|$ so that the total change in charge density $\Delta\rho_{el} + \Delta\rho_{ion}$ will have the opposite sign compared to the imposed electronic charge density (Fig. V.1(b)). This means that at frequencies just below resonance the ions can *over-screen* the imposed electron modulation due to the resonance behavior. A test electron will now be attracted to regions of large imposed electron density rather than repelled, so the effective electron-electron interaction is attractive. This is the interesting regime in regard to the interactions that drive superconductivity. One way to think about this regime is that the electron interaction can be attractive when two electrons are separated in time -- i.e., by an integer multiple of a relevant phonon period -- even if we think about the interaction occurring at a specific position. (Thanks to Brad Ramshaw for pointing this out to me.)

In our simple model, if the imposed frequency $\omega$ is larger than the resonance frequency of the oscillator, $\omega_0$, then the ions oscillate out of phase with their driving force with decreasing amplitude as $\omega$ is increased. This means in this regime that $\Delta\rho_{ion}$ will have the same sign as $\Delta\rho_{el}$, and the ions will anti-screen the imposed electronic charge density (Fig. V.1(c)). The effective electron-electron interaction here will be repulsive, with strength even greater than the bare electron-electron interaction.

Adding a bare electron-electron interaction to the effects of the ionic screening, the final result for the effective electron-electron interaction in our Einstein model will be something like Fig. V.3. The effective electron-electron interaction will be repulsive near DC and at frequencies above $\omega_0$, but for a range of frequencies below $\omega_0$ the overscreening from the ions can provide an effective attractive interaction.

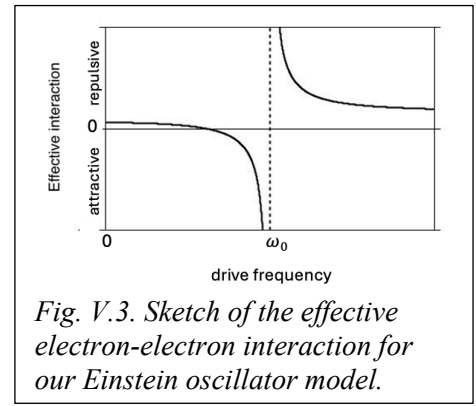

Fig. V.3. Sketch of the effective electron-electron interaction for our Einstein oscillator model.

In a more realistic calculation there are lots of phonon modes rather than a single Einstein frequency, and one must add up over all modes. The effective interaction entering the pair-scattering matrix elements for the superconducting ground state can be calculated via 2nd-order perturbation theory to give a result with a resonance structure very similar to our simple Einstein model.

$$V_{pairing} = \sum_{k,q} \left( V_{\text{bare}} + |M_q|^2 \frac{\hbar\omega_q}{(\varepsilon_k - \varepsilon_{k+q})^2 - (\hbar\omega_q)^2} \right) c^\dagger_{k+q} c^\dagger_{-k-q} c_{-k} c_k. \tag{5.4}$$

Here $\hbar\omega_q$ is the energy of the phonon mode with wavevector $q$ and $M_q$ is a matrix element. If the difference in energy between the two single-electron states involved in a pair scattering event, $|\varepsilon_k - \varepsilon_{k+q}|$ is less than the phonon energy corresponding to the scattering wavevector, $\hbar\omega_q$, then term involving the phonon screening is negative and can contribute to an attractive interaction. Whether or not the final total effective interaction is attractive or repulsive depends on whether or not the phonon-mediated interactions overcome the bare interaction, so that the final sign can be positive or negative depending on the material.

For simplicity, the condition that $|\varepsilon_k - \varepsilon_{k+q}|$ should be small in order to have an attractive interaction is usually accounted for approximately by requiring that separately both $|\varepsilon_k - \varepsilon_F|$ and $|\varepsilon_{k+q} - \varepsilon_F|$ must be small -- less than some cut-off of order a characteristic phonon energy $(\hbar\omega_D)$ – and it is also typically assumed that all of the non-zero matrix elements for the pairing interaction within these cut-offs are equal to the same constant $(-\lambda_{BCS})$, independent of which single-electron states are involved. The final model BCS Hamiltonian that we will consider is therefore

$$H = \sum_{k,s} \epsilon^0_{k,s} c^\dagger_{k,s} c_{k,s} - \lambda_{BCS} \sum_{k \neq j} c^\dagger_{j\uparrow} c^\dagger_{-j\downarrow} c_{-k\downarrow} c_{k\uparrow}, \tag{5.5}$$

where the second sum is restricted to states within a cut-off range around the Fermi level, with $|\varepsilon_k - \varepsilon_F| \leq \hbar\omega_D$ and $|\varepsilon_j - \varepsilon_F| \leq \hbar\omega_D$. For attractive interactions, by our conventions $\lambda_{BCS}$ is positive. With the assumptions that we have used in reaching the model BCS Hamiltonian, we have arrived back to an expression that is the same as our previous "universal Hamiltonian" model except that the exchange interaction is not included. The exchange interaction will not enter into our discussion of superconductivity because almost all the states for which we wish to compare energies will have the same total spin $S = 0$ (which is already the state favored by exchange when electron-electron interactions are attractive).

It is possible to do better than the approximations used in the model BCS Hamiltonian -- to take into account the details of the phonon spectrum for a particular material and how they effect the phonon-mediated electron-electron interaction, and to calculate the strength of the coupling. This is called the Migdal-Eliashberg theory. We will nevertheless stick with the BCS approximation because it works well for capturing the qualitative features of superconductivity. The full theory is necessary for an accurate description when the effective electron-electron interaction is strong enough that one needs to go beyond "weak coupling." Superconducting lead (Pb) is an example where this can be important.

Screening mediated by excitations other than phonons can also give rise to effectively attractive electron-electron interactions in some materials. For example, spin fluctuations associated with a nearby antiferromagnetic phase are thought to be the mechanism for some classes of unconventional superconductors. For all-electronic mechanisms like spin fluctuations, the frequency-dependent resonant behavior that is relevant for generating overscreening by phonons is unlikely to apply, because electronic dynamics are much faster compared to phonons. Adaptations of Migdal-Eliashberg theory can nevertheless still be useful in modeling such interactions and the associated unconventional superconductors. We will touch upon the issues of non-phononic mechanisms and unconventional superconductors in the final lecture.

**Instability of the Fermi Sea (The Cooper Argument)**

Leon Cooper introduced a beautiful and simple argument which shows that the Fermi liquid state is unstable in the presence of attractive electron-electron interactions. The idea is to consider two extra electrons outside a "frozen" Fermi sphere for a simple free-electron metal bandstructure (i.e., plane-wave states). By "frozen", I mean that we'll imagine that the electrons inside the Fermi sphere stay inert – they will just block the scattering of electrons into the states with energies lower than the Fermi energy. But for the two electrons outside the Fermi sphere, we can form a wavefunction from a superposition of any of the empty states. By solving this 2-electron problem explicitly for an attractive interaction, Cooper showed that it is possible to form a two-electron bound state with energy less than $2\varepsilon_F$ (i.e., twice the Fermi energy) using only single-electron states with energy greater than or equal to $\varepsilon_F$. It follows that the Fermi liquid is unstable – you can take a full Fermi sphere, take two electrons out of states at $\varepsilon_F$, put them in a bound state to lower the total energy, and repeat. As long as the bound states don't block occupation of too large a fraction of the states with energy greater than $\varepsilon_F$ you can keep making more bound states. In the end you'll get a correlated state with total energy less than the Fermi liquid state. This is somewhat surprising because two electrons alone in empty space don't form a bound state in 3 dimensions unless an attractive interaction is stronger than some threshold. However, Cooper-pair bound states form even for arbitrarily weak attractive interactions in 3 dimensions.

Let's see how this works. The Schrodinger equation for two interacting electrons is

$$-\frac{\hbar^2}{2m}(\nabla_1^2 + \nabla_2^2)\Phi(r_1, r_2, \sigma_1, \sigma_2) + V(r_2 - r_1)\Phi(r_1, r_2, \sigma_1, \sigma_2) = E\Phi(r_1, r_2, \sigma_1, \sigma_2). \quad (5.6)$$

We can solve this 2-body problem using the standard trick of separating of variables into center-of-mass and relative coordinates: $\Phi(r_1, r_2, \sigma_1, \sigma_2) = \psi(\vec{R})\varphi(\vec{r})$ with $\vec{R} = (\vec{r}_1 + \vec{r}_2)/2$ and $\vec{r} = \vec{r}_2 - \vec{r}_1$. If there is no magnetic field applied then for the ground state the center-of-mass part of the wavefunction is just a constant, so we'll ignore that part for now and focus on the relative-coordinate part of the wavefunction. (We'll come back to the center-of-mass part when we consider current flow in a superconductor.)

It is then straightforward to solve for the relative-coordinate part of the wavefunction as a superposition of unfilled paired plane-wave states $\{\vec{k}\uparrow, -\vec{k}\downarrow\}$ in the product wavefunctions $e^{i\vec{k}\cdot\vec{r}_2}e^{-i\vec{k}\cdot\vec{r}_1} = e^{i\vec{k}\cdot\vec{r}}$; this superposition happens to be simply a Fourier series summed over wavevectors $\vec{k}$ outside the frozen Fermi sphere.

$$\varphi(\vec{r}) = \sum_{|\vec{k}|>k_F} a_{\vec{k}}\, e^{i\vec{k}\cdot\vec{r}}. \quad (5.7)$$

The procedure is to substitute this trial wavefunction into the 2-electron Schrodinger equation and solve for the coefficients $a_{\vec{k}}$ and the energy eigenvalue $E$.

$$\sum_{|\vec{k}|>k_F} a_{\vec{k}}\, 2\frac{\hbar^2 k^2}{2m} e^{i\vec{k}\cdot\vec{r}} + V(\vec{r})\sum_{|\vec{k}|>k_F} a_{\vec{k}}\, e^{i\vec{k}\cdot\vec{r}} = E\sum_{|\vec{k}|>k_F} a_{\vec{k}}\, e^{i\vec{k}\cdot\vec{r}}. \quad (5.8)$$

I will define $\varepsilon_k = \frac{\hbar^2 k^2}{2m}$ to save writing. Now without loss of generality we can express also the interaction $V(\vec{r})$ as a Fourier transform $V(\vec{r}) = \sum_{\vec{q}} V_{\vec{q}} e^{i\vec{q}\cdot\vec{r}}$ and manipulate the resulting double-sum in the second term of the Schrodinger equation

$$\sum_{\vec{q}} V_{\vec{q}} e^{i\vec{q}\cdot\vec{r}} \sum_{|\vec{k}|>k_F} a_{\vec{k}} e^{i\vec{k}\cdot\vec{r}} = \sum_{\vec{k}',|\vec{k}|>k_F} V_{\vec{k}'-\vec{k}} a_{\vec{k}} e^{i\vec{k}'\cdot\vec{r}} = \sum_{\vec{k},|\vec{k}'|>k_F} V_{\vec{k}-\vec{k}'} a_{\vec{k}'} e^{i\vec{k}\cdot\vec{r}} \quad (5.9)$$

where I have relabeled summation indices twice, $\vec{k}' = \vec{k} + \vec{q}$ in the middle sum and swapping $\vec{k}'$ and $\vec{k}$ in the final sum. Now (given that the center-of-mass part of the wavefunction can be taken to be a constant), the interaction coefficient

$$V_{\vec{k}-\vec{k}'} = \int d^3R \, d^3r \, e^{-i\vec{k}'\cdot\vec{r}_2} e^{i\vec{k}'\cdot\vec{r}_1} V(\vec{r}_2 - \vec{r}_1) e^{i\vec{k}\cdot\vec{r}_2} e^{-i\vec{k}\cdot\vec{r}_1} \quad (5.10)$$

corresponds precisely to the matrix element for scattering from the pair state $\{\vec{k}, -\vec{k}\}$ to $\{\vec{k}', -\vec{k}'\}$. By our assumption of a frozen Fermi surface, we must have both $|\vec{k}| > k_F$ and $|\vec{k}'| > k_F$, and so we can further restrict the third sum to addition over $|\vec{k}| > k_F$ and combine all three terms in the Schrodinger equation into one big sum

$$\sum_{|\vec{k}|>k_F} \left\{ 2\varepsilon_k a_{\vec{k}} + \sum_{|\vec{k}'|>k_F} V_{\vec{k}-\vec{k}'} a_{\vec{k}'} - E a_{\vec{k}} \right\} e^{i\vec{k}\cdot\vec{r}} = 0. \quad (5.11)$$

Since every term in a Fourier series must be zero if the total of the Fourier series is zero, this requires

$$(2\varepsilon_k - E) a_{\vec{k}} = -\sum_{|\vec{k}'|>k_F} V_{\vec{k}-\vec{k}'} a_{\vec{k}'}. \quad (5.12)$$

You can view this as a formula to diagonalize the Hamiltonian, including pair scattering.

At this stage we use our BCS approximation for the interaction matrix elements $V_{\vec{k}-\vec{k}'} = -\lambda_{BCS}$ for $|\varepsilon_k - \varepsilon_F| \leq \hbar\omega_D$ and $|\varepsilon_{k'} - \varepsilon_F| \leq \hbar\omega_D$, and 0 otherwise, to reach:

$$(2\varepsilon_k - E) a_{\vec{k}} = \lambda_{BCS} \sum_{k_F < |\vec{k}'| < k_D} a_{\vec{k}'}. \quad (5.13)$$

(Here $k_D$ is defined so that $\varepsilon_{k_D} = \varepsilon_F + \hbar\omega_D$.) The sum over coefficients on the right-hand side will just be some number. Let us call it $A$. Therefore we can now solve for the coefficients in terms of the eigenvalue energy $E$.

$$a_{\vec{k}} = \frac{\lambda_{BCS} A}{2\varepsilon_k - E}. \quad (5.14)$$

We still don't know the energy $E$ yet, but we can determine this simply by summing both sides of this equation over $k_F < |\vec{k}| < k_D$:

$$\sum_{k_F < |\vec{k}| < k_D} a_{\vec{k}} = \lambda_{BCS} A \sum_{k_F < |\vec{k}| < k_D} \frac{1}{2\varepsilon_k - E}. \quad (5.15)$$

The sum on the left is just $A$ again, so we have

$$1 = \lambda_{BCS} \sum_{k_F < |\vec{k}| < k_D} \frac{1}{2\varepsilon_k - E} \quad \text{or} \quad 1 = \lambda_{BCS} \sum_{\varepsilon_F < \varepsilon_k < \varepsilon_F + \hbar\omega_D} \frac{1}{2\varepsilon_k - E}. \quad (5.16)$$

This is an implicit equation for the energy $E$. I'll replace the sum by an integral over energies near the Fermi level, assuming an approximately constant density of orbital levels $N(\varepsilon)$ over the range of integration (not counting spin)

$$\frac{1}{\lambda_{BCS}} = N(\varepsilon_F) \int_{\varepsilon_F}^{\varepsilon_F + \hbar\omega_D} \frac{d\varepsilon}{2\varepsilon - E} = \frac{N(\varepsilon_F)}{2} \ln\left[\frac{2(\varepsilon_F + \hbar\omega_D) - E}{2\varepsilon_F - E}\right] \tag{5.19}$$

or

$$E = 2\varepsilon_F - \frac{2\hbar\omega_D}{e^{\frac{2}{\lambda_{BCS}N(\varepsilon_F)}} - 1}. \tag{5.18}$$

This is our answer for the energy eigenstate. If we have an attractive interaction so that $\lambda_{BCS} > 0$, the denominator is positive so that this result gives $E < 2\varepsilon_F$! This means that as long as the electron-electron interaction is attractive, no matter how weak, the Fermi liquid is unstable by the argument discussed above.

In the limit of weak coupling (meaning $\lambda_{BCS} N(\varepsilon_F) \ll 1$), we can also write a slightly more compact expression

$$E = 2\varepsilon_F - 2\hbar\omega_D e^{-\frac{2}{\lambda_{BCS}N(\varepsilon_F)}} \tag{5.19}$$

The binding energy per Cooper pair, $2\Delta_{\text{Coop}} \equiv 2\hbar\omega_D e^{-\frac{2}{\lambda_{BCS}N(\varepsilon_F)}}$, is generally of order about (1 K)$k_B$, which sets the order-of-magnitude temperature scale for a typical superconductor. (The actual critical temperature for a conventional superconductor can range from much less than 1 K to 40 K at ambient pressure and up to 250 K at high pressure, depending on material. Some unconventional superconductors reach critical temperatures over 100 K at ambient pressure, but the formula for $\Delta_{\text{Coop}}$ does not apply for them since their attractive interactions are generally not phonon-mediated.)

You will note that the binding energy is a non-analytic function of the coupling parameter $\lambda_{BCS}$. As a consequence, perturbation theory in the parameter $\lambda_{BCS}$ is incapable of describing the pair-binding process, even in the limit of weak coupling. As a historical note, I've been told that this fact likely slowed the progress in understanding superconductivity by decades.

**Form of the Cooper-Pair Wavefunction**

More than just giving the energy eigenvalue, the Cooper argument provides the full form of the Cooper pair wavefunction.

$$\varphi(\vec{r}) = \sum_{|\vec{k}|>k_F} a_{\vec{k}}\, e^{i\vec{k}\cdot\vec{r}} = \sum_{|\vec{k}|>k_F} \frac{C}{2\varepsilon_k - E} e^{i\vec{k}\cdot\vec{r}} = \sum_{|\vec{k}|>k_F} \frac{C}{(2\varepsilon_k - 2\varepsilon_F) + 2\Delta_{\text{Coop}}} e^{i\vec{k}\cdot\vec{r}} \tag{5.20}$$

where the constant $C$ can be picked for normalization. The Fourier coefficients here depend on the wavevector only through the magnitude $k$, so the wavefunction coefficients are spherically symmetric in momentum space and (after Fourier transforming) the wavefunction $\varphi(\vec{r})$ is spherically symmetric in real space. Since $\varphi(\vec{r}) = +\varphi(-\vec{r})$ the spatial wavefunction is symmetric under exchange of electron coordinates, so in order to have a totally antisymmetric wavefunction we must have antisymmetric $S=0$ spin wavefunction. The Cooper pair solution therefore corresponds to an "s-wave" (meaning spherically symmetric orbital) spin-singlet bound state of the two electrons.

The origin of the spherical symmetry for the Cooper-pair wavefunction lies in our assumption that the interaction matrix elements $V_{\vec{k}-\vec{k}'}$ are all equal to the same constant, independent of angle for $\vec{k} - \vec{k}'$. This applies to good accuracy for phonon-mediated pairing interactions, because phonon properties are roughly isotropic. However, the concept of a Cooper pair is more general than just for such conventional superconductors. For unconventional superconductors in which the mechanism behind the attractive interactions is different from overscreening by phonons, the interaction can be angle-dependent and this can lead to different symmetries of the Cooper-pair wavefunction (e.g., *p*-wave, *d*-wave, etc.) More on this later.

One final lesson to be drawn from the Cooper argument concerns the spatial range of the Cooper-pair wavefunction. The characteristic energy range of the large components in the Fourier expansion for $\varphi(\vec{r})$ is $\varepsilon_k - \varepsilon_F \sim \Delta_{\text{Coop}}$, or $\Delta\varepsilon \sim \frac{\hbar^2 k_F}{m}|\Delta k| \sim \Delta_{\text{Coop}}$ (see Fig. V.4), so the characteristic wave-vector scale for the Fourier components is

$$|\Delta k| \sim \frac{m \Delta_{\text{Coop}}}{\hbar^2 k_F}. \tag{5.21}$$

By the property of Fourier transforms, the spatial range in real space is

$$|\Delta r| \sim \frac{1}{|\Delta k|} \sim \frac{\hbar^2 k_F}{m \Delta_{\text{Coop}}}. \tag{5.22}$$

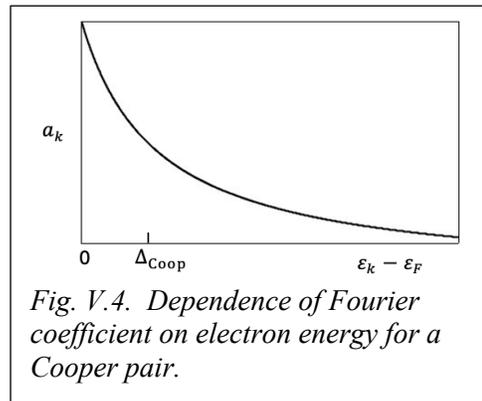

*Fig. V.4. Dependence of Fourier coefficient on electron energy for a Cooper pair.*

Putting in typical values for a superconductor with a critical temperature on the few Kelvin scale, the length scale is about 100 nm. This is big! It means that there are lots of interpenetrating Cooper pairs in a typical superconductor. One consequence of this is that a mean-field approximation works surprisingly well for superconductors. The BCS theory that we will discuss next time is at its heart a mean-field theory. This fortunate behavior is in contrast to ferromagnets, for which mean-field approximations generally work surprisingly poorly.

The Cooper argument shows that the Fermi Liquid is unstable when the effective electron-electron interaction is attractive, but it doesn't say exactly what state is stable. The process of promoting two electrons at a time from the Fermi sphere and letting them undergo pair scattering to form a bound state can't go on forever for an unlimited number of electrons. The pair scattering will eventually be blocked by the partial occupation of states above the Fermi energy by the previous pairs. In the next lecture we'll consider a few different alternative ways of thinking about the true ground state, and why it can be a superconductor with zero resistance.

**What have we learned?:**
- The resonance response of a phonon mode can lead to over-screening of variations in electronic charge density if the relevant variation frequency is slightly less than the phonon resonance frequency. This can produce an effectively attractive electron-electron interaction. Because the attractive interaction depends on a resonance response of the phonons, it is limited to pair scattering between electronic states in the immediate vicinity of the Fermi energy (differing from the Fermi energy by less than an energy cut-off associated with a characteristic phonon energy).

- The Cooper argument shows that when the electron-electron interaction is attractive (no matter how weakly attractive), two electrons outside of a frozen Fermi sea can form a bound state with energy less than $2\varepsilon_F$. It follows that the Fermi liquid state is unstable in the presence of an attractive interaction.

- The wavefunction for the Cooper-pair bound state consists of superpositions of states in each of which the two electrons simultaneously occupy a pair of time-reversal-related single-electron states $\{\vec{k}\uparrow, -\vec{k}\downarrow\}$. Electron-electron interactions couple these different Kramers pair states.

- The Cooper-pair wavefunction for a conventional phonon-mediated BCS superconductor is a spherically-symmetric (*s*-wave), spin-singlet state with a large spatial extent, typically of order 100 nm. Unconventional superconductors also have Cooper pairs, but the pair wavefunctions can have different symmetries and length scales.

## Appendix to Section V. Analysis of Pair-Scattering Matrix Elements Using Bloch States as the Single-Electron Basis States

In our previous "thought calculation" about the universal Hamiltonian for metals, we saw that the pair-scattering matrix elements that can give rise to superconductivity come simply from evaluating the screened Coulomb interaction between Slater determinant states. Previously, we constructed these Slater determinants using randomly fluctuating standing-wave single-electron basis states (appropriate, for example, to a disordered finite metal "quantum dot"). We found that the pair-scattering terms resulted from taking the general expression for the 2-body-interaction matrix element:

$$V_2 = \frac{1}{2} \sum_{\substack{j,k,m,n \\ s,\sigma}} V_{jkmn} c^\dagger_{k,\sigma} c^\dagger_{j,s} c_{m,s} c_{n,\sigma} \tag{5.23}$$

with

$$V_{jkmn} = \int d^3r_1 d^3r_2 \psi_k^*(r_2) \psi_j^*(r_1) V_{e-e}(r_1 - r_2) \psi_m(r_1) \psi_n(r_2). \tag{5.24}$$

For simplicity, we ignored spin-orbit scattering so that the standing-wave wavefunctions inside the disordered quantum dot could be chosen to be real-valued. This made it easy to see how the electrons could be chosen pairwise so that the integrand in Eq. (5.24) has a uniform sign at all positions for a short-ranged screened Coulomb interaction so that the overall integral is thus large in magnitude (instead of having an integrand that fluctuates both positive and negative as a function of position so that the overall integral is small). The pair scattering term is this case corresponds to a process in which two electrons both in the orbital state $\psi_n(r)$ (with opposite spins) scatter to a different orbital state $\psi_j(r)$ (also with opposite spins).

If, instead of starting with single-electron basis states that are standing waves, we instead start with a Bloch-wave basis for the single-electron states, one can work through the same type of logic to see which interaction matrix elements are large. One reaches the standard BCS form of the pairing interaction this way, but there are a few subtleties.

Let us from this point on interpret the subscript letters on the wavefunctions as the wave vectors of Bloch states, such that the orbital Bloch states have the form

$$\psi_k(r) = e^{ik \cdot r} u_k(r). \tag{5.25}$$

For simplicity we will continue to ignore spin-orbit coupling, so each of these states can be occupied spin up and spin down. In this case the orbital Bloch states $\psi_k(r)$ and $\psi_{-k}(r)$ will be related by time-reversal symmetry such that one can define wavefunctions in the form $\psi_k^*(r) = e^{-ik \cdot r} u_k^*(r) = \psi_{-k}(r) = e^{-ik \cdot r} u_{-k}(r)$ so that $u_{-k}(r) = u_k^*(r)$. (Under the time-reversal operation the spin will be flipped as well.) If we evaluate the integral in Eq. (5.24) for Bloch-state wavefunctions, assuming the $\delta$-function approximation for the screened Coulomb interaction, we have

$$V_{jkmn} = V_0 \int d^3r \, e^{i(m+n-k-j) \cdot r} u_k^*(r) u_j^*(r) u_m(r) u_n(r). \tag{5.26}$$

From this we see that the very largest matrix elements for the pairing interaction come from choosing $k=-j$ and $n=-m$, to give an integrand of the form $|u_j(r)|^2 |u_m(r)|^2$ with a definite sign. This just corresponds to choosing the states to be matched into the time-reversed pairs of Bloch states $\psi_k(r)$ and $\psi_{-k}(r)$ (where $k$ is the wave vector). However, the periodic parts $u(r)$ of the Bloch functions are generally not very highly oscillatory, so that it is also possible to get sizable matrix elements just by making the more general choice $m + n - k - j = 0$, so that the exponential term in Eq. (5.26) does not oscillate. This corresponds to a scattering process in which two electrons with total wavevector $m + n = p$, scatter into two other states with the same total wavevector $k + j = p$. If we define a scattering wavevector $q$ such that $j = m + q$ and $k = n - q$, then these contributions to these pairing matrix elements can be written

$$V_{pairing} = \frac{1}{2} \sum_{\substack{m,n,q \\ s,\sigma}} V_{m+q,n-q,m,n} c^\dagger_{n-q,\sigma} c^\dagger_{m+q,s} c_{m,s} c_{n,\sigma} \tag{5.27}$$

with (for a $\delta$-function interaction)

$$V_{m+q,n-q,m,n} = V_0 \int d^3r \, u^*_{n-q}(r) u^*_{m+q}(r) u_m(r) u_n(r). \tag{5.28}$$

In the presence of lattice translation symmetry, the many-body energy eigenstates will be eigenstates of the total wave vector, and it follows that the Slater determinants that contribute to a given many-body energy eigenstate can all be chosen to each have the same definite value of the total wavevector. In the ground state this total wavevector must be zero in order to minimize the overall kinetic energy. Therefore each of the Slater determinants that contribute to the ground state must have a total wavevector of zero. This is achieved by constructing each Slater determinant state so it consists of paired occupation of single-electron states (with the total wavevector $m + n = 0$ for each of these pairs). In this case the subset of the important pair-scattering matrix elements that can contribute to the ground state are also the ones in which one pair with total wavevector 0 scatters into a different pair with total wavevector 0. If we define the wavevectors $a$ and $b$ so that $b = m = -n$ and $a = m + q = -(n - q)$ the subset of the pair-scattering matrix elements that contribute to the ground state energy is

$$V_{pairing} = \frac{1}{2} \sum_{\substack{a,b \\ s,\sigma}} V_{a,-a,b,-b} c^\dagger_{-a,\sigma} c^\dagger_{a,s} c_{b,s} c_{-b,\sigma}. \tag{5.29}$$

with (for a $\delta$-function interaction)

$$V_{a,-a,b,-b} = V_0 \int d^3r \, u^*_{-a}(r) u^*_a(r) u_b(r) u_{-b}(r) = V_0 \int d^3r \, |u_a(r)|^2 |u_b(r)|^2. \tag{5.30}$$

The last equality follows since $u^*_{-a}(r) = u_a(r)$ and $u_{-b}(r) = u^*_b(r)$ in the absence of spin-orbit coupling. For a $\delta$-function interaction we also have $V_{a,-a,b,-b} = V_{-a,a,b,-b} = V_{a,-a,-b,b} = V_{-a,a,-b,b}$ (we will use this below).

It is also possible to consider many-body states with a non-zero value of the total wavevector. In this case, the lowest-energy state can be constructed by choosing all electrons to be occupied in pairs with the same non-zero wavevector sum, and then to incorporate pair scattering between these states. The many-body state in this case corresponds to a moving superconducting condensate with a non-zero supercurrent. (We will consider states of this form later when we analyze the phenomenon of zero resistance.)

For the assumption of a $\delta$-function interaction, the pairing matrix elements in Eq. (5.29) can be simplified even a bit more. If we expand Eq. (5.29) in this case,

$$V_{pairing} = -\frac{1}{2}\sum_{a,b}V_{a,-a,b,-b}\left(c^\dagger_{-a\uparrow}c^\dagger_{a\uparrow}c_{b\uparrow}c_{-b\uparrow} + c^\dagger_{-a\downarrow}c^\dagger_{a\downarrow}c_{b\downarrow}c_{-b\downarrow} + c^\dagger_{-a\uparrow}c^\dagger_{a\downarrow}c_{b\downarrow}c_{-b\uparrow} + c^\dagger_{-a\downarrow}c^\dagger_{a\uparrow}c_{b\uparrow}c_{-b\downarrow}\right). \quad (5.31)$$

The first two contributions on the right hand side indicate scattering from a pair with spin 1 (rather than spin 0) to another pair with spin 1. In our previous thought calculation (with standing-wave states) these terms could not appear because in that context a spin-1 pair violates the Pauli exclusions principle, e.g. two spin ups in the same orbital state. That is not the case for the pairs in Eq. (5.31) because the orbital states within each pair are different – Bloch states with opposite wavevectors. For a $\delta$-function interaction, however, when we sum over $a$ or $b$ these spin-1 pair terms will still add to zero. For example, given (as determined above) that for a $\delta$-function interaction we have $V_{a,-a,b,-b} = V_{-a,a,b,-b}$, if we sum the first term over $a$ and consider the contributions from both $a_0$ and $-a_0$,

$$V_{a_0,-a_0,b,-b}c^\dagger_{-a_0\uparrow}c^\dagger_{a_0\uparrow}c_{b\uparrow}c_{-b\downarrow} + V_{-a_0,a_0,b,-b}c^\dagger_{a_0\uparrow}c^\dagger_{-a_0\uparrow}c_{b\uparrow}c_{-b\downarrow}$$
$$= V_{a_0,-a_0,b,-b}\left(-c^\dagger_{a_0\uparrow}c^\dagger_{-a_0\uparrow}c_{b\uparrow}c_{-b\downarrow} + c^\dagger_{a_0\uparrow}c^\dagger_{-a_0\uparrow}c_{b\uparrow}c_{-b\downarrow}\right) = 0. \quad (5.32)$$

The minus sign comes from the anticommutation of the raising operators.

For the 4$^{th}$ term in Eq. (5.31), $c^\dagger_{a\downarrow}c^\dagger_{-a\uparrow}c_{b\uparrow}c_{-b\downarrow} = c^\dagger_{-a\uparrow}c^\dagger_{a\downarrow}c_{-b\downarrow}c_{b\uparrow}$ by two anticommutations. For a $\delta$-function interaction after summing over all values of $a$ and $b$ this will give exactly the same contribution as the third term in Eq. (5.31). Therefore we can cancel the factor of 1/2 in front of Eq. (5.31) and write the form of the pairing interaction (for a $\delta$-function screened Coulomb interaction and for states with zero total crystal momentum) simply as

$$\boxed{V_{pairing} = \sum_{a,b}V_{a,-a,b,-b}c^\dagger_{-a\downarrow}c^\dagger_{a\uparrow}c_{b\uparrow}c_{-b\downarrow}.} \quad (5.33)$$

The usual starting point for the BCS model of superconductivity is to make the approximation that $V_{a,-a,b,-b} = -\lambda_{BCS}$, a constant independent of the choice of basis states $a$ and $b$, at least for states near the Fermi level. This contribution takes two electrons that are originally in a combination of time-reversed states $(b\uparrow,-b\downarrow)$ and scatters them into a different combination of time-reversed states $(a\uparrow,-a\downarrow)$. When the effective electron-electron interaction is attractive (corresponding to $V_{a,-a,b,-b} < 0$ or $\lambda_{BCS} > 0$), this type of pair scattering can destabilize the Fermi liquid state and produce "$s$-wave" (or spherically symmetric) spin-singlet superconductivity. If instead of a constant coefficient $-\lambda_{BCS}$, one assumes a coefficient with angular dependence as a function of the difference of wavevectors $\vec{b} - \vec{a}$, this can result in spin-singlet superconductivity in which the superconducting wavefunction has angular dependence in real space, either anisotropic $s$-wave superconductivity if the pair wavefunction never changes sign, or $d$-wave, $f$-wave (etc.) superconductivity if the pair wavefunction has sign reversals.

We have ignored spin-orbit coupling in this discussion, but the argument can be generalized to include spin-orbit coupling, in which case the important pair-scattering happens between pairs of states related by time-reversal symmetry.

If one goes beyond the assumption of a $\delta$-function interaction, then it is no longer true that $V_{a,-a,b,-b} = V_{-a,a,b,-b}$:

$$V_{a,-a,b,-b} = \int d^3r_1 d^3r_2 \; u^*_{-a}(r_2) u_{-b}(r_2) V_{ee}(r_2 - r_1) u^*_a(r_1) u_b(r_1)$$
$$V_{-a,a,b,-b} = \int d^3r_1 d^3r_2 \; u^*_a(r_2) u_{-b}(r_2) V_{ee}(r_2 - r_1) u_b(r_1) u^*_{-a}(r_1). \tag{5.34}$$

In this case the cancellations that eliminate the first two terms on the right in Eq. (5.31) are no longer exact, and a few more terms can survive to contribute to the pairing interaction. The coefficient in the usual BCS expression Eq. (5.33) for scattering between $S = 0$ doublets takes the form $(V_{a,-a,b,-b} + V_{-a,a,-b,b})/2$ and in addition to Eq. (5.33) there will be non-zero triplet pair-scattering terms with coefficients $\propto V_{a,-a,b,-b} - V_{-a,a,b,-b}$ that correspond to the scattering of spin-1 combinations of electrons from orbital states $(b,-b)$ to spin-1 combinations of $(a,-a)$. However, the absolute value $|V_{a,-a,b,-b} - V_{-a,a,b,-b}|$ will usually be much smaller than the main coefficients in Eq. (5.33) (since it goes to zero in the approximation of a $\delta$-function interaction), and so one can expect these triplet-pairing terms usually to be unimportant relative to the singlet-pairing terms that are the main focus of the BCS Hamiltonian. One exception is when the singlet-pairing coefficients are positive, $(V_{a,-a,b,-b} + V_{-a,a,-b,b})/2 > 0$, corresponding to a repulsive effective interaction in this scattering channel, while the matrix elements for the triplet scattering have the sign corresponding to an effectively attractive interaction. In this case, the primary BCS coupling has the wrong sign to induce a superconducting instability, and it may be possible to have a triplet superconductivity many-body state in which the Cooper pairs have total spin 1 rather than spin 0. However, our model is almost certainly too simple to properly consider such cases – the interactions will likely be energy dependent rather than a constant, and different band structures and spin-orbit interactions can play a big role.

The triplet pairing terms did not appear in our previous treatment when we were using standing wave single-electron basis states because the triplet-pairing state requires lattice translation to be a good symmetry (one can say that simple defect scattering can "break pairs"), and the assumption of real-valued standing wave states does not respect this symmetry.

## VI. Superconductivity: Basic Properties

*Topics: 1. Intuitive pictures for understanding the superconducting ground state and its low-energy excited states*

*2. Some of the ways in which superconductors differ from normal-state Fermi liquids – energy gap, condensation energy, zero resistance, and the Meissner effect.*

The rigorous solution to the Bardeen-Cooper-Schrieffer (BCS) model of superconductivity is generally covered in a different Cornell course, so my goal in this lecture will be less rigorous and more qualitative. I'll discuss three different (but equivalent) ways of thinking about the superconducting ground state, each of which can provide some useful insights. Then I'll use these pictures to discuss some of the ways in which superconductivity produces dramatic changes in properties compared to a normal-state metal: (a) the existence of an energy gap to excitations, (b) zero electrical resistance, and (c) the Meissner effect by which a superconductor will expel an applied magnetic field up to some critical field. My overall goal today is to give you a running start if your research work will involve superconducting materials. In the next lecture I will go into more detail about some applications of superconductivity and a brief introduction to unconventional (non-*s*-wave) superconductors.

### **Three pictures of the superconducting ground state for the BCS model Hamiltonian**:

**1**. All electrons are described by one macroscopic quantum wavefunction which corresponds to a bound pair state: $\Phi_{\text{BCS}}(\vec{r}_1\sigma_1, \vec{r}_2\sigma_2) = \psi((\vec{r}_1 + \vec{r}_2)/2)\varphi(\vec{r}_2 - \vec{r}_1)$(spin wavefunction). This state is similar to the solution of the Cooper pair argument except that it involves partial occupation of single-electron states both above and below the Fermi energy, rather than just partial occupation of states above a frozen Fermi sea. In principle, the total many-electron wavefunction will follow from putting electrons pairwise into this state and antisymmetrizing at the end to make an appropriate fermionic wavefunction:

$$\Psi_{\text{BCS}}(\vec{r}_1\sigma_1, \ldots, \vec{r}_N\sigma_N) = \mathcal{A}[\Phi_{\text{BCS}}(\vec{r}_1\sigma_1, \vec{r}_2\sigma_2) \cdots \Phi_{\text{BCS}}(\vec{r}_{N-1}\sigma_{N-1}, \vec{r}_N\sigma_N)] \quad (6.1)$$

where I use the symbol $\mathcal{A}$ to mean antisymmetrize. If we tried to force all the electrons into the same single-electron state (rather than the same pair state), the antisymmetrization would of course give zero because of the Pauli exclusion principle. However, antisymmetrization of a product of the same pair state does not reduce to zero. Formally, the appropriate pair state $\Phi_{\text{BCS}}(\vec{r}_1\sigma_1, \vec{r}_2\sigma_2)$ could be determined by using a variational technique to minimize the total energy.

**2**. Form a ground state as a superposition of Slater determinants made by pairwise occupation of Kramers doublets of single-electron states, and minimize the total energy considering both the single-electron energies and the BCS pair-scattering interaction term. I'll explain this pictorially.

Let's take the ket below as a cartoon representation of the Slater determinant for a Fermi liquid state, where each horizontal line indicates a Kramers doublet of single-particle states related by time-reversal symmetry $\{\vec{k}\uparrow, -\vec{k}\downarrow\}$ that are occupied within the Slater determinant up to some Fermi level at zero temperature.

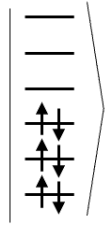

This state provides no energy benefit from the pair scattering terms in the Hamiltonian. We can evaluate this explicitly

$$\left\langle \cdots \middle| -\lambda_{BCS} \sum_{k \neq j} c^\dagger_{j\uparrow} c^\dagger_{-j\downarrow} c_{-k\downarrow} c_{k\uparrow} \middle| \cdots \right\rangle$$

This expectation value is identically equal to zero since in each term of the sum the interaction depopulates one occupied pair state and populates a different previously-empty pair state, so the result will be orthogonal to the original state.

To gain an energy benefit from the pair-scattering terms, we must have a superposition of different Slater determinants. The superconducting ground state can be written in the form

$$\Psi_{BCS} = c_1 \left| \cdots \right\rangle + c_2 \left| \cdots \right\rangle + c_3 \left| \cdots \right\rangle + c_4 \left| \cdots \right\rangle + \cdots$$

For this state the action of the pair scattering operator on the wavefunction now does not lead to an orthogonal state. Pair scattering on a single Slater determinant in the superposition results in another of the Slater determinants already in the superposition so that the final expectation value of the pair-scattering operators is non-zero. With this type of superposition there can be an energy benefit from pair scattering that reduces the total energy below that of the Fermi-liquid ground state energy. Occupancy of single-electron doublets above the original Fermi energy will cost some extra kinetic energy, but this can be more than compensated by the energy gain from the pair-scattering contribution (similar to what happened in the Cooper argument).

A variational calculation can be used to determine the coefficients $\{c_i\}$ that minimize the total energy. The result for the average occupation of the single-electron pair states within the superposition looks like the curve in Fig. VI.1:

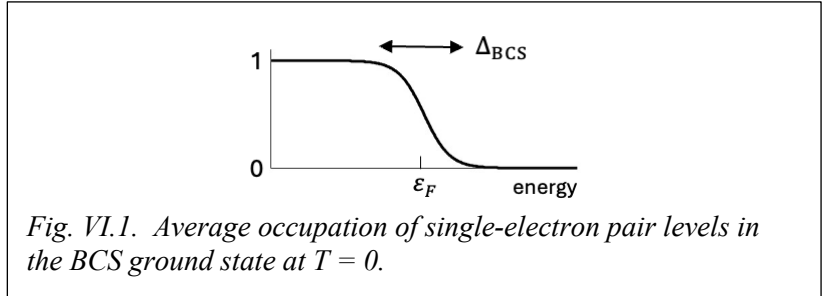

*Fig. VI.1. Average occupation of single-electron pair levels in the BCS ground state at T = 0.*

The curve has a shape similar to a Fermi function at some finite temperature, but this nevertheless corresponds to a zero-temperature state. The width of partial occupation is similar to what I will shortly define as the superconducting gap, $\Delta_{BCS}$.

This model was solved exactly by R. W. Richardson in the 1960's, several years after the original 1957 Bardeen-Cooper-Schrieffer solution that I will discuss next. It is trickier than the BCS approach because it assumes a fixed total number of electrons, and this causes the occupation probabilities of the different single-electron levels to be correlated, a constraint that is not straightforward to implement.

**3**. (Mathematically the easiest approach) Bardeen, Cooper, and Schrieffer avoided having to deal with the constraint of fixed electron number by postulating a state with an indefinite number of electrons (similar to a grand canonical ensemble)

$$\Psi_{BCS} = \prod_{\vec{k}} \left( u_{\vec{k}} \ |\vec{k} \uparrow \text{ and } -\vec{k} \downarrow \text{ empty} > + v_{\vec{k}} \ |\vec{k} \uparrow \text{ and } -\vec{k} \downarrow \text{ full} > \right)$$

$$= \prod_{\vec{k}} \left( u_{\vec{k}} + v_{\vec{k}} \ c^{\dagger}_{\vec{k}\uparrow} c^{\dagger}_{-\vec{k}\downarrow} \right) |\text{vacuum}>. \qquad (6.2)$$

(Here |vacuum> corresponds to a state with no electronic states occupied.) They then used a variational technique to solve for the coefficients $\{u_{\vec{k}}, v_{\vec{k}}\}$ that give the lowest energy, subject to appropriate normalization, for a given average number of electrons. This allows for a simple calculation. You can project onto a given number of electrons $N$ at the end, if you want, but for a large number of electrons this grand canonical approach gives results that are indistinguishable from our picture #2 fixed-$N$ calculation. These two approaches begin to differ appreciably only when the number of electrons is small enough that the single-electron level spacing is greater than or equal to about half the superconducting gap $\Delta_{BCS}$. There is a nice discussion of the projection procedure in chapter 3 of Tinkham's *Introduction to Superconductivity* book and a comparison of the two approaches is also reviewed in ref. 7.

**Properties of Superconductors**

Next I will discuss some of the main ways that superconductors differ from normal metals. I will use whichever of our 3 pictures is most appropriate for illustrating a particular point, switching between them as convenient.

*Energy gap for low-energy excitations*

How much energy does it take to "break a pair"? (Say, to make a total spin $S = 1$ state rather than a total spin $S = 0$ state if you have an even number of electrons in your sample.)

An $S = 1$ state requires that two single-electron pair doublets $\{\vec{k}\uparrow, -\vec{k}\downarrow\}$ in each Slater determinant are occupied by a single electron, rather than being doubly-occupied or empty as in the BCS ground state. This will block those two doublets from participating in the pair-scattering process, with the consequence that the $S = 1$ state will receive less energy benefit from pair-scattering. It turns out that the lowest-energy option for doing this is to have the same two doublets singly occupied in all of Slater determinants in the superposition, as this has the least impact on the blocking of the pair scattering

$$\Psi_{\text{excited}}(S=1) = d_1 \left| \cdots \right\rangle + d_2 \left| \cdots \right\rangle + d_3 \left| \cdots \right\rangle + d_4 \left| \cdots \right\rangle + \cdots$$

This state will be higher in energy than the ground state both because of a (small) increase in the single-electron energy and a (larger) increase due to the interaction energy because the having two Kramers doublets singly occupied reduces the amount of beneficial pair-scattering. The result in a BCS calculation is that the energy to break a pair as the level spacing goes to zero and for weak coupling ($\lambda_{BCS} N(\varepsilon_F) \ll 1$) is

$$2\Delta_{\text{BCS}} = 4\hbar\omega_D e^{-\frac{1}{\lambda_{BCS} N(\varepsilon_F)}} \tag{6.3}$$

The same energy cost will be incurred by blocking two Kramers doublets with single unpaired electrons of any spin; this is not a spin-dependent energy cost. In fact, at a fixed electron number there are no single-electron excitations which have energy less than $2\Delta_{\text{BCS}}$. The excitations are therefore gapped by this amount, a situation very different from the ungapped excitations of a normal Fermi liquid. (You will note that $\Delta_{\text{BCS}}$ is a little different from the quantity $\Delta_{\text{Coop}}$ that resulted from the Cooper argument – this is because the BCS calculation involves a self-consistent pair occupation distribution, not a frozen Fermi sea.)

Adding a single electron to an otherwise fully-paired superconducting ground state will also block one single-electron doublet from participating in pair scattering. This situation comes up when one attempts to tunnel single electrons onto or off a sample of superconductor. For example the energy required to tunnel an electron onto a fully-paired superconductor in the BCS theory is $\varepsilon_F + \Delta_{\text{BCS}}$ rather than just $\varepsilon_F$ as for a Fermi-liquid state.

There are many important consequences of the fact that a superconductor has this energy gap for excitations. The following consequences hold for conventional superconductors, in which the gap function is spherically symmetric, with no nodes.
- The superconducting state is durable at temperatures much less than $2\Delta_{\text{BCS}}/k_B$ because the excitations that might degrade the superconductivity are frozen out. The size of the energy gap is directly related to how high in temperature the superconducting state remains stable – as $k_B T$ increases to approach $\Delta_{\text{BCS}}$, more and more pairs are broken by thermal excitation and eventually the superconductivity is destabilized. In the BCS theory, the critical temperature is given by $T_c = 1.764\Delta_{\text{BCS}}/k_B$. Measured values of $T_c$ vary greatly between different superconducting materials, from well below 1 K to above 100 K. The highest-

$T_c$'s measured in well-confirmed experiments at ambient pressure are above 130 K in some of the high-$T_c$ cuprate superconductors.
- Specific heat: The specific heat well below the superconducting critical temperature reflects the gap, at low temperature typically

$$C \propto e^{-\frac{2\Delta_{BCS}}{k_B T}} + AT^3 \qquad (6.4)$$

where the cubic term is due to phonons.
- Thermal conductivity: At temperatures well below $2\Delta_{BCS}/k_B$ the low-energy excitations are frozen out and the superconducting state corresponds to a well-defined single quantum state with zero entropy. It is not capable of absorbing small amounts of energy (less than $2\Delta_{BCS}$) and therefore it cannot carry thermal energy from place to place. Its electronic thermal conductivity well below $T_c$ is zero. (This is closely related to the previous point, as the electronic thermal conductivity is proportional to the electronic specific heat.)
- Microwave absorption: Consistent with energy conservation, there is no absorption by the superconducting ground state when $\hbar\omega < 2\Delta_{BCS}$, while there is lots of absorption for higher frequencies. This applies to both electromagnetic radiation and sound.
- Single-electron tunneling between a normal metal and a superconductor: The tunneling probability as a function of electron energy $\varepsilon - \varepsilon_F$ to add an electron to a superconductor above the Fermi energy or remove an electron from below the Fermi energy is shown in Fig. VI.2. If $|\varepsilon - \varepsilon_F| < \Delta_{BCS}$ there is no way for a single electron to tunnel onto or out of the superconductor while conserving energy, so the tunneling probability is zero. When $|\varepsilon - \varepsilon_F| > \Delta_{BCS}$ single-electron tunneling is allowed, and furthermore there is a sum rule so that the tunneling probability lost from within the gap moves to produce peaks in the tunneling probability near the gap edges.

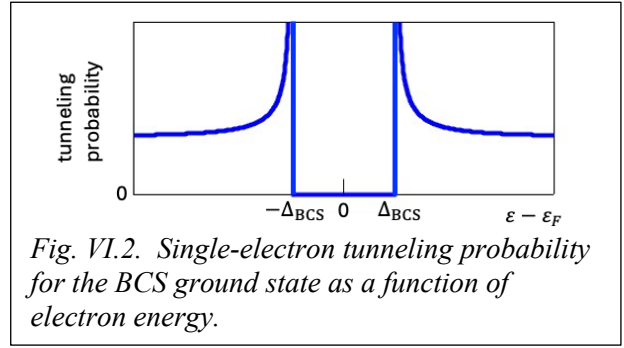

Fig. VI.2. Single-electron tunneling probability for the BCS ground state as a function of electron energy.

   A pair of electrons tunneling simultaneously can still be added at low energy, because that causes no blocking of phase space for pair scattering. This is the basis of the Josephson effect and SQUIDs (superconducting quantum interference devices) that I will discuss next time.
- Isotope effect: The excitation gap and therefore the critical temperature predicted by the BCS theory are proportional to a cut-off energy $\hbar\omega_D$ associated with a phonon energy scale, if the attractive interaction is due to overscreening by phonons (see Eq. (6.3)). If the atoms involved in the overscreening have their mass $M$ changed by isotope substitution, $\Delta_{BCS}$ should therefore be proportional to $1/\sqrt{M}$, as long as $\lambda_{BCS}$ is unchanged. The observation of this isotope effect provided confirmation of the involvement of phonons in the pairing mechanism for early classical superconductors.

*Condensation energy*

The ground-state energy of a superconductor $E_S$ is lower than the energy of a normal Fermi liquid state, $E_N$. By how much? What is $E_N - E_S$?

You can measure this condensation energy using specific heat. The extra energy in the normal state is released as a sample is cooled through the superconducting transition temperature, resulting in a peak in the specific heat near the transition temperature. Integration over this peak gives $E_N - E_S$. Alternatively, for the type I superconductors we will discuss next time the condensation energy can also be determined from the value of the critical magnetic field that destroys superconductivity.

Within the picture of Cooper pairs, one might guess that if the energy to break one Cooper pair is $2\Delta_{BCS}$ and with a total of $N$ electrons in a sample there are $N/2$ Cooper pairs in the superconducting state, then the total condensation energy is $N\Delta_{BCS}$. However this is incorrect by orders of magnitude. The actual value is of order $N\Delta_{BCS}(\frac{\Delta_{BCS}}{\varepsilon_F})$, smaller by about a power of $10^4$ for a typical superconductor.

There are a couple of useful ways to think about this. One correct viewpoint is that breaking up one Cooper pair means that two of the single-electron doublets will be singly-occupied, blocking their use for pair scattering in the formation of other Cooper pairs. This means that breaking one pair weakens the binding energy of all the others, so that breaking the second Cooper pair requires slightly less energy than $\Delta_{BCS}$. Breaking the third pair is even easier, etc., so that the total condensation energy does not simply scale like $N\Delta_{BCS}$. A second correct viewpoint is that only single-electron states within an energy scale of about $\Delta_{BCS}$ around the Fermi level have their average occupation changed in the superconducting state relative to the normal state (recall Fig. VI.1), which means that only these single-electron states contribute significantly to the energy gained by pair scattering. The number of these states is of order $N(\frac{\Delta_{BCS}}{\varepsilon_F})$ if $N$ is the total number of electrons.

This situation means that the condensation energy per electron in a superconductor is actually remarkably small – for a superconductor like Al which has a transition temperature near 2 K the condensation energy per electron is about $(4\text{ mK})k_B$. This makes clear the importance of collective effects in stabilizing superconductivity.

*Zero Resistance*

Consider the full two-electron pair wavefunction, including both the center-of-mass and relative-coordinate parts: $\Phi_{BCS} = \psi(\vec{R})\varphi(\vec{r})$. For the ground state in the absence of any applied electrical or magnetic fields, the center of mass part is just $\psi(\vec{R}) =$ constant, corresponding to a center of mass at rest. When we considered low-energy excitations previously, we therefore only considered excitations within the relative-coordinate part of the wavefunction. However, in an applied field the center-of-mass part of the wavefunction can also change. Since all of the electron pairs are described by the *same* macroscopic wavefunction, an applied electric field will cause all of the electrons to be accelerated together, so that the center-of-mass part of the wavefunction gains a non-zero momentum $\vec{p}$ to take the form

$$\psi(\vec{R}) \rightarrow e^{i\vec{p}\cdot\vec{R}/\hbar} \tag{6.5}$$

(with the relative-coordinate part not significantly changed). Crucially, all of the electron pairs must have the *same* total momentum $\vec{p}$. Microscopically, the electronic basis states contributing to the pair scattering must now have the form $\{\vec{k}+\frac{\vec{p}}{2}\uparrow, -\vec{k}+\frac{\vec{p}}{2}\downarrow\}$, all with the same value of $\vec{p}$, instead of the $\{\vec{k}\uparrow, -\vec{k}\downarrow\}$ combinations that participate in pair-scattering to construct the at-rest ground state. When the superconducting condensate is moving, any scattering process which might change the momentum of an electron (or the total momentum of a pair) would mean that there is at least one single electron state, e.g. $\vec{q}\uparrow$, that becomes occupied while its pair-partner state $(-\vec{q}+\vec{p})\downarrow$ is not. As a consequence, scattering would cause those single-electron states to be blocked from participating in the pair-scattering process and the condensation energy would have to be reduced. In other words, any scattering process must necessarily "break a Cooper pair," and as long as the kinetic energy of each Cooper pair remains much less than $2\Delta_{BCS}$ there is no way to break pairs and conserve energy. With no possibility of scattering, the moving electrons can do nothing other than keep moving with no degradation in the current, which means that the electrical resistance is zero.

The ultimate limit in principle for the superconducting current density within a material is when the kinetic energy per Cooper pair becomes large enough to be comparable to $2\Delta_{BCS}$. This is the limit that applies in microscopic wires. However, in macroscopic superconducting wires, usually a lower limiting condition occurs first – that the Oersted magnetic field generated by the supercurrent flow exceeds the critical magnetic field, and that is enough to begin to destroy the superconductivity. (See the discussions of critical magnetic fields to follow.)

*Meissner Effect*

The Meissner effect is the expulsion of magnetic field from inside a superconductor, if the magnetic field is not too strong. One aspect of the Meissner effect is a relatively trivial consequence of the existence of zero resistance inside a superconductor. If you ramp up a magnetic field applied to any metal, the electromotive force generated by the changing magnetic field will generate screening currents at the metal's surface. These screening currents in turn will generate a magnetic field that opposes the applied field, by Lenz's Law. If the material has zero resistance, these screening currents will produce perfect diamagnetism that does not decay with time, in that the magnetic field produced by the screening currents will exactly cancel the applied magnetic field, leaving no net magnetic field inside the material.

The Meissner effect in a superconductor has an additional aspect that goes beyond this consequence of zero resistance, however. Suppose you apply a magnetic field to a piece of material starting at a temperature above its superconducting critical temperature so that the magnetic field penetrates through the material, and then you cool the material through the critical temperature. In a superconductor, the magnetic field will be expelled from the interior of the material as it becomes superconducting. If the only physics at work were the creation of zero resistance upon cooling, the magnetic field would instead in this case be locked in place upon cooling rather than expelled from the interior of the material, because zero-resistance currents would act to resist any change in the magnetic field penetrating the sample.

What is the microscopic reason for the Meissner Effect? Recall from our "thought calculation" about electron-electron interactions that to get the large pair-scattering matrix elements which produce superconductivity, it was necessary to select the wavefunctions entering into the integral for the electron-electron matrix element in a particular way -- pairwise. For the Bloch-state basis, these large pair-scattering matrix elements correspond to scattering two electrons from one Kramers doublet state $\{\vec{k}\uparrow, -\vec{k}\downarrow\}$ to another Kramers doublet state $\{\vec{j}\uparrow, -\vec{j}\downarrow\}$. The critical factor to get a large element is that the single-electron states within a Kramers doublet are related by time reversal symmetry, so that $\psi_{-\vec{k}}(r) = \psi_{\vec{k}}^*(r)$. With this relationship, the amplitude of the pair-scattering matrix element take the form (assuming the screened electron-electron interaction can be approximated as a delta function, as we did in the "thought calculation")

$$V_{\text{pairing}} \approx V_0 \int d^3r_1 d^3r_1 \psi_{\vec{j}}^*(r_1)\psi_{-\vec{j}}^*(r_2)\delta(r_2 - r_1)\psi_{\vec{k}}(r_2)\psi_{-\vec{k}}(r_1)$$
$$\approx V_0 \int d^3r |\psi_{\vec{j}}(r)|^2 |\psi_{\vec{k}}(r)|^2 \approx -\lambda_{\text{BCS}}. \tag{6.6}$$

Choosing the wavefunctions pairwise as Kramers doublets related by time-reversal symmetry was essential to get an integrand that was positive-definite and therefore to get a large value for the integral. If an applied magnetic field penetrates through a material it will break the time reversal symmetry of the single-electron states. The peaks and troughs in the single-electron wavefunctions get out of alignment, so there are no pairs of basis states with a relationship analogous to $\psi_{\vec{k}}(r)\psi_{-\vec{k}}(r) = |\psi_{\vec{k}}(r)|^2$. There is therefore no way to choose basis-state wavefunctions in pairs to get a positive-definite integrand for the interaction integral, and no way to get large pairing matrix elements.

Since the BCS gap goes like $\Delta_{\text{BCS}} = 2\hbar\omega_D e^{-\frac{1}{\lambda_{\text{BCS}}N(\varepsilon_F)}}$, if the matrix element $\lambda_{\text{BCS}}$ is reduced in magnitude by breaking time reversal symmetry and messing up the alignment of the basis-state wavefunctions, then $\Delta_{\text{BCS}} \to 0$ in an exponential fashion. Therefore BCS superconductivity would quickly be killed if a magnetic field were successful in penetrating through a superconducting region.

As a consequence, if a sample in the presence of an applied magnetic field is cooled through its superconducting critical temperature, in the process of cooling it can be energetically favorable for the sample to generate loops of screening supercurrent at its surface so that the net magnetic field in the bulk of the sample remains zero. This maximizes the attractive pairing interaction and allows the energy of the sample to be lowered by the superconducting condensation energy. This is the origin of the Meissner effect. Generating these screening currents does come with an energy cost, however, which can be calculated based on the energy density associated with the cancelling magnetic field: $\Delta E_{\text{screening}} = B^2(\text{sample volume})/(2\mu_0)$. Under conditions when the screening energy cost would become larger than the condensation energy benefit, the applied magnetic field becomes incompatible with spatially-uniform superconductivity. This defines a critical magnetic field corresponding to the condition $B_c^2$(sample volume)$/(2\mu_0) = E_\text{N} - E_\text{S}$. For a superconductor with a transition temperature of order 1 K, a typical value of $B_c$ is of order a few times 0.01 Tesla, or a few times 100 Gauss.

Next time I will discuss the interactions of superconductors with magnetic fields in more detail, and we will see that there are two different types of superconductors with qualitatively different properties for applied magnetic fields near $B_c$. In type I superconductors, magnetic fields above $B_c$ simply destroy superconductivity, and the sample undergoes a first-order transition to a normal state. In type-II superconductors, however, a magnetic field is able to penetrate a superconductor in the form of quantized fluxons which make the superconductor spatially non-uniform but do not destroy it utterly. (More on this next time.)

**What have we learned?:**

- The superconducting state can be viewed as a macroscopic quantum wavefunction which results from a particular superposition of Slater determinant states. For a superconductor at rest, in each of the Slater determinant states the Kramers doublets $\{\vec{k}\uparrow, -\vec{k}\downarrow\}$ are either doubly-occupied by electrons or empty, and pair scattering of electrons between these Kramers doublets couples the different Slater determinants and allows the total energy to be lower than the energy of a conventional Fermi liquid. This energy lowering is known as the condensation energy.

- If a Kramers doublet is occupied by a single electron (rather than 0 or 2 electrons) this blocks that doublet from participating in the pair-scattering process and reduces the superconducting condensation energy. As a consequence, there is an energy gap to any low-energy excitation in a superconductor that "breaks a pair," given by $2\Delta_\text{BCS}$.

- Zero resistance in a superconductor is a natural consequence of a macroscopic wavefunction that can move in response to an applied electric field, together with an energy gap for pair-breaking excitations. If there is no way for an electron to scatter without breaking a pair, and pair-scattering is energetically forbidden because of the energy gap, then there is no way for superconducting electrons to relax their momentum or energy, and an electrical current once started will continue to flow without any resistance.

- (The Meissner effect) For low magnetic fields, it is energetically favorable for a BCS superconductor to expel an applied magnetic field (using screening by loops of supercurrent at its surface) in order to preserve time-reversal symmetry in the interior of the superconductor and thereby preserve large pair-scatting matrix elements for the attractive electron-electron interaction.

## VII. More About Superconductors + Their Applications

*Topics: 1. How magnetic-flux quantization in a superconducting loop follows from the requirement that the center-of-mass part of the superconducting wavefunction be single-valued.*

*2. How the physics of quantized fluxons produces differences between type-I and type-II superconductors.*

*3. The current-phase relationship of a superconducting Josephson junction and two applications of Josephson junctions, as superconducting qubits and as magnetic-field detectors (SQUIDs).*

*4. Brief introduction to unconventional superconductors.*

In the last lecture we discussed that the pair wavefunction in a superconductor can be considered the product of a center-of-mass part and a relative-coordinate part:

$$\Phi_{BCS} = \psi(\vec{R})\varphi(\vec{r}). \tag{7.1}$$

We have focused so far on the relative-coordinate part of the wavefunction when analyzing the condensation energy and energy gaps to excitations. For these cases the center of mass of the electrons stays essentially at rest so that we can simply take $\psi(\vec{R}) =$ constant. Today we will change gears to focus primarily on the center-of-mass part of the superconducting wavefunction. The quantum mechanics of the center-of-mass part is known as the Ginzburg-Landau Theory of superconductivity. This theory describes in an elegant way how the flow of supercurrents is related to gradients in the phase of $\psi(\vec{R})$ and how supercurrents interact with applied magnetic fields, among other subjects. Strictly speaking, Ginzburg-Landau Theory is equivalent to the full BCS theory only in the regime near and below the superconducting critical temperature, but it can give qualitative and semi-quantitative understanding more generally.

It is conventional to parameterize the center-of-mass part of the pair wavefunction as $\psi(R) = \sqrt{\rho(R)}e^{i\phi(R)}$. In regard to the amplitude of the wavefunction, we will mostly care only about relative amplitudes, but choosing to call the amplitude $\sqrt{\rho(R)}$ allows one to identify $|\psi(R)|^2$ with a total local density of superconducting electrons. In a uniformly superconducting material, typically $\rho(R)$ is constant as a function of position except near sample boundaries (and near cores of the fluxons I will discuss later), so outside of these cases the important spatial dependence generally comes from the phase $\phi(R)$.

Within the Ginzburg-Landau theory, $\psi(R)$ satisfies an equation very similar to, but not exactly the same, as the usual time-independent Schrodinger equation (in addition to a kinetic energy term the Ginzburg-Landau expression also contains a nonlinear contribution $\propto |\psi(R)|^2\psi(R)$, by logic analogous to the Landau theory of phase transitions). Supercurrent flow is associated with the flow of probability density for $\psi(R)$ (times $2|e|$). In the case when a magnetic field is present, this flow can be written in terms of the usual expectation value of the kinetic momentum operator for a charged particle in quantum mechanics, which incorporates coupling to a magnetic field by inclusion of the vector potential $\vec{A}(R)$:

$$\vec{j}_s = \frac{q}{m_p} Re\left\{\psi^*\left(\frac{\hbar}{i}\vec{\nabla} - q\vec{A}\right)\psi\right\} = \frac{e\rho}{m}(\hbar\vec{\nabla}\phi + 2e\vec{A}). \tag{7.2}$$

Here for a Cooper pair $m_p$ is equal to twice the electron mass $m$ and $q = -2e$, and $Re$ indicates to take the real part. In the last step I am assuming $\rho(R) = $ constant.

## Magnetic flux quantization within superconducting loops

Let's first apply this framework to analyze a superconducting sample with a hole which is threaded by a magnetic field (Fig. VII.1). If you pick a circular loop path going around the hole and sufficiently deep into the superconductor that the surface screening currents at the position of the loop are negligible, then $\vec{j}_s = 0$ everywhere on the loop. Therefore, taking a line integral over the loop,

$$\oint \vec{j}_s \cdot d\vec{R} = 0 = \frac{e\rho\hbar}{m} \oint \vec{\nabla}\phi \cdot d\vec{R} + \frac{2e^2\rho}{m} \oint \vec{A} \cdot d\vec{R} \qquad (7.3)$$

or

$$\oint \vec{\nabla}\phi \cdot d\vec{R} = -\frac{2e}{\hbar} \oint \vec{A} \cdot d\vec{R}. \qquad (7.4)$$

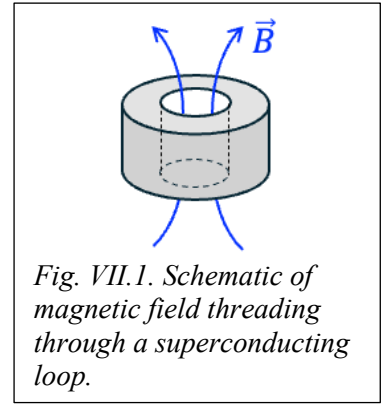

Fig. VII.1. Schematic of magnetic field threading through a superconducting loop.

Now, $\oint \vec{\nabla}\phi \cdot d\vec{R} = \phi(\vec{R}_2) - \phi(\vec{R}_1)$ where $\vec{R}_1$ and $\vec{R}_2$ are the beginning and end points of the path. For a closed loop these are the same points, and in order for the wavefunction to be single valued we must have $\phi(\vec{R}_2) = \phi(\vec{R}_1)$ modulo $2\pi$. Therefore $\oint \vec{\nabla}\phi \cdot d\vec{R} = 2\pi(\text{integer})$.

We can evaluate the right-hand side of Eq. (7.4) as $\oint \vec{A} \cdot d\vec{R} = \int (\vec{\nabla} \times \vec{A}) \cdot d(\text{Area})$ by Stokes theorem $= \int \vec{B} \cdot d(\text{Area}) = $ magnetic flux enclosed by the loop.

Therefore, putting everything together, the flux enclosed within any fully superconducting loop (away from any screening currents) must have a quantized magnitude given by

$$\text{Magnetic flux} = \frac{h}{2e}(\text{integer}) \qquad (7.5)$$

where the unit of quantization $h/2e = 2 \times 10^{-15}$ Tesla-m², or $2 \times 10^{-7}$ Gauss-cm².

## Type-I and Type-II Superconductors

Consider the boundary of a superconductor with a small magnetic field applied parallel to the interface. The field will not penetrate far into the superconductor because of the superconducting screening currents, but it will penetrate slightly with a decay length $\lambda$ (known as the penetration depth) set by the spatial distribution of screening currents (Fig. VII.2). This exponential decay can be described by what is known as the London Equations, which give a value $\lambda \propto n_s^{-1/2}$, where $n_s$ is the superfluid density. Depending on the material, $\lambda$ generally ranges from 15 to a few 100 nm in conventional superconductors. In some superconductors $n_s$ can be very small because the electron density itself is very small, so the penetration depth can be even longer.

There is also a 2nd important length scale in a superconductor, called the coherence length, $\xi$ (Fig. VII.2). This is the distance over which the superconducting properties can vary.

You can think of it qualitatively as the size of a Cooper pair. Depending on the material, $\xi$ can range from a few nm to about 1000 nm.

Suppose that you apply a magnetic field with strength close to the critical magnetic field $B_c$ and oriented parallel to the surface of a superconductor. (From last time, $\frac{B_c^2}{2\mu_0} = \frac{(E_N - E_S)}{\text{(unit volume)}}$.) The penetration of the magnetic field and the spatial variations of the condensation energy can occur over different length scales. We can consider the energy balance at the interface. There will be an energy benefit of letting the magnetic field penetrate $\sim \frac{B_c^2}{2\mu_0} \lambda/$(unit area) and at the same time a cost of lost condensation energy $\sim \frac{B_c^2}{2\mu_0} \xi/$(unit area). The sum corresponds to an effective surface free energy $\sim \frac{B_c^2}{2\mu_0} (\xi - \lambda)/$(unit area) that can be either positive or negative. This difference in sign defines Type-I and Type-II superconductors. (A more rigorous calculation shows that the actual crossover in the sign of the surface free energy occurs at $\xi/\lambda = \sqrt{2}$ rather than $\xi/\lambda = 1$.)

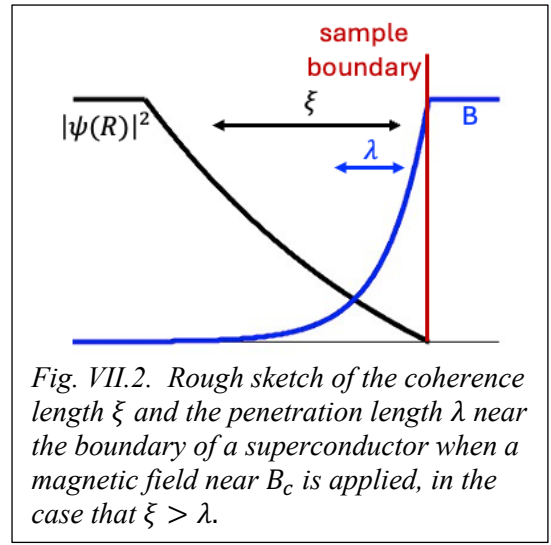

Fig. VII.2. Rough sketch of the coherence length $\xi$ and the penetration length $\lambda$ near the boundary of a superconductor when a magnetic field near $B_c$ is applied, in the case that $\xi > \lambda$.

Type-I superconductors have $\xi > \sqrt{2}\lambda$ and a positive surface energy when an applied magnetic field approaches the critical field $B_c$. This means that under this condition the optimum-energy situation is to minimize the surface area in contact with the magnetic field. The sample tries to exclude the magnetic field at best it can, but at the critical field $B_c$ that is no longer possible and there is a first-order transition to the normal state in which the magnetic field penetrates everywhere throughout the material.

Type-II superconductors have $\xi < \sqrt{2}\lambda$ and a negative surface energy. When the applied magnetic field approaches $B_c$, the optimum configuration for these materials is to maximize the surface area corresponding to the boundary of the superconductor with the magnetic field. This can be achieved by having the magnetic field penetrate into the bulk of the material in the form of individual quantized units of magnetic flux, known as fluxons or Abrikosov vortices. The end-result is that the material is no longer a spatially-uniform superconductor, but instead it is threaded by thin filaments of normal material each containing a quantized unit of magnetic flux surrounded by a vortex of supercurrent that screens the flux from going everywhere else within the superconductor. The fluxons interact with each other and can form lattice structures.

The ability of Type-II superconductors to host magnetic fluxons without having the superconductivity fully destroyed allows them to remain superconducting to much larger applied magnetic fields than typical Type-I superconductors – up to several tens (or even hundreds) of Tesla for some materials, which is of order 1000 times a typical Type-I superconductor. (Cartoons of typical Meissner-effect curves for Type-I and Type-II superconductors are shown in Fig. VII.3.) Type-II superconductors are therefore good for making high-field superconducting magnets. However, if the fluxons move around they can dissipate energy and cause electrical resistance and the decay of supercurrents. So the design of superconducting magnet materials requires careful consideration of how to keep the fluxons pinned so that they are not able to shift position.

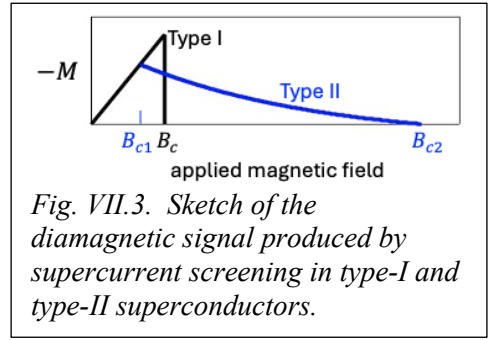

*Fig. VII.3. Sketch of the diamagnetic signal produced by supercurrent screening in type-I and type-II superconductors.*

## Superconductivity Limited by Spin Zeeman Splitting in a Magnetic Field

In addition to Type-II superconductivity, there is another strategy by which superconducting samples can maintain superconductivity to large magnetic fields by avoiding having to pay the energy cost of generating screening supercurrents. In this case, there is a different mechanism by which superconductivity can ultimately be destroyed at very large magnetic fields, limited not by the orbital motion of supercurrents but by a spin-based Zeeman energy that can disrupt the Cooper-pair binding energy.

If a superconducting sample is much thinner than the magnetic-field penetration depth in at least one direction, then the magnetic field can fully penetrate the sample without the build-up of significant screening currents. This can happen for a superconducting film with nanoscale thickness with a magnetic field applied very-accurately parallel to the film plane, one-dimensional nanoscale wires, or nanoparticles. In these cases, the magnetic flux through the sample is also generally small enough that the time-reversal symmetry of the single-electron basis states is barely affected, and the pair scattering that drives superconductivity remains strong. The end result is that the superconducting condensation energy remains large even in the presence of the magnetic field up to field scales much larger than would be possible in a bulk sample given the energy costs of the Meissner effect. Ultimately, though, a magnetic field still destabilizes BCS superconductivity by a different mechanism – superconductivity is destroyed in a spin-singlet superconductor if the spin Zeeman energy becomes comparable to the superconducting gap, so that the energy splitting between the spins in a Kramers doublet becomes comparable to the binding energy of a Cooper pair. For a material with weak spin-orbit scattering, this physics yields the "Clogston-Chandrasekhar" or "Pauli paramagnetism" limit for a depairing magnetic field of $B_p = \Delta_{\text{BCS}}/(\mu_B \sqrt{2})$, which as a rule of thumb works out to be about $(1.8 \text{ Tesla}/K)T_c$ where $T_c$ is the transition temperature of the material. This field scale is about 100 times larger than the $B_c$ limit for a bulk Type-I superconductor. This is still not a fundamental limit, however, because materials with large spin-orbit coupling can have greatly reduced Zeeman splitting compared to samples without spin-orbit coupling, so that the Zeeman splitting does not become comparable to the superconducting gap until even larger magnetic fields. This occurs in the van der Waals superconductor NbSe$_2$ (which has a $T_c$ for monolayer samples about 3 K ranging up to 6 K for trilayers) in which there is a strong spin-orbit field oriented perpendicular to the plane and so superconductivity is stable against an in-plane-oriented magnetic field to above 30 Tesla (X. Xi et al., *Nature Physics* **12**, 139 (2016)).  It also occurs in thin superconducting films containing heavy metals, in which the single-electron basis states have an almost equal superposition of spin-up and spin-down states and so have reduced sensitivity to Zeeman splitting.  Entirely different physics applies for spin-triplet superconductors, where the Zeeman effect does not disrupt pairing, and in a few cases it appears that a large magnetic field can help to stabilize this phase.

**Superconducting Devices -- Josephson Junctions**

(For more on this topic, see a nice discussion in the Feynman lectures.)

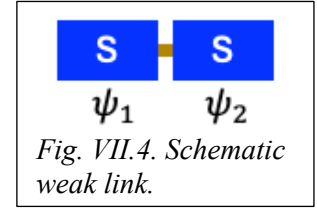

Fig. VII.4. Schematic weak link.

Suppose you have a "weak link" between two superconductors (Fig. VII.4).  This could be a thin normal-metal layer, a tunnel junction, or a narrow constriction connecting two larger pieces of superconductor.  By "weak" I mean that the superconductor pair wavefunctions from each side decay evanescently within the weak link. These types of structure are known as a Josephson junctions, and as we will see soon they are critical components in both superconducting quantum interference devices (SQUIDs) used for magnetic-field detection and for superconducting qubits used in the development of quantum computation.  A simple analysis will show that a supercurrent can pass through a Josephson junction, with a value that depends on the difference in superconducting phase on the two sides of the junction.  (From a microscopic viewpoint, this supercurrent consists of pairs of electrons tunneling simultaneously.  Pair tunneling does not block any Kramers doublets from participating in pair scattering, so pair tunneling is not impeded by the same excitation gap as single-electron tunneling.)

Let the center-of-mass parts of the superconducting wavefunctions in the two electrodes immediately adjacent to the weak link have the values $\psi_1 = \sqrt{\rho_1}e^{i\phi_1}$ and $\psi_2 = \sqrt{\rho_2}e^{i\phi_2}$.  Also suppose there might be a voltage $V$ across the junction (with the reference point for $V = 0$ at the midpoint of the junction for convenience).  We can easily write down two Schrodinger equations for $\psi_1$ and $\psi_2$ accounting for the potential-energy difference due to the applied voltage and also for the simplest possible model for coupling between the wavefunctions on the two sides:

$$i\hbar \frac{\partial \psi_1}{\partial t} = -2e\left(\frac{V}{2}\right)\psi_1 + K\psi_2$$
$$i\hbar \frac{\partial \psi_2}{\partial t} = +2e\left(\frac{V}{2}\right)\psi_2 + K\psi_1 \tag{7.6}$$

where $K$ is a coupling constant.  If you substitute in $\psi_1 = \sqrt{\rho_1}e^{i\phi_1}$ and $\psi_2 = \sqrt{\rho_2}e^{i\phi_2}$ and then separate the Schrodinger equations into real and imaginary parts, the result is

$$\frac{d\rho_1}{dt} = \frac{2K}{\hbar}\sqrt{\rho_1\rho_2}\sin(\phi_2 - \phi_1) = -\frac{d\rho_2}{dt} \tag{7.7}$$

$$\frac{d\phi_1}{dt} = -\frac{K}{\hbar}\sqrt{\frac{\rho_2}{\rho_1}}\cos(\phi_2 - \phi_1) + \frac{eV}{\hbar} \tag{7.8}$$

$$\frac{d\phi_2}{dt} = -\frac{K}{\hbar}\sqrt{\frac{\rho_1}{\rho_2}}\cos(\phi_2 - \phi_1) - \frac{eV}{\hbar}. \tag{7.9}$$

For a symmetric weak link with $\rho_1 = \rho_2$, this means the time dependence of the phase difference is

$$\frac{d}{dt}(\phi_2 - \phi_1) = -\frac{2eV}{\hbar}. \tag{7.10}$$

We can draw several conclusions from this analysis. First, the quantity $(-2e)d\rho_1/dt$ can be interpreted as the value of an instantaneous charge current across the junction. This current is generally non-zero even with no applied voltage across the junction, and its value depends on the phase difference. At $V = 0$, this supercurrent has the simple form

$$I = I_0 \sin(\phi_2 - \phi_1) \tag{7.11}$$

where $I_0$ is the maximum supercurrent for a particular junction, and at $V = 0$ the phase difference is constant in time.

If a voltage is present, integration of the equation for $d(\phi_2 - \phi_1)/dt$ gives $\phi_2 - \phi_1 = -\frac{2eVt}{\hbar} + \text{constant}$, so that the supercurrent flow is now time dependent, $I(t) = I_0 \sin\left(-\frac{2eVt}{\hbar} + \text{constant}\right)$. If a constant voltage is present, the supercurrent will therefore be oscillatory with a frequency $\omega = 2\pi f = 2eV/\hbar$. For $eV$ of order a typical superconducting gap energy, this frequency is in the microwave range.

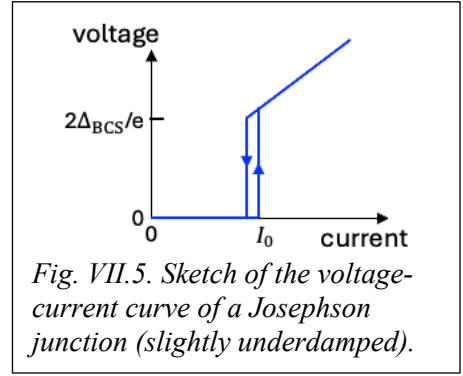

Fig. VII.5. Sketch of the voltage-current curve of a Josephson junction (slightly underdamped).

Experimentally, it is more common to pass a definite current through a Josephson junction rather than a applying a definite voltage, because of the vanishing impedance of the junction when only supercurrent flows. (Applying a definite current is called a current bias.) A typical voltage-current curve then looks something like Fig. VII.5. For applied currents less than $I_0$, the maximum possible supercurrent value of for the junction being measured, there will be a supercurrent with no voltage generated, and the phase difference will adjust to whatever constant value is needed to be consistent with the applied current. For applied currents larger than $I_0$, the junction is not capable of carrying all of the applied current as supercurrent, so it will need to transmit a normal current by single-electron tunneling. Single-electron tunneling requires that the voltage jump to a value $|eV| \geq 2\Delta_{BCS}$ in order that the applied voltage can pay the energy-gap costs to end up with an unpaired electron in both electrodes. Given that the applied voltage is non-zero, the supercurrent in this regime will be oscillatory. (The time-dependent part is not shown in Fig VII.5.) Some hysteresis is possible upon sweeping up in current versus sweeping down, due for example due to heating.

If you apply a current with both dc and ac components, the fact that the current-phase relationship $I = I_0 \sin(\phi_2 - \phi_1)$ is nonlinear allows for phase locking between the applied ac current and the intrinsic oscillatory dynamics of the junction. The result is known as Shapiro

steps – the dc component of the voltage can exhibit quantized steps with values $= nhf/(2e)$ even in the regime $|I| < I_0$. Here $f$ is the applied frequency and $n$ is an integer.

## Superconducting Devices – Superconducting Quantum Interference Devices (SQUIDs)

A dc SQUID consists of two Josephson junctions connected in parallel using bulk superconducting wires (Fig. VII.6). If a magnetic flux penetrates through the loop connecting the junctions, we will see that this changes the relative value of the phase changes across the two Josephson junctions. This tunes the sum of the supercurrents through the junctions between constructive and destructive interference, so that the total critical supercurrent changes, and this change can be used as a measure of the strength of magnetic field passing through the loop. The sensitivity for magnetic-field detection grows with the area of the loop, so this provides a way to make extremely sensitive measurements of magnetic field.

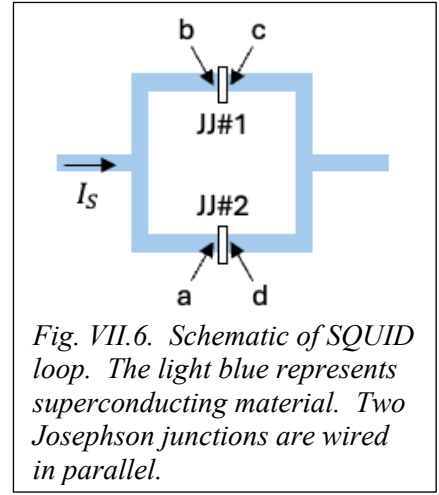

Fig. VII.6. Schematic of SQUID loop. The light blue represents superconducting material. Two Josephson junctions are wired in parallel.

The details about how a SQUID works can be understood based on what we know now about superconductor phases, vector potentials, and Josephson junctions. Let us define the superconducting phase change across junction 1 as $\delta_1 = \phi_c - \phi_b$ and the phase change across junction 2 as $\delta_2 = \phi_d - \phi_a$ so that the total current through the SQUID is $I_{\text{total}} = I_0(sin\delta_1 + sin\delta_2)$, assuming for simplicity that the two junctions are identical with the same maximum supercurrent $I_0$. Because the supercurrent flow through the SQUID is limited by the Josephson junctions to values much less than the superconducting wires could carry by themselves, to a good approximation the supercurrent density in the bulk wires is $j_s \approx 0$. By Eq. (7.2), the spatial variation of the superconducting phase within the wires follows $\vec{\nabla}\phi = -\frac{2e}{\hbar}\vec{A}$. By evaluating line integrals for $\vec{A}$ along the superconducting wires between points a and b and between points c and d, we can compute the changes in superconducting phase along these paths

$$\phi_b - \phi_a = -\frac{2e}{\hbar}\int_a^b \vec{A}\cdot d\vec{R}; \quad \phi_d - \phi_c = -\frac{2e}{\hbar}\int_c^d \vec{A}\cdot d\vec{R}. \qquad (7.12)$$

Adding these two expressions gives

$$\phi_d - \phi_a - (\phi_c - \phi_b) = -\frac{2e}{\hbar}\left[\int_a^b \vec{A}\cdot d\vec{R} + \int_c^d \vec{A}\cdot d\vec{R}\right]. \qquad (7.13)$$

Now the widths of Josephson junctions are negligible on the scale of the SQUID's macroscopic loop and the vector potential is necessarily continuous, so we can take $\vec{A}(c) = \vec{A}(b)$ and $\vec{A}(d) = \vec{A}(a)$. With this identification, the sum of the line integrals becomes a line integral of $\vec{A}$ over a closed loop, and we have already seen via Stoke's theorem this integral is equal to the magnetic flux enclosed within the loop, which I will call $\Phi_{\text{flux}}$. Our final result then relates the difference between the phase changes across the two Josephson junctions to the magnetic flux enclosed within the loop

$$\delta_2 - \delta_1 = -\frac{2e}{\hbar}\Phi_{\text{flux}}. \qquad (7.14)$$

Now imagine measuring the maximum total supercurrent flow through the SQUID as a junction of increasing the applied magnetic field, starting at $\Phi_{flux} = 0$. Assuming identical junctions, the initial current flow will be divided equally between them, so that $\delta_1 = \delta_2 \equiv \delta_0$. As the flux increases, this will cause the two phase changes to diverge:

$$\delta_1 = \delta_0 + \frac{e}{\hbar}\Phi_{flux}; \quad \delta_2 = \delta_0 - \frac{e}{\hbar}\Phi_{flux}. \tag{7.15}$$

The total supercurrent through the two Josephson junctions is then

$$I_{total} = I_0[sin\delta_1 + sin\delta_2] = I_0\left[sin\left(\delta_0 + \frac{e}{\hbar}\Phi_{flux}\right) + sin\left(\delta_0 + \frac{e}{\hbar}\Phi_{flux}\right)\right]$$
$$= 2I_0 sin\delta_0 \, cos\left[\frac{e}{\hbar}\Phi_{flux}\right]. \tag{7.16}$$

The oscillatory dependence on $\Phi_{flux}$ shows that the magnetic flux can tune the magnitude of the maximum supercurrent between zero and twice the supercurrent of the individual junctions. The zero current case corresponds to equal and opposite supercurrents through the two junctions so that the total supercurrent is zero.

To use a SQUID to measure a magnetic field, you can either measure the total critical current directly, in which case the result is $\propto cos\left[\frac{e}{\hbar}\Phi_{flux}\right]$, or you can apply a current greater than the maximum critical current and you will get a non-zero resistance that varies periodically with the enclosed magnetic flux. SQUID loops for magnetic field detection can be used in either a fixed geometry or they can be fabricated as part of scanning-probe microscopes to measure microscopic patterns of magnetic field as a function of position. Scanning SQUIDs have played important roles in understanding the physics of magnetic fluxons, unconventional superconductors, and the patterns of current flow within topological materials.

**Superconducting Devices – Qubits**

A Josephson junction acts as a phase-dependent inductor, in that by a minor re-writing of the formulas we derived previously (Eqs. (7.10) and (7.11)) the voltage across the junction takes the form $V = -L(\delta)dI/dt$ with

$$L(\delta) = \frac{\hbar}{2e}\frac{1}{I_0 cos\delta} \tag{7.17}$$

where $\delta = \phi_2 - \phi_1$ is the phase difference across the junction. This inductance is the result of the kinetic energy of the supercurrent rather than a magnetic-field energy, and it is therefore called a kinetic inductance. If a Josephson junction is connected in parallel with a capacitor using superconducting wires, the combination acts as a dissipationless LC oscillator. At low temperature, the quantum-mechanical energy levels of this oscillator are quantized. Because the inductance is phase dependent rather than simply being a constant, the energy levels turn out not to be perfectly evenly spaced as in a simple-harmonic oscillator, but to have unequal level spacings. This allows the occupation of different states and their quantum-mechanical superpositions to be initialized, controlled, and read out with somewhat amazing precision using electrical gates and applied microwave signals, thus providing all of the ingredients needed to make qubits for the development of quantum computation.

In recent years, several different designs of superconducting qubits have been studied, depending on the relative values of capacitance, inductance, Josephson current, and (in designs incorporating isolated superconducting islands) the charging energy of the island. Excellent progress has been made in making qubits insensitive to various sources of noise which can produce quantum decoherence and degradation of the quantum computation, although further improvement in this area remains a pressing research need in order to realize useful computation.

For those of you interested in details, there are nice reviews of this field given in P. Krantz et al., "A quantum engineer's guide to superconducting qubits," *Appl. Phys. Rev*. **6**, 021318 (2019) and M. Kjaergaard et al., "Superconducting Qubits: Current State of Play," *Annu. Rev. Condens. Matter Phys*. 11, 369 (2020).

**<u>Superconducting Devices – Detectors for Astronomy and Cosmology</u>**

Superconducting devices play a central role in astronomical and cosmological studies by providing sensitive detectors for photons across a wide frequency range. The detectors in use are generally of two types, transition edge sensors and kinetic inductance detectors.

Transition edge sensors consist of superconducting films controlled so that their temperature is very close to the superconducting critical temperature and their resistance is therefore in a narrow transition region between the normal-state resistance and zero. At this bias point, absorbed radiation with frequency above $2\Delta_{BCS}/\hbar$ can heat the sample very slightly, and this is enough to produce a large change in the resistance that lasts momentarily before the detector cools back down.

Kinetic inductance detectors consist of a Josephson junction combined with a capacitor to make a superconducting LC oscillator (much like certain qubit designs). Incoming radiation with frequency above $2\Delta_{BCS}/\hbar$ can break Cooper pairs in the vicinity of the Josephson junction, reducing the superconducting carrier density (and hence $I_0$) and thereby increasing the kinetic inductance momentarily. This changes the resonance frequency of the LC oscillator, which can be measured very precisely because the resonances of low-dissipation superconducting LC oscillators can have very high quality factors. Kinetic inductance detectors are attractive because they can be easily multiplexed by designing the different LC oscillators to have different oscillation frequencies, which has the potential to simplify the read-out electronics when large arrays of detectors are employed.

**<u>Unconventional Superconductors</u>**

In some materials, an attractive electron-electron interactions and associated superconductivity can arise from mechanisms different than phonon-mediated overscreening. This is a field in which much remains to be understood, but it appears that all-electronic mechanisms can play a central role (e.g., interactions mediated by antiferromagnetic fluctuations, ferromagnetic fluctuations, or fluctuations in some other electronic order parameter). The dynamics of electronic excitations are generally much faster than phonons so electronic interactions are unlikely to produce the frequency-dependent overscreening that allows phonon-mediated processes to produce attractive interactions between two electrons at the same position but separated in time. Instead, the all-electronic mechanisms can evade the short-range repulsion between like-charged electrons by generating effectively-attractive interactions between electrons separated in space (thank you to Brad Ramshaw for explaining this insight to me). A $\delta$-function interaction is therefore not an appropriate approximation anymore. The consequence for superconductivity is that an all-electronic mechanism generally does not produce spherically-symmetric *s*-wave pair wavefunctions of the type we derived for the BCS approximation (where the maximum probability for the relative-coordinate part of the wavefunction occurs for $r = 0$), but rather they give pair wavefunctions with nodes at $r = 0$ (i.e., *p*-wave, *d*-wave, etc.).

I will call superconductors "unconventional" when the pairing mechanism is not phonon mediated or the symmetry of the pair wavefunction is not a spin-singlet *s*-wave. Typically, these two aspects go together – as it would take a somewhat pathological situation for a phonon-mediated electron-electron interaction to favor a non-*s*-wave pair wavefunction. Despite the term "unconventional", many of the properties of these superconductors are still very similar to conventional superconductors. Their superconductivity can still be understood as an instability to the formation of Cooper pairs, with an energy gap to excitations (over at least most of momentum space), a Meissner effect, and zero resistance. The electrons can still be described by a single quantum-coherent macroscopic center-of-mass wavefunction governed by a Schrodinger-like equation within Ginzburg-Landau theory, and all of the various properties that follow from having such a coherent macroscopic wavefunction remain: flux quantization, fluxons, Josephson effect, etc. The main differences from conventional superconductors are in the symmetry of the pairing wavefunction, and quantitative (but not qualitative) differences for parameters such as the coherence length and penetration depth.

The formation of unconventional pair wavefunctions can be understood within a generalization of the Cooper argument. For the BCS model Hamiltonian that we used when we first discussed the Cooper argument, we assumed that the pair-scattering matrix element $V_{\vec{k}-\vec{j}}$ had the same value $= -\lambda_{BCS}$ between any two single-electron doublets $\{\vec{k}\uparrow, -\vec{k}\downarrow\}$ and $\{\vec{j}\uparrow, -\vec{j}\downarrow\}$, as long as the single-electron states involved are sufficiently close to the Fermi energy (within a cutoff $\hbar\omega_D$). But suppose that is not the case, and $V_{\vec{k}-\vec{j}}$ depends on the scattering angle between $\vec{k}$ and $\vec{j}$. In this case it is possible to expand both the interaction matrix element and the pair wavefunction in angle-dependent spherical harmonics (or alternatively, whatever irreducible representations are appropriate for a specific crystal structure). With the spherical harmonics one can use the expansions

$$V_{\vec{k}-\vec{j}} = \sum_\ell V_\ell P_\ell(cos\theta) \tag{7.18}$$
$$\varphi(\vec{r}) = \sum_{\vec{k},\ell,m} \chi_\ell(|\vec{k}|) Y_{\ell m}(\hat{k}) \tag{7.19}$$

where $P_\ell(cos\theta)$ are the Legendre polynomials and $cos\theta = \vec{k}\cdot\vec{j}$. When you substitute these expressions into the Schrodinger Equation and solve as we did for the Cooper argument, the different angular momentum components decouple. The end result is the same as we found in the Cooper argument but with a separate, decoupled equation for each angular momentum component.

$$(2\varepsilon_k - E)\chi_\ell(|\vec{k}|) = -V_\ell \sum_{k_F<|\vec{k}'|<k_D} \chi_\ell(|\vec{k}'|). \tag{7.20}$$

(Compare to Eq. (5.13).) If any of the spherical harmonic components of the interaction, $V_\ell$, are negative, corresponding to an attractive interaction in that channel, there is an instability to a superconducting state. In general, the most negative coefficient should dominate.

In non-*s*-wave superconductors, the pair wavefunction is no longer spherically symmetric in real space and the associated excitation gap is not spherically symmetric in crystal-momentum space. For example, the high-temperature superconducting cuprate materials have been shown to be quasi-2-dimensional *d*-wave superconductors, so that the relative-coordinate part of the pair wavefunction has the form of a $d_{x^2-y^2}$ orbital in real space (Fig. VII.7):

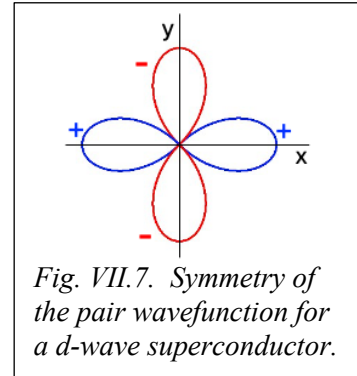

Fig. VII.7. Symmetry of the pair wavefunction for a d-wave superconductor.

$$\varphi(\vec{r}) = f(r)(cos^2\phi - sin^2\phi) \qquad (7.22)$$

If a Josephson junction is made so that it contains more than one crystalline facet, so that more than one orientation of the pair wavefunction comes into play, the extra internal angular dependence within the pair wavefunction can alter the magnetic-field response of the junction. The form of flux quantization for superconducting loops containing such junctions can also be modified (e.g., half-quantized fluxons are possible). In crystal-momentum space the excitation gap Δ has nodes where the gap goes to zero in certain directions, with important consequences for all of the experimental properties related to this gap.

*p*-wave superconducting wavefunctions are odd under the exchange of electron positions so the spin part of the pair wavefunction is necessarily a spin triplet rather than *S* = 0. Superfluid He$^3$ was long ago established as the analog of a *p*-wave superconductor, but realizations in electronic materials have been more slow in coming. Superconducting Sr$_2$RuO$_4$ was for 25 years thought to be a strong candidate for being spin triplet, but newer experiments have recently ruled this out (A. Pustogow et al., *Nature* **574**, 72 (2019); K. Ishida et al., *J. Phys. Soc. Jpn* **89**, 034712 (2020)). UTe$_2$, β-Bi$_2$Pd, and CeRh$_2$As$_2$ are currently candidate spin-triplet superconductors, and these appear to be on firmer footing.

Non-centrosymmetric crystals with significant spin-orbit coupling cannot be purely spin-singlet or spin-triplet (even when the pairing mechanism is conventional). Instead their superconducting states necessarily consist of a singlet/triplet admixture. This allows for investigation of spin effects not present in purely spin-singlet superconductors.

Some classes of unconventional superconductors can be topological. The idea is similar to our previous discussion of topology in non-superconducting insulators. Superconductivity causes an electron band to be gapped, and integration over the band gives topological invariants that can be either topologically trivial or non-trivial. The topologically non-trivial superconductors necessarily have surface states with zero energy in the middle of the superconducting gap. For proposed applications in quantum computing, the most interesting topological superconductors are those in which the surface states act as Majorana fermions – quasiparticles which are their own antiparticle. These might potentially be used for low-decoherence quantum computation by braiding – rotating the positions of individual Majoranas relative to the others. Proposals for realizing Majorana quasiparticles include the use of a time-reversal-symmetry-breaking spin-triplet $p_x + ip_y$ superconductor or a superconductor with the combination of strong Rashba spin-orbit coupling and an applied magnetic field providing a spin Zeeman interaction.

Another current research topic of significant interest is "superconducting diodes" – structures which provide a superconducting critical current that differs in magnitude depending on whether the current flows forward or backward. These would be useful for providing a directionality (i.e., "non-reciprocity") in superconducting logic circuits. To be implemented within an individual material, superconducting diode behavior requires a material with both broken inversion symmetry and broken time-reversal symmetry.

**What have we learned?:**

- Magnetic flux is quantized within any unbroken superconducting loop when that loop contains a path away from the superconducting screening currents.

- Individually-quantized units of magnetic flux called fluxons can penetrate Type-II superconductors and enable them to retain their superconductivity to higher values of applied magnetic field than Type-I superconductors. Type-I superconductors pay a high price in screening currents to keep a magnetic field from penetrating into the bulk of the superconductor.

- "Weak links" in superconducting circuits act as Josephson junctions through which a supercurrent flows with an amplitude dependent on the superconducting phase difference across the junction, $I = I_0 sin(\phi_2 - \phi_1)$. This behavior corresponds to a nonlinear inductance which allows for control of superconducting qubits.

- dc SQUIDs are superconducting loops containing two Josephson junctions. They allow for sensitive magnetic-field detection.

- Unconventional superconductors: Pairing mechanisms different from phonon-mediated electronic interactions can generate superconductors in which the symmetry of the pair wavefunction is different from a spherically-symmetric *s*-wave state. Most of the superconducting properties are still similar to conventional superconductors (Cooper pairs, Meissner effect, zero resistance, Josephson junctions, etc.), but the existence of sign changes in the pair wavefunction (for *p*-wave, *d*-wave, etc. states) can modify some properties of Josephson junctions and flux quantization, and lead to nodes in the excitation gap for certain directions of momentum space.

**Homework Problems**:

**1. The Hubbard Model and the Mott Transition**: (Adapted from Ashcroft & Mermin problem 32.5) If we limit our Hilbert space to contain only a few orbital basis states, we can explicitly carry through a form of the "thought calculation" discussed in class for determining the energy eigenstates of an interacting electron system. In this problem, we will examine the physics of half-filled electron bands (1 electron per unit cell in a crystal) and we will find that for strong enough interactions the metallic Fermi liquid state is unstable in a way that is different from either the magnetic or superconducting instabilities (the two non-Fermi liquid states that we discuss in class). In order to make things tractable, we will deal with the "Hubbard Model". This is a 1-dimensional tight-binding Hamiltonian, with the addition of an "on-site" energy cost U for two electrons to occupy the same site. That is, on account of screening, we make a crude approximation that the Coulomb repulsion between electrons can be ignored unless two electrons occupy the same cell in our tight binding chain. Explicitly, the Hamiltonian we will use is
$$H = H_{TB} + H_c.$$
If we label spatial sites in a (one-dimensional) tight binding chain by the index n, and the electron spin by the index s, we can write $H_{TB}$ either in terms of bras and kets:
$$H_{TB} = -t \sum_{n,s} \left[ |n+1,s\rangle\langle n,s| + |n,s\rangle\langle n+1,s| \right]$$
or (equivalently) in second-quantized notation
$$H_{TB} = -t \sum_{n,s} \left[ c^{\dagger}_{n+1,s} c_{n,s} + c^{\dagger}_{n,s} c_{n+1,s} \right].$$
The hopping matrix element, *t*, gives rise to a kinetic energy for the electrons that is lowered if the wavefunction delocalizes into long-wavelength plane waves. The electron spin does not flip during a hopping transition.
The Coulomb energy cost for double-occupancy of a position state has the form
$$H_c = U \sum_n c^{\dagger}_{n,\uparrow} c_{n,\uparrow} c^{\dagger}_{n,\downarrow} c_{n,\downarrow}.$$
Recall that the creation and annihilation operators for fermions obey anticommutation relationships:
$$\{c_{n,s}, c^{\dagger}_{n',s'}\} = \delta_{n,n'} \delta_{s,s'}, \quad \{c^{\dagger}_{n,s}, c^{\dagger}_{n',s'}\} = \{c_{n,s}, c_{n',s'}\} = 0.$$

(a) We will begin by considering a two-site chain, n=1,2 (without periodic boundary conditions). Considering for the moment only $H_{TB}$ (set *U*=0), show that the single-electron eigenstates of the problem can be written as
$$c^{\dagger}_{S,s}|0\rangle \text{ and } c^{\dagger}_{A,s}|0\rangle, \text{ where } c^{\dagger}_{S,s} = \left(c^{\dagger}_{1,s} + c^{\dagger}_{2,s}\right)/\sqrt{2} \text{ and } c^{\dagger}_{A,s} = \left(c^{\dagger}_{1,s} - c^{\dagger}_{2,s}\right)/\sqrt{2},$$
and where $|0\rangle$ denotes the vacuum state. (You have done this many times before, using a slightly different language.) Note that the eigenstates created by $c^{\dagger}_{S,s}$ and $c^{\dagger}_{A,s}$ have the Bloch form, with *k*=0 and π/*a*, respectively, if the lattice constant is *a*. There are 6 possible 2-electron Slater-determinant states which can be formed from these single-electron states. List them and the energy of each (still assuming U=0).
(Note that using second-quantized notation makes this task very easy -- for example, the lowest-energy 2-electron Slater-determinant eigenstate of $H_{TB}$ can be written $c^{\dagger}_{S,\uparrow} c^{\dagger}_{S,\downarrow}|0\rangle$, with energy -2t.)

(b) Now consider only the Coulomb term, without $H_{TB}$ (assume $t=0$, but $U \neq 0$.) With no hopping, the ordinary position states are energy eigenstates. By making Slater determinants from the position states (i.e., with $c^\dagger_{1,s}$ and $c^\dagger_{2,s}$), find the six 2-electron eigenstates of $H_c$, and list their energies. Note that the eigenstates of $H_{TB}$ and $H_c$ are not the same.

(c) Let's begin the process of finding the energy levels and eigenstates of the full Hamiltonian for two interacting electrons. Working in the basis of the 2-electron Slater-determinants formed from $c^\dagger_{S,s}$ and $c^\dagger_{A,s}$ (not $c^\dagger_{1,s}$ and $c^\dagger_{2,s}$!), write down the 6 × 6 Hamiltonian matrix including the contributions from both $H_{TB}$ and $H_c$. (The contributions from $H_{TB}$ are trivial from part (a). For $H_c$, I found the easiest method was to solve for $c_{1,s}$, $c_{2,s}$, $c^\dagger_{1,s}$ and $c^\dagger_{2,s}$ in terms of $c_{S,s}$, $c_{A,s}$, $c^\dagger_{S,s}$ and $c^\dagger_{A,s}$, and then rewrite $H_c$ as a sum over products of the symmetric and antisymmetric state operators rather than the position state operators. There will be a bit of algebra, and a lot of terms. But the non-zero matrix elements between Slater determinant states can then be read off directly.)

(d) The energy expectation value of the 2-electron ground state of the non-interacting problem ($c^\dagger_{S,\uparrow} c^\dagger_{S,\downarrow} |0\rangle$) becomes $(U/2)-2t$ under the influence of $H_{TB} + H_c$. (This should be one of your matrix elements.) Why is there a Coulomb energy cost $U/2$ associated with this state? Show the reason explicitly by writing out this state in terms of the positions states 1 and 2, and determining the probability that both electrons may be found at the same position. Because of this energy cost, the ground state of the full Hubbard Hamiltonian may be very different from the ground state of just $H_{TB}$.

(e) Diagonalize your 6 by 6 Hamiltonian matrix from part (c). Show that the 6 energy eigenvalues are

$$\frac{U}{2} - \sqrt{\frac{U^2}{4} + 4t^2},\ 0,\ 0,\ 0,\ U,\ \frac{U}{2} + \sqrt{\frac{U^2}{4} + 4t^2}.$$

(Check: are the two limits $t \ll U$ and $t \gg U$ correct?)

(f) In the limit of a large repulsive $U$ (i.e., $U \gg t$), write out the ground eigenstate in terms of the position states 1 and 2 and explain how nature conspires to avoid paying the Coulomb charging energy cost.

(g) Consider also the somewhat non-intuitive case of a large attractive (i.e., negative) $U$, with $|U| \gg t$. Express the ground state wavefunction in both (i) the basis of symmetric and antisymmetric states and (ii) in the basis of the individual position states. In both cases you should find that the ground state has the form of a superposition of Slater determinant states in which spin-up and spin-down electrons always occur in pairs, occupying the same orbital states. These types of "electron-pair" wavefunctions are the key to understanding superconductivity, which comes about when there is effectively an attractive electron-electron interaction.

**Applications** (think about these ideas, but do not hand in)

**(i)** You will note that the ground state listed in (e) is non-degenerate, while the second energy levels are 3-fold degenerate. This can be understood in terms of the total spin of the eigenstates: The ground state is a spin singlet and the second energy levels are an $S=1$ triplet. In fact, in the regime where $U \gg t$, where we can think of each orbital site as occupied by one electron, the lowest 4 energy levels of the system can be described by an effective Hamiltonian of the Heisenberg form:

$$H_{eff} = E_0 + J\vec{S}_1 \bullet \vec{S}_2$$

where $E_0$ is a constant energy offset, $S_1$ and $S_2$ are the electron spins at the two sites, and $J$ is the "exchange constant". In the limit $U \gg t$ the exchange constant is $4t^2/(U\hbar^2)$ and is positive, so that to minimize the energy the effective spin interaction leads to antiferromagnetic spin alignment. If you feel like working this out yourself, it is convenient to express the operator $\vec{S}_1 \bullet \vec{S}_2$ either as $\vec{S}_1 \bullet \vec{S}_2 = (\vec{S}_1 + \vec{S}_2)^2 - \vec{S}_1^2 - \vec{S}_2^2$ or $\vec{S}_1 \bullet \vec{S}_2 = S_{1,z}S_{2,z} + \frac{1}{2}(S_{1,+}S_{2,-} + S_{1,-}S_{2,+})$.)

**(ii)** It is not much more difficult to think of generalizing this 2-state problem to a longer tight-binding chain, with $2M$ electrons on $2M$ sites, with periodic boundary conditions. (Going even to 3 dimensions is possible.) Ignoring $H_c$, the non-interacting electron eigenstates are the Bloch states, associated with the creation operators

$$\psi^{\dagger}_{k,s} = \frac{1}{\sqrt{2M}}\sum_n e^{inak}c^{\dagger}_{n,s}.$$

The ground state for $2M$ non-interacting electrons is formed by making a Slater determinant of the lowest-energy $M$ Bloch states, each occupied with spin up and spin down:

$$\Psi_{ind} = \psi^{\dagger}_{k_0\uparrow}\psi^{\dagger}_{k_0\downarrow}\psi^{\dagger}_{k_1\uparrow}\psi^{\dagger}_{k_1\downarrow}...|0\rangle.$$

The band structure for this non-interacting problem is just the usual 1-D tight binding form $\varepsilon(k) = -2t\cos(ka)$. The minimum energy required to excite an electron from the ground state and generate a current carrying state is $-2t\{\cos[(\pi/2)+(2\pi a/2Ma)] - \cos(\pi/2)\} = 2t\sin(\pi/M)$. For large $M$, this is a very small fraction of the bandwidth ($4t$), as we expect for a good metal. However, if one expands out this independent-electron ground state in terms of the position states, it is clear (just as in the 2-state case) that it involves configurations in which position states are doubly occupied. A somewhat messy calculation shows that $\langle \Psi_{ind}|H_c|\Psi_{ind}\rangle \propto MU$. This means that if $U \gg t$, the independent-electron ground state has quite a high energy ($> 0$), and it is more advantageous for the $2M$ electrons to exist in a different Slater determinant state of the form

$$\Psi_{Mott} = c^{\dagger}_{1,s_1}c^{\dagger}_{2,s_2}...c^{\dagger}_{2M,s_{2M}}|0\rangle,$$

which avoids double occupancy. It is easy to see that $\Psi_{Mott}$ has a total energy expectation value $= 0$, and degeneracy $2^{2M}$ due to spin. An analysis of off-diagonal matrix elements indicates that the true ground state has an antiferromagnetic arrangement of the spins. The neat thing is that $\Psi_{Mott}$ is an insulator. (In fact, an antiferromagnetic insulator, like the high-$T_c$ superconductor compounds in the absence of doping.) In order to form a current carrying state, an electron must be excited from one position state to another, but all the others are already occupied, so this costs at least $U$ in energy. When $U \gg t$, this is a prohibitive amount. Therefore we have a situation in which there is a metal insulator-transition (the "Mott Transition") as the interaction strength $U$ is varied with respect to *t*. For *U/t* small, the independent-electron ground state is a good approximation to the true ground state, and we have a metal. However, for *U/t* large, the localized-electron (and therefore insulating) Mott state is the more correct description.

**2. Magnetic Domain Walls.** Consider a 1-dimensional model of a ferromagnetic domain wall as a function of position like the one drawn:

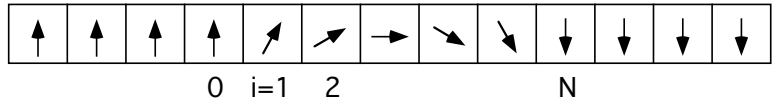

Spins to the left and right are aligned ferromagnetically in opposite directions, and the domain wall consists of a uniform rotation by an angle $\pi$ over a fixed number N of discretized volume elements, so that the angle between the magnetic moments of neighboring volume elements is $\theta = \pi/N$.

**(a)** Suppose that the exchange energy of this chain is modeled by a Heisenberg Hamiltonian between nearest-neighbor spins of the form

$$H_{ex} = -J \sum_i \vec{S}_i \cdot \vec{S}_{i+1}$$

(with $J > 0$ for a ferromagnetic interaction). Treating the moments as classical spins of fixed magnitude $S$, what is the excess exchange energy associated with the presence of the domain wall of width $N$, relative to the case of ferromagnetic alignment for all the spins? (Be careful to subtract off the energy corresponding to fully aligned spins. Do not assume that $N \gg 1$ yet.)

**(b)** Let the z-direction be the direction that the left-most spin is pointing in the figure. Suppose that the spins have an easy axis in this z-direction, so that there is a magnetocrystalline anisotropy energy which takes the form:

$$H_{anis} = -K \sum_i \left( \vec{S}_i \cdot \hat{z} \right)^2,$$

and favors alignment along $+\hat{z}$ or $-\hat{z}$ for $K > 0$, with higher energies for other orientations. What is the excess anisotropy energy associated with the presence of the domain wall of width $N$, relative to the case of ferromagnetic alignment for all the spins? (You may assume here that $N \gg 1$ and freely convert any sums to integrals. Again, be careful to subtract off the energy for fully aligned spins.)

**(c)** Assuming that the magnetic boundary conditions force a domain wall to exist, and ignoring possible contributions from magnetic dipole-dipole energy, does the domain wall have an optimum width ($N$)? If so, determine the value. You may assume here that $N \gg 1$.

(Note for experts: You can get an even-better optimized energy by allowing $d\theta/dx$ to be greater near the middle of the domain wall (where there is the greatest cost in anisotropy energy) compared to near the edges, rather than having a constant value as we have assumed here. A better approximation for the functional form of the domain wall is $S_z = -\tanh((x-x_0)/w)$, where $x_0$ is the position of the center of the domain wall and $w$ is the domain-wall width.)

**3. Switching of a single-domain magnetic particle**. (Adapted from Kittel ch. 15, problem 5) Consider a small spherical single-domain particle of a uniaxial ferromagnet, such as cobalt. Assume that the magnitude M of the total magnetization vector $\vec{M}$ stays constant in the presence of applied magnetic fields, but the direction of the vector can switch or rotate. Let the anisotropy energy be $U_{anis} = (\beta M^2 \sin^2\theta)/2$, where $\theta$ is the angle between the magnetization vector and the easy axis (which is in the z-direction), and $\beta$ is a positive constant. Let $U_{field} = -\vec{H} \cdot \vec{M} = -H_z M \cos\theta - H_x M \sin\theta$ be the interaction energy with the external field H. Define $U_{total} = U_{anis} + U_{field}$.

(a) Assume first that $\vec{M}$ is aligned with an easy axis and a magnetic field directed in the opposite direction is gradually increased in strength. Show that the reversed field required to switch the magnetization is $H_{sw} = \beta M$, in the units being used. (There is a further hint in Kittel.) Plot the magnetization curve $M\cos\theta$ vs. H for a field parallel to the easy axis (for both increasing and decreasing H, extending over a range larger than from $-\beta M$ to $\beta M$), to exhibit the presence or absence of hysteresis.

(b) For the case of simple uniaxial anisotropy described above, determine the values of the switching field for an arbitrary angle between the applied field and the easy axis (that is, determine at what sets of coordinates $H_x$ and $H_z$ switching will occur). The problem has cylindrical symmetry around the easy axis, so there is no need to worry about the y-direction. From part (a) you should understand that the switching field occurs when the local minimum of the magnetic energy (a value of $\theta$ where $dU_{total}/d\theta = 0$) becomes an inflection point (meaning that $d^2 U_{total}/d\theta^2 = 0$, as well). [Hint: you can write the first condition in the form

$$\left[ \frac{H_x}{\sin\theta} - \frac{H_z}{\cos\theta} - \beta M \right] (\text{angular factors}) = 0,$$

where the "angular factors" will not matter in the 2nd condition. From the second condition, you can then eliminate $\sin\theta$, leaving yourself with a simple expression relating $H_x$, $H_z$, and $\beta M$ at the switching condition.] Plot out this curve for all possible angles (both positive and negative values of $H_x$ and $H_z$) – it is known as the "Stoner-Wohlfarth astroid".

**4. Bloch's Theorem and Spin Waves**. The low-energy electronic excited states of ferromagnets include spin wave excitations in addition to the particle-hole excitations present in Fermi-liquid (non-magnetic) metals. As a simple model of spin waves, consider a 1-dimensional chain of $N$ spins each with magnitude $S$, subject to periodic boundary conditions, in which nearest-neighbor spins interact in such a way as to favor parallel alignment. One way to model this is with a simple "Heisenberg" dot-product Hamiltonian of the form

$$H = -J \sum_i \vec{S}_i \cdot \vec{S}_{i+1} = -J \sum_i \left[ S_i^z S_{i+1}^z + \frac{1}{2}\left( S_i^+ S_{i+1}^- + S_i^- S_{i+1}^+ \right) \right].$$

where $S_i^+ = S_i^x + iS_i^y$ and $S_i^- = S_i^x - iS_i^y$ are the spin raising and lowering operators at site i on the chain -- which act to change the eigenvalue of $S_i^z$ by $\pm 1$. Let the ferromagnetic ground state, with all spins aligned "up" be called $|F\rangle$ (so that $S_i^z|F\rangle = +S|F\rangle$, where S without indices is a number, not an operator). You may use the identities that $S_i^+|F\rangle = 0$ and $S_i^+ S_i^-|F\rangle = 2S|F\rangle$ for all sites i, and that spin operators corresponding to different sites commute.

(a) For a one-dimensional chain with $N$ sites and periodic boundary conditions, show that the state $|F\rangle$ is an eigenstate of the Hamiltonian with energy $-NJS^2$.

(b) Next consider states in which the ferromagnetic ground state has been altered by decreasing $S_z$ of the spin at site $j$ by one unit (so that the total z-component of spin for the system is NS-1). With the proper normalization factor, you can write this state in the form $|j\rangle \equiv (1/\sqrt{2S})S_j^-|F\rangle$. This is not an eigenstate of H. Show that for the Heisenberg Hamiltonian,
$$H|j\rangle = [-NJS^2 + 2JS]|j\rangle - SJ|j-1\rangle - SJ|j+1\rangle. \tag{1}$$

(c) Using Bloch's Theorem, we can write down immediately the correct form of the energy eigenstates with total z-component of spin = NS-1. The true energy eigenstates can be written in a plane wave form very analogous to phonons or a tight-binding electron eigenstate:
$$\Psi_k = \sum_j e^{ikja}|j\rangle$$
Since each state $|j\rangle$ has a total value of $S_z$=NS-1, then so does $\Psi_k$.

Using a wavefunction of this form and Eq. (1), calculate the spin-wave dispersion curve -- the difference between the energy eigenvalue of the Bloch state with wave vector $k$ and the ground state energy $-NJS^2$. Compare/contrast this spin-wave dispersion curve with a typical 1-d acoustic phonon dispersion curve, including the power-law of the dispersion curve for small $|k|$. (Hint: In evaluating $H\Psi_k$, it will help to relabel the counting index in some of your sums so that in all the sums you are summing over $|j\rangle$ rather than $|j+1\rangle$ or $|j-1\rangle$.)

(d) For 3-d metallic ferromagnet, the dispersion curve will scale the same way as a function of $|k|$ as your answer in part (c), and each spin wave will have boson statistics, contributing an average energy $\approx k_B T$ for each mode with $\hbar\omega(k) < k_B T$ (so that it is not frozen out at low temperature). Given this information, rank the size of the low-temperature contributions to the specific heat (from smallest to biggest) for a 3-d metallic ferromagnet due to electronic (electron-hole) excitations, spin-waves, and phonons. Explain.

### 5. Effects on the Cooper argument of blocking a single-electron doublet from participating in the pair scattering.

Recall the Cooper argument: that if there exist two electrons excited above a frozen Fermi sea and they are subject to an attractive interaction, then it is possible for them to form a bound state with energy $E = 2\varepsilon_F - 2\Delta_{\text{Coop}}$ which is less than their starting independent-electron energy $2\varepsilon_F$, where $2\Delta_{\text{Coop}}$ is the binding energy (see Eq. (5.16))

$$1 = \lambda_{BCS} \sum_{\varepsilon_F < \varepsilon_k < \varepsilon_F + \hbar\omega_D} \frac{1}{2\varepsilon_k - (2\varepsilon_F - 2\Delta_{\text{Coop}})}$$

Here $\lambda_{BCS}$ (> 0) is the magnitude of the attractive electron-electron interaction and $\hbar\omega_D$ is an energy cut-off beyond which the interaction ceases to be attractive. In this problem we will assume for simplicity that the single-electron doublets are equally spaced with a single-electron level spacing $\delta$ and we will label the doublets as indicated in the figure to the right.

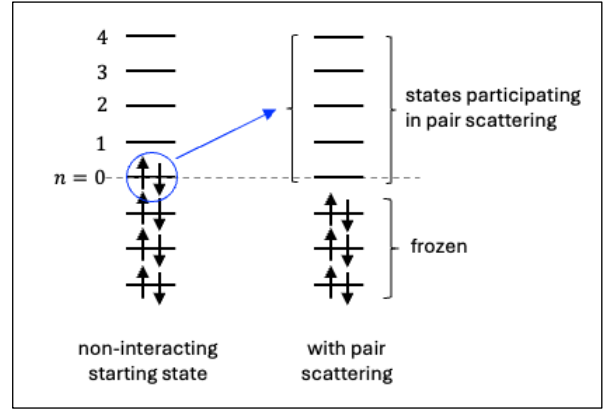

(a) Let's first check the Cooper argument numerically for a particle with an even number of electrons. Assume the electrons in all energy levels lower than the one I have marked as $n = 0$ are "frozen." In the Fermi liquid state, the highest-energy pair of electrons occupy the single-electron doublet labeled $n = 0$, and we'll call the total energy of these two electrons $2\varepsilon_0$. In the Cooper argument, these two electrons can use the $n = 0$ doublet and all the higher-lying doublets for pair scattering. These participating doublets have single-electron energies $\varepsilon_0 + n\delta$. The equation for the binding energy becomes

$$1 = \frac{\lambda_{BCS}}{2\delta} \sum_{n=0}^{\frac{\hbar\omega_D}{\delta}} \frac{1}{n + \frac{\Delta_{Coop}}{\delta}}$$

Assume the parameters $\frac{\lambda_{BCS}}{2\delta} = 0.25$ and $\frac{\hbar\omega_D}{\delta} = 120$. Evaluate the finite sum and determine the positive-valued solution for $\frac{\Delta_{Coop}}{\delta}$ numerically. Compare to the analytic solution in which the sum is approximated by an integral.

(b) Imagine that the $n = 1$ level is blocked from participating in the pair scattering because it is occupied by a single unpaired electron level (as it would be if the metal particle were occupied by an odd number of electrons). The energy doublets available for pair scattering in a Cooper-like argument are now n = 0, 2, 3, 4,… (but not $n = 1$). By evaluating the sum above numerically excluding the $n = 1$ term, determine the how much the Cooper-pair binding energy is reduced by having one single-electron doublet blocked from the pair scattering.

(c) Is it optimal to keep the unpaired single-electron in the $n = 1$ level, or would it be better to block the $n = 0$ or $n = 2$ levels from the pair-scattering, instead? Be sure to consider both the single-electron energies and the pair-binding energies when comparing these different possible ground states.

**Acknowledgments:**

These notes are shared as a contribution to the broader impact activities of NSF grant DMR-2104268. I am grateful to the many years of students in Phys. 7635 for their contributions to the course. I also thank Chuck Black, Jan von Delft, and Brad Ramshaw for discussions and insights.